\documentclass[11pt,a4paper]{article}
\pdfoutput=1
\usepackage{jheppub}
\usepackage{amsmath,amssymb,mathtools}

\usepackage{xcolor}

\newcommand{\zt}{{\tilde{z}}}
\newcommand{\Mt}{\widetilde{M}}

\newcommand{\sigg}{\Sigma_{\mathfrak{g}}}
\newcommand{\mg}{\mathfrak{g}}
\newcommand{\mfrk}{\mathfrak}

\usepackage{tensor}
\newcommand{\ind}[1]{\indices{#1}}

\newcommand{\wt}[1]{\widetilde{#1}}

\title{Euclidean Black Saddles and AdS$_4$ Black Holes}
\author{Nikolay Bobev,}
	\author{Anthony M. Charles,}
\author{and Vincent S. Min}

\affiliation{Institute for Theoretical Physics, KU Leuven, \\
Celestijnenlaan 200D, B-3001 Leuven, Belgium}

\emailAdd{nikolay.bobev@kuleuven.be}
\emailAdd{anthony.charles@kuleuven.be}
\emailAdd{vincent.min@kuleuven.be}

\abstract{We find new asymptotically locally AdS$_4$ Euclidean supersymmetric solutions of the STU model in four-dimensional gauged supergravity. These ``black saddles'' have an $S^1\times \Sigma_{\mathfrak{g}}$ boundary at asymptotic infinity and cap off smoothly in the interior. The solutions can be uplifted to eleven dimensions and are holographically dual to the topologically twisted ABJM theory on $S^1\times \Sigma_{\mathfrak{g}}$. We show explicitly that the on-shell action of the black saddle solutions agrees exactly with the topologically twisted index of the ABJM theory in the planar limit for general values of the magnetic fluxes, flavor fugacities, and real masses. This agreement relies on a careful holographic renormalization analysis combined with a novel UV/IR holographic relation between supergravity parameters and field theory sources. The Euclidean black saddle solution space contains special points that can be Wick-rotated to regular Lorentzian supergravity backgrounds that correspond to the well-known supersymmetric dyonic AdS$_4$ black holes in the STU model.}

\begin{document}

\maketitle

\section{Introduction}
\label{sec:Introduction}

A fundamental entry in the holographic dictionary is the map between the on-shell action of an asymptotically locally AdS gravitational solution and the path integral of the dual quantum field theory. This relation offers insights both in the physics of strongly coupled quantum field theory as well as the structure of quantum gravity. Testing precisely this duality between gravitational on-shell actions and quantum field theory path integrals in the context of string and M-theory is facilitated by the rapid developments in supersymmetric localization for strongly coupled gauge theories. Supersymmetric localization can often be used to reduce the path integral of a QFT on a compact curved manifold to a matrix integral, see \cite{Pestun:2016zxk} for a review. One can then use matrix model techniques to solve these integrals in the planar limit of the gauge theory. This in turn leads to a panoply of exact results for QFTs with a holographic dual description that can be used either as precision tests of the AdS/CFT correspondence or as a new window into the structure of supergravity and string theory. 

Of central interest in this work is the path integral of a partially topologically twisted three-dimensional $\mathcal{N}=2$ SCFT placed on $S^1\times \Sigma_{\mathfrak{g}}$ where $\Sigma_{\mathfrak{g}}$ is a compact Riemann surface of genus $\mathfrak{g}$ and the $S^1$ has radius $\beta$. This partition function is often referred to as a \textit{topologically twisted index} and was computed by supersymmetric localization for a broad class of $\mathcal{N}=2$ QFTs in \cite{Benini:2015noa}, see also \cite{Benini:2016hjo,Closset:2016arn}. This topologically twisted index, $Z_\text{CFT}(u,\mathfrak{p})$, preserves two real supercharges and depends on the parameters $u$ and $\mfrk{p}$ associated to the flavor symmetries of the $\mathcal{N}=2$ SCFT. The real constants, collectively denoted by $\mathfrak{p}$, specify quantized magnetic fluxes on $\Sigma_{\mathfrak{g}}$ for the background gauge fields that couple to the flavor symmetry of the SCFT. The parameters $u = \Delta + i\beta \sigma $ are complex and are determined by the ``real masses'' $\sigma$ and the electric chemical potentials $\Delta$, which can also be thought of as Wilson lines for the Cartan subalgebra of the flavor symmetry. To study the topologically twisted index in a holographic setting we focus on the ABJM theory which has a well-known holographic dual in terms of the AdS$_4\times S^7/\mathbb{Z}_k$ solution of M-theory \cite{Aharony:2008ug}. It was shown in \cite{Benini:2015eyy} that in the large $N$ limit the topologically twisted index of this theory at Chern-Simons level $k=1$ takes the form
\begin{equation}
	\log Z_\text{CFT}(u^I, \mathfrak{p}^I) = \frac{\sqrt{2} N^{3/2}}{3} \sqrt{u^0 u^1 u^2 u^3} \sum_I \frac{\mathfrak{p}^I}{u^I}\,,
\label{eq:Ztopintro}
\end{equation}
where $I=0,1,2,3$ labels the Cartan generators of the global symmetry algebra. 

A natural question in the context of holography is to find a supergravity dual solution to the deformed ABJM theory on $S^1\times \Sigma_{\mathfrak{g}}$ and compute the partition function \eqref{eq:Ztopintro} in terms of its on-shell action. This question was studied in detail in \cite{Benini:2015eyy} where it was shown that the topologically twisted index \eqref{eq:Ztopintro} accounts for the Bekenstein-Hawking entropy of the  supersymmetric asymptotically AdS$_4$ static black hole solutions of gauged supergravity found in \cite{Cacciatori:2009iz}.\footnote{See  \cite{Zaffaroni:2019dhb} for a review, a comprehensive account of many of the recent developments, and a more complete list of references.} In order to achieve this feat for a given black hole one has to fix the magnetic charges $\mathfrak{p}^I$ and extremize the index in \eqref{eq:Ztopintro} over the parameters $u^{I}$.\footnote{If the black hole is dyonic one also has to perform a Legendre transformation of the topologically twisted index with respect to $u^{I}$, before extremization, in order to introduce the electric charges $\mathfrak{q}^I$ \cite{Benini:2016rke}.} This $\mathcal{I}$-extremization procedure is associated with the emergence of the AdS$_2$ geometry in the near-horizon region of the supersymmetric black hole and the so-called attractor mechanism for the supergravity scalar fields \cite{DallAgata:2010ejj,Hristov:2010ri}. From the perspective of the ABJM theory on $S^1\times \Sigma_{\mathfrak{g}}$ it can also be understood as a prescription to fix the unique superconformal R-symmetry of the effective one-dimensional quantum mechanics on $S^1$ which controls the dynamics of the theory at energies much lower than those set by the scale of $\Sigma_{\mathfrak{g}}$. 

While the results of \cite{Benini:2015eyy,Benini:2016rke} and the subsequent developments on the interplay between the topologically twisted index and holography summarized in \cite{Zaffaroni:2019dhb} offer important insights into the microscopic physics of AdS$_4$ black holes, there are still several open questions related to the holographic description of the topologically twisted index in \eqref{eq:Ztopintro}. Our goal in this paper is to shed light on these questions and construct supergravity solutions which are the holographic dual of the partially twisted ABJM theory on $S^1\times \Sigma_{\mathfrak{g}}$ for \textit{arbitrary} values of the deformations parameters $(u,\mathfrak{p})$ and whose regularized on-shell action agrees with the topologically twisted index \eqref{eq:Ztopintro}. In particular, our supergravity solutions exist for values of the deformation parameters $u^{I}$ which do not extremize the topologically twisted index. Thus, these ``black saddle'' geometries can be viewed as a vast generalization of the AdS$_4$ black hole solutions in \cite{Cacciatori:2009iz}.

We construct the supersymmetric black saddle solutions in the STU model of four-dimensional gauged supergravity. This theory arises as a consistent truncation of eleven-dimensional supergravity on $S^7$ and therefore the solutions we construct can be explicitly embedded in M-theory \cite{Azizi:2016noi}. For generic values of the parameters the black saddles are Euclidean solutions and do not admit an analytic continuation to regular Lorentzian backgrounds. To construct these Euclidean solutions we follow the procedure utilized in \cite{Freedman:2013oja}, see also \cite{Bobev:2013cja,Bobev:2016nua,Bobev:2018wbt}, to study Euclidean supergravity backgrounds in a holographic context. Namely, we start with the supersymmetry variations of the Lorentzian supergravity theory and a suitable ansatz for the bosonic fields. We then analytically continue this setup to Euclidean signature and solve the resulting BPS equations.\footnote{ We comment on the relation between Euclidean supergravity solutions constructed by this method and the Euclidean supergravity approach proposed in \cite{deWit:2017cle} in Section~\ref{sec:Supergravity}.} The black saddles, as depicted in Figure~\ref{fig:h2vsr2}, are asymptotically locally $\mathbb{H}^4$ (i.e. Euclidean AdS$_4$) solutions with $S^1 \times \Sigma_{\mathfrak{g}}$ boundary geometry. The metric on the compact Riemann surface $\Sigma_{\mathfrak{g}}$ is of constant curvature and the size of the $S^1$ is finite. In the bulk of the four-dimensional space the solutions are smooth with a regular metric in the infrared (IR) region where the $S^1$ shrinks to zero size and the geometry is $\mathbb{R}^2\times \Sigma_{\mathfrak{g}}$. In addition to the metric, the solutions have non-trivial profiles for the $U(1)$ gauge fields and the scalars in the supergravity theory. The asymptotic values of these bosonic fields near the AdS$_4$ boundary are related to the deformation parameters $(u,\mathfrak{p})$ in the topologically twisted index. In the IR region of the geometry the scalar fields are not determined uniquely which in turn provides the necessary free parameters in the black saddle solutions to account for the general values of $u^{I}$ in the topologically twisted index. The black saddles are smooth solutions with generically finite values of the size $\beta$ of the $S^1$ in the ultraviolet (UV) region of the geometry. In the limit $\beta \to \infty $ the supergravity background takes a special form which admits analytic continuation to Lorentzian signature and the black saddles reduce to the supersymmetric black hole solutions in \cite{Cacciatori:2009iz}. The BPS equations for the general Euclidean black saddle solutions do not admit analytic solutions and we resort to analyzing them explicitly in several limits as well as using numerical techniques.

To establish that the black saddle supergravity solutions provide the holographic dual description of the topologically twisted ABJM theory on $S^1 \times \Sigma_{\mathfrak{g}}$ for general values of $(u,\mathfrak{p})$ we show that their regularized on-shell action agrees with the supersymmetric localization result in \eqref{eq:Ztopintro}. Importantly, we are able to evaluate this on-shell action without the need for explicit analytic black saddle solutions. This is possible because the on-shell action reduces to a total derivative and thus receives contributions only from the asymptotic UV and IR regions of the geometry, which in turn can be analyzed explicitly. 

In the process of evaluating the black saddle on-shell action we need to address three important subtleties. First, we show that the on-shell action receives non-trivial contributions from the IR region of the geometry for general finite values of $\beta$. This calculation can be viewed as providing a supersymmetric regularization for the on-shell action of the black hole solutions \cite{Cacciatori:2009iz}. This in turn resolves the subtleties associated to the contribution from the near-horizon AdS$_2$ region in the on-shell action of extremal black holes, see for example \cite{Halmagyi:2017hmw,Cabo-Bizet:2017xdr}. Second, in addition to the standard infinite counterterms that render the on-shell action finite in the holographic renormalization procedure, we also need to add finite counterterms built out of the scalar fields in the STU model. Some of these finite counterterms were studied previously in \cite{Freedman:2013oja,Freedman:2016yue,Bobev:2018wbt} and we use these results to fix the counterterm coefficients. We furthermore exhibit the utility of the finite counterterms by deriving explicitly a supersymmetric Ward identity in the ABJM theory using the holographic dictionary and black saddle on-shell action. Third, after evaluating the finite on-shell action we need to take into account the alternate quantization needed for the scalars of the STU model by performing an appropriate Legendre transformation. This is again similar to previous results in a different holographic context \cite{Freedman:2013oja,Freedman:2016yue,Bobev:2018wbt}. The end result of this careful application of holographic renormalization is that the on-shell action of the black saddle supergravity solutions agrees \textit{precisely} with the supersymmetric result for the topologically twisted index of the ABJM theory in \eqref{eq:Ztopintro}. 

The outline of this paper is as follows.  In Section~\ref{sec:BlackSaddles} we summarize the supersymmetric localization results for the topologically twisted index of the ABJM theory on $S^1\times \Sigma_{\mathfrak{g}}$ and present the simplest black saddle solutions. In Section~\ref{sec:Supergravity} we discuss general Euclidean black saddles in the STU model of four-dimensional gauged supergravity and present in detail the calculation of their on-shell action. We exhibit several explicit analytical and numerical black saddle solutions in Section~\ref{sec:solutions}. A discussion of our results and some avenues for future work is presented in Section~\ref{sec:discussions}. In the appendices we summarize our supergravity notation and conventions and present the details of the derivation of the BPS equations and the calculation of the on-shell action of the black saddle solutions.

\section{The topologically twisted index and black saddles}
\label{sec:BlackSaddles}

The goal of this section is to introduce the main characters in our story: the topologically twisted index of three-dimensional $\mathcal{N}=2$ SCFTs and the Euclidean supergravity solutions which we call black saddles. To be concrete we focus on the particular example of the ABJM theory \cite{Aharony:2008ug} and a simple supersymmetric gravitational solution constructed in \cite{Romans:1991nq}, see also~\cite{Caldarelli:1998hg}.

\subsection{Field theory}
\label{sec:fieldtheorypartfunc}

We consider the ABJM theory which describes the low-energy physics of $N$ M2-branes probing a $\mathbb{C}^4/\mathbb{Z}_k$ transverse space in M-theory. This is a Chern-Simons matter theory with $U(N)_{k}\times U(N)_{-k}$ gauge group and four bi-fundamental chiral multiplets. Here the integer $k$ is the Chern-Simons level and while many of our results generalize to general values of $k$ we fix $k=1$ in this paper. The theory with $k=1$ preserves $\mathcal{N}=8$ supersymmetry and has an $SO(8)$ R-symmetry. It is often useful to formulate the theory in the language of $\mathcal{N}=2$ supersymmetry, where only $SU(2) \times SU(2) \times U(1)_b \times U(1)_R$ subgroup of the global symmetry is manifest. Here $U(1)_R$ is the $\mathcal{N}=2$ superconformal R-symmetry and $SU(2) \times SU(2) \times U(1)_b$ plays the role of a flavor symmetry.

We are interested in placing the theory on the compact manifold $S^1 \times \Sigma_{\mathfrak{g}}$ with metric
\begin{equation}\label{eq:metQFT}
ds^2 = \beta^2 d\tau^2+ds_{\Sigma_\mg}^2~,
\end{equation}
where $\tau \sim \tau+1$, $\beta$ determines the circumference of the $S^1$, and $\Sigma_{\mathfrak{g}}$ is the constant curvature metric on a smooth compact Riemann surface with Ricci scalar $\kappa=-1$ for $\mg>1$, $\kappa=0$ for $\mg=1$, and $\kappa=1$ for $\mg=0$. In order to preserve supersymmetry we perform a partial topological twist by turning on a background gauge field for the $U(1)_R$ R-symmetry. We are also free to deform the theory further, while still preserving supersymmetry, by turning on non-trivial expectations values for the fields in the background vector multiplets that couple to the flavor currents $J_F^\alpha$ ($\alpha=1,2,3$) in the Cartan subalgebra of the flavor symmetry. These additional deformation parameters are the real masses $\sigma^\alpha_F$, the electric chemical potentials $\Delta^\alpha_F$, and the background magnetic fluxes $\mathfrak{p}^\alpha_F$. For general values of these parameters the deformed Lagrangian of the ABJM theory preserves two real supercharges. We note that supersymmetry fixes the values of the background magnetic flux and the chemical potential for the $U(1)_R$ symmetry as follows:
\begin{equation}\begin{aligned}
	\mathfrak{p}_R &= \mathfrak{g}-1~, \\
	\Delta_R &= n\pi~, \quad n \in \mathbb{Z}~.
\label{eq:toptwistcft}
\end{aligned}\end{equation}
We will mostly focus on the choice $n=1$ in this work but will offer some comments for general values of $n$ in due course.

A central object of interest is the partition function of this model for general values of the deformation parameters $Z_\text{CFT}(\beta ,\Delta^\alpha_F, \sigma^\alpha_F, \mathfrak{p}^\alpha_F)$. As shown in \cite{Benini:2015noa} and discussed further in \cite{Benini:2015eyy,Benini:2016hjo,Closset:2016arn}, this path integral can be computed by supersymmetric localization and is often referred to as a \textit{topologically twisted index}, since it can be rewritten as a Witten index of the schematic form
\begin{equation}\label{eq:topinddef}
	Z_\text{CFT}(\beta,\Delta^\alpha_F, \sigma^\alpha_F, \mathfrak{p}^\alpha_F) = \text{Tr}\,(-1)^F e^{i \Delta^\alpha_F J_F^\alpha} e^{-\beta H}~.
\end{equation}
Here $F$ is the fermion number and the Hamiltonian $H$ is a function of the real masses and magnetic charges, i.e. $H = H(\sigma^\alpha_F, \mathfrak{p}^\alpha_F)$.  Note that the partition function is computed in an ensemble that is canonical with respect to the magnetic charges but grand canonical with respect to the electric charges.  This is why it depends on the magnetic charges and the electric chemical potentials, but not on electric charges.

Our main interest in this work is to compute the topologically twisted index of the ABJM theory in \eqref{eq:topinddef} using holography. To this end it is convenient to establish some further notation. We can go into the ``democratic basis'' $(\Delta^I, \sigma^I, \mathfrak{p}^I)$, where $I = 0,\ldots,3$ enumerates the four $U(1)$ global  symmetries, by defining 
\begin{equation}\begin{aligned}
\label{eq:basis}
	\Delta^0 &= \frac{1}{2}\left(\Delta_R + \Delta^1_F + \Delta^2_F + \Delta^3_F\right)~, \quad &\Delta^1 &= \frac{1}{2}\left(\Delta_R + \Delta^1_F - \Delta^2_F - \Delta^3_F\right)~, \\
	\sigma^0 &= \frac{1}{2}\left(\sigma^1_F + \sigma^2_F + \sigma^3_F\right)~, \quad &\sigma^1 &= \frac{1}{2}\left(\sigma^1_F - \sigma^2_F - \sigma^3_F\right)~, \\
	\mathfrak{p}^0 &= \frac{1}{2}\left(\mathfrak{p}_R + \mathfrak{p}^1_F + \mathfrak{p}^2_F + \mathfrak{p}^3_F\right)~, \quad &\mathfrak{p}^1 &= \frac{1}{2}\left(\mathfrak{p}_R + \mathfrak{p}^1_F - \mathfrak{p}^2_F - \mathfrak{p}^3_F\right)~, \\
\end{aligned}\end{equation}
plus cyclic permutations for $\Delta^{2,3}$, $\sigma^{2,3}$, and $\mathfrak{p}^{2,3}$.  As a consequence of the supersymmetry constraints in \eqref{eq:toptwistcft}, these quantities satisfy the following constraints:
\begin{equation}
	\sum_I \Delta^I = 2 \Delta_R = 2\pi~,  \qquad	\sum_I \sigma^I = 0~, \qquad \sum_I \mathfrak{p}^I = 2 \mathfrak{p}_R = 2(\mathfrak{g}-1)~.
\label{eq:cftconst}
\end{equation}
It is also convenient to define the ``complex fugacities'' $u^I$ as follows:
\begin{equation}\label{eq:uIdefQFT}
	u^I = \Delta^I + i \beta \sigma^I~.
\end{equation}
Given the constraints \eqref{eq:cftconst}, we immediately find that these complex fugacities satisfy
\begin{equation}\label{eq:uIQFTconstr}
	\sum_I u^I = 2\pi~.
\end{equation}
As shown in detail in \cite{Benini:2015eyy} the partition function \eqref{eq:topinddef} can be computed explicitly in the large $N$ limit by using supersymmetric localization and matrix model techniques. The result is
\begin{equation}
	\log Z_\text{CFT}(u^I, \mathfrak{p}^I) = \frac{\sqrt{2} N^{3/2}}{3} \sqrt{u^0 u^1 u^2 u^3} \sum_I \frac{\mathfrak{p}^I}{u^I}~.
\label{eq:cftpart2}
\end{equation}
Note that not all $u^I$ and $\mathfrak{p}^I$ in \eqref{eq:cftpart2} can be fixed independently; they must satisfy the constraints (\ref{eq:cftconst}) and \eqref{eq:uIQFTconstr}.  Note also that the complex fugacities should be thought of as functions of the deformation parameters, i.e.  $u^I = u^I(\beta, \Delta^I, \sigma^I)$, so the partition function in \eqref{eq:cftpart2} is still in the same ensemble as in \eqref{eq:topinddef} where we fix the same ten parameters $(\beta,\Delta^\alpha_F, \sigma^\alpha_F, \mathfrak{p}^\alpha_F)$.

It was emphasized in \cite{Benini:2015eyy,Benini:2016rke} that the topologically twisted index in \eqref{eq:cftpart2} can be used to compute the entropy of supersymmetric static asymptotically AdS$_4$ black holes. To achieve this one has to perform the so-called $\mathcal{I}$-extremization procedure. Namely, one has to extremize $\mathcal{I} \equiv \log Z_\text{CFT}(u^I, \mathfrak{p}^I) - i u^I \mathfrak{q}_I$ over the complex fugacities $u^{I}$ obeying the constraint in \eqref{eq:uIQFTconstr} for fixed magnetic charges $\mathfrak{p}^I$ and electric charges $\mathfrak{q}_I$.  In analogy with similar extremization results in other dimensions (see e.g.~\cite{Intriligator:2003jj,Jafferis:2010un,Benini:2012cz}), the interpretation of this procedure is that it selects the unique superconformal R-symmetry of the one-dimensional effective quantum mechanics theory obtained at low energies from the topologically twisted three-dimensional theory on $S^1\times \Sigma_{\mathfrak{g}}$. This interpretation is in harmony with the presence of the AdS$_2\times \Sigma_{\mathfrak{g}}$ near-horizon geometry in the supersymmetric black hole solutions. It is important to stress, however, that the $\mathcal{I}$-extremization procedure is not necessary in the QFT; the topologically twisted index in \eqref{eq:cftpart2} is a perfectly good supersymmetric observable for general values of the deformation parameters obeying the constraints \eqref{eq:cftconst} and \eqref{eq:uIQFTconstr}. This in turn leads to the natural goal of finding the holographic dual description of the partition function in \eqref{eq:cftpart2} without employing $\mathcal{I}$-extremization. 

As we discuss in detail below, the topologically twisted index in \eqref{eq:cftpart2} can indeed be computed in supergravity in terms of an on-shell action of Euclidean solutions which we call black saddles. The supersymmetric black holes solutions with AdS$_2\times \Sigma_{\mathfrak{g}}$ near-horizon geometry which realize holographically the $\mathcal{I}$-extremization procedure then emerge as a special limit of these black saddles.  We also note that, to the best of our knowledge, the result in \eqref{eq:cftpart2} has not been rigorously established for general complex values of the parameters $u^{I}$ in the large $N$ limit of the matrix model arising from supersymmetric localization. Based on the results in~\cite{Benini:2015eyy}, however, there is strong evidence for the validity of this result. Our supergravity black saddle results below serve as additional confirmation that \eqref{eq:cftpart2} indeed holds in general.

To illustrate the utility of the black saddle solutions we focus on a special limit of the topologically twisted index in which all $u^{I}$ are equal and the flavor magnetic fluxes are switched off.\footnote{Due to the quantization of magnetic flux on the Riemann surface, the universal twist of the ABJM theory for $k=1$ is possible only if $\mathfrak{g}$ is an odd integer. This restriction can be removed by taking $k>1$.} The topologically twisted index then takes the simple form
\begin{equation}\label{eq:univlogZ}
	\log Z_\text{CFT}^{\rm univ} = \frac{\sqrt{2}\pi}{3}(\mathfrak{g}-1) N^{3/2}~.
\end{equation}
This corresponds to the so-called \textit{universal twist} studied in \cite{Azzurli:2017kxo,Bobev:2017uzs}. The result in \eqref{eq:univlogZ} actually goes beyond ABJM and extends to a large class of three-dimensional $\mathcal{N}=2$ SCFTs with a gravity dual for which one can show that in the planar limit there is a relation between the topologically twisted index and the supersymmetric partition function on the round $S^3$
\begin{equation}\label{eq:univlogZFS3}
	\log Z_\text{CFT}^{\rm univ} =- (\mathfrak{g}-1)\log Z_{S^3} = (\mathfrak{g}-1)F_{S^3} ~.
\end{equation}
Therefore, all results we discuss below for this universal twist are valid for more general theories than the specific ABJM example. Finally, we note that the result in \eqref{eq:univlogZ} is derived \cite{Benini:2015eyy,Hosseini:2016tor,Azzurli:2017kxo} assuming that the topologically twisted index scales as $N^{3/2}$ and is thus valid for $\mathfrak{g}\neq1$. 

\subsection{Gravity}
\label{subsec:grav}

To illustrate the supergravity solutions of interest in this work we start by a review of the poster child of this class of solutions first constructed by Romans in \cite{Romans:1991nq} and later studied in \cite{Caldarelli:1998hg}.

The solution studied in \cite{Romans:1991nq,Caldarelli:1998hg} is an asymptotically locally AdS$_4$ background of minimal $\mathcal{N}=2$ gauged supergravity. The bosonic content of this theory consists of the metric and a $U(1)$ gauge field known as the graviphoton. The theory is a limit of the STU model discussed in Section~\ref{sec:Supergravity} where all four $U(1)$ gauge fields are taken to be equal and all scalars vanish. Using this notation the Lorentzian black hole solution constructed in \cite{Romans:1991nq,Caldarelli:1998hg} takes the simple form
\begin{equation}\begin{aligned}
	ds^2 &= - U(r) dt^2 + \frac{dr^2}{U(r)} + r^2 ds_{\sigg}^2~, \\
	U(r) &= \left(\sqrt{2} g r - \frac{1}{2\sqrt{2} g r} \right)^2~, \\
	F^I &= \frac{1}{4\xi g}\,V_{\sigg}~, \\
	z^\alpha &= 0~,
\label{eq:univBH}
\end{aligned}\end{equation}
where $\mathfrak{g}>1$ is the genus of the Riemann surface with constant curvature metric $ds_{\sigg}^2$ and volume form $V_{\sigg}$ and $g$ is related to the scale $L$ of AdS$_4$ as $g=1/\sqrt{2}L$. The magnetic charges are specified by $p^I = \frac{1}{4\xi g}$, where the parameter $\xi$ takes values $\pm 1$ depending on whether we want the magnetic charges to be negative or positive. The magnitude of the magnetic flux through the Riemann surface is fixed by supersymmetry to be related to the AdS radius. This is the supergravity manifestation of the topological twist condition (\ref{eq:toptwistcft}). It was shown in~\cite{Romans:1991nq,Caldarelli:1998hg} that it is not possible to add electric charges to this solution while preserving both supersymmetry and regularity. By the same token, it is not possible to construct similar regular solutions with spherical or toroidal horizons, i.e. for $\mathfrak{g} = 0$ or $\mathfrak{g} = 1$.  The magnetic black hole (\ref{eq:univBH}) is hence the unique static regular 1/4-BPS black hole solution for any given Riemann surface with constant curvature metric and genus $\mathfrak{g}$.\footnote{Generalizations of this solutions to Riemann surfaces with general metrics were recently studied in \cite{Bobev:2020jlb}.}  

This black hole solution interpolates between an asymptotically locally AdS$_4$ spacetime at $r \to \infty$ and the near-horizon AdS$_2 \times \sigg$ region at $r = \frac{1}{2g}$.  The Bekenstein-Hawking entropy of this extremal black hole is proportional to the area of the horizon and is given by
\begin{equation}
	S_\text{BH} = \frac{A}{4 G_N} = \frac{\pi}{4 g^2 G_N} (\mfrk{g}-1)~,
\label{eq:univsbhsec2}
\end{equation}
where $G_N$ denotes Newton's constant for the four-dimensional gravity theory.  Using the standard holographic relation $\frac{1}{4g^2G_N} = \frac{\sqrt{2}}{3}N^{3/2}$ we find that the black hole entropy agrees with the topologically twisted index of the ABJM theory in \eqref{eq:univlogZ} \cite{Benini:2015eyy}. Moreover, this simple black hole solution can be embedded in string theory and M-theory in a variety of different ways and the black hole entropy always agrees with the large $N$ limit of the topologically twisted index \cite{Azzurli:2017kxo}.

At this point it is tempting to declare victory since we have an agreement between the field theory partition function and the entropy of a black hole. However, there are at least two important subtleties to take into account. The first one stems from the fact that the AdS/CFT dictionary relates the partition function of the bulk gravity theory and the dual field theory. The gravitational partition function in the semi-classical approximation is determined by the regularized on-shell action of the solution at hand. Computing the on-shell action of the black hole solution in \eqref{eq:univBH} is subtle due to the fact that the temperature of the solution vanishes, or alternatively due to infinitely long throat in the near-horizon AdS$_2$ region.\footnote{It can be shown that indeed the on-shell action of the solution \eqref{eq:univBH} is equal to the entropy in \eqref{eq:univsbhsec2} by considering the extremal black hole as the zero-temperature limit of a family of non-extremal black holes \cite{Azzurli:2017kxo}. This approach to regularizing the on-shell action has the disadvantage of breaking supersymmetry.} The second subtlety arises from the dual field theory localization calculation of the topologically twisted index. The path integral of the partially topologically twisted ABJM theory on $S^1\times\sigg$ is convergent when the size $\beta$ of the $S^1$ is finite. Importantly, the end result for the topologically twisted index \eqref{eq:univlogZ} is independent of this size. The black hole solution on the other hand is Lorentzian and has vanishing temperature which implies that the supergravity value of $\beta$ is infinite.

\begin{figure}\begin{center}
	\includegraphics[width=0.7\textwidth]{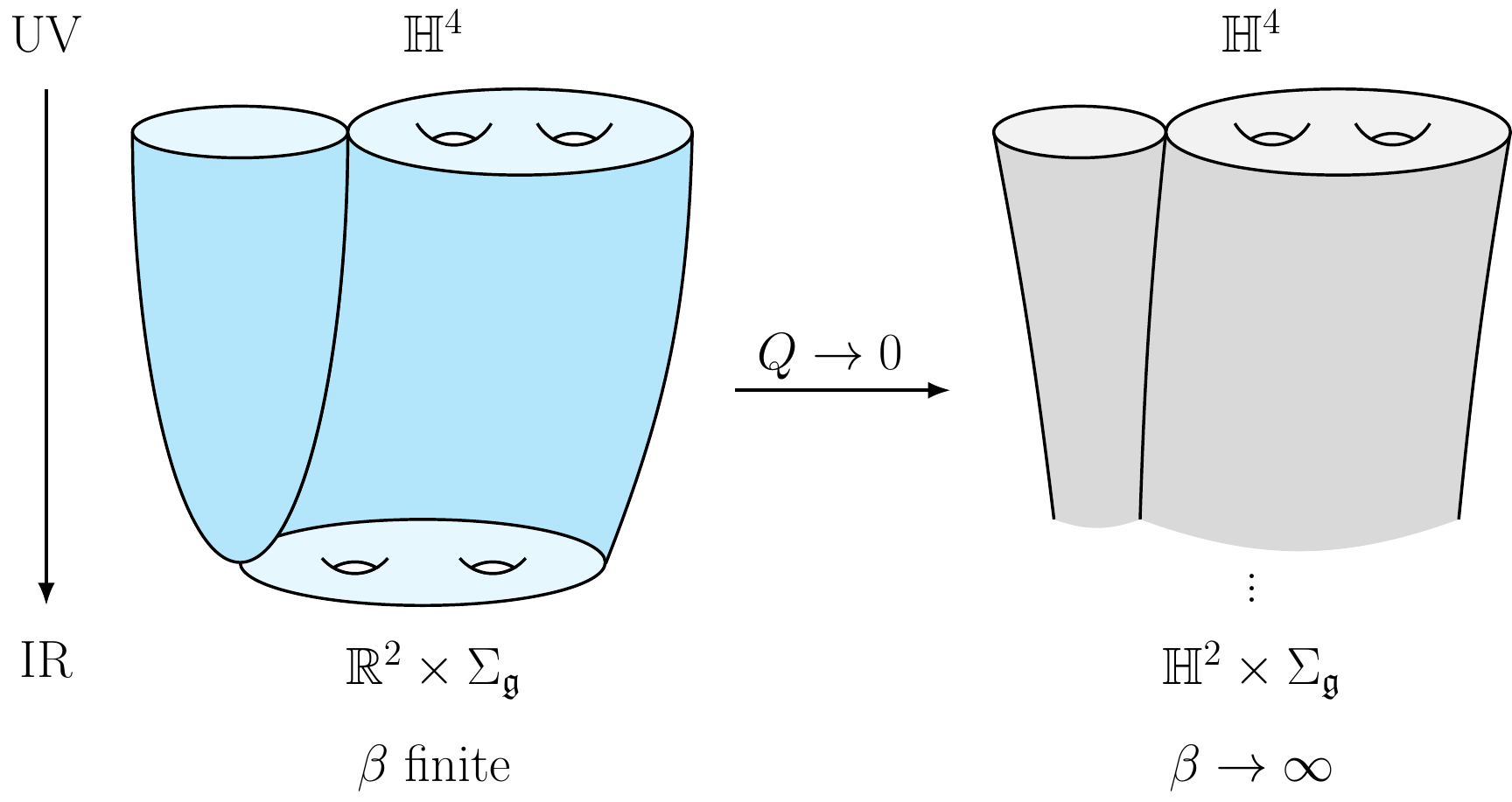}
	\caption{A schematic depiction of the Euclidean black saddle solution \eqref{eq:univBS} and the universal black hole \eqref{eq:univBH}.  In the limit  $Q\to 0$ the periodicity of the $\tau$ coordinate becomes infinite and the IR region acquires the metric on $\mathbb{H}^2\times \Sigma_{\mathfrak{g}}$ which can then be analytically continued to the near-horizon AdS$_2 \times \sigg$ metric of the black hole.  Both classes of solutions are asymptotically locally $\mathbb{H}^4$ in the UV region with $S^1 \times \sigg$ boundary.}
	\label{fig:h2vsr2}
\end{center}\end{figure}

Both subtleties summarized above are related to the fact that the topologically twisted ABJM theory and the accompanying supersymmetric localization calculation of the topologically twisted index are properly defined in Euclidean signature with finite $\beta$. This in turns points to their resolution, namely we should think of the black hole solution in \eqref{eq:univBH} as a member of a family of Euclidean supersymmetric saddle points of the supergravity equations of motion.  Indeed, as we show in detail in Sections~\ref{sec:Supergravity} and \ref{sec:solutions}, it is possible to construct explicitly such a family of supersymmetric Euclidean solutions which takes the simple form 
\begin{equation}\label{eq:univBS}\begin{aligned}
	ds^2 &= U(r) d\tau^2 + \frac{dr^2}{U(r)} + r^2 ds_{\sigg}^2~, \\
	U(r) &= \left(\sqrt{2} g r - \frac{1}{2\sqrt{2} g r} \right)^2 - \frac{Q^2}{8 r^2}~, \\
	F^I &= \frac{1}{4\xi g}\,V_{\sigg} + \frac{Q}{4r^2} d\tau\wedge dr~, \\
	z^\alpha &= \tilde{z}^{\alpha}= 0~,
\end{aligned}\end{equation}
where again we take $\mathfrak{g} > 1$, and $\xi = \pm 1$. This is a one-parameter family of Euclidean solutions labeled by the electric charge $Q$.  The metric function $U(r)$ has two zeroes $r_\pm$, given by
\begin{equation}\label{eq:rpmunivBS}
	r_\pm = \frac{\sqrt{1 \pm g |Q|}}{2 g}~.
\end{equation}
Note that $r_+$ is real and positive for any value of the charge.  Additionally, even if $r_-$ is real, $r_+$ is always bigger, and so the spacetime has a cap-off at $r = r_+$.  In fact, this is a smooth cap-off, since in the limit where $r \to r_+$ the metric becomes locally $\mathbb{R}^2 \times \sigg$. The absence of conical singularities on $\mathbb{R}^2$ determines that the coordinate $\tau$ is periodic, such that $\tau \sim \tau +\beta_\tau$, where the periodicity is
\begin{equation}\label{eq:betaunivBS}
	\beta_\tau = \frac{\pi\sqrt{1 + g |Q|}}{g^2 |Q|}~.
\end{equation}
Thus, by turning on a finite value of $Q$ for the Euclidean solutions, we have a finite periodicity $\beta_\tau$ for the $\tau$ coordinate.  This is precisely the bulk dual to the size $\beta$ of the $S^1$ in the dual field theory description.  In the limit where $Q \to 0$, the periodicity in \eqref{eq:betaunivBS} becomes infinite, $\beta_\tau \to \infty$, and we recover a Euclideanized version of the black hole solution in \eqref{eq:univBH}.  We should stress that it is only in the $Q \to 0$ limit that there is a sensible Lorentzian version of the solution, i.e. one where the gauge field is real and there are no naked singularities. This is in harmony with the results in \cite{Romans:1991nq,Caldarelli:1998hg} where it was shown that there are no regular supersymmetric Lorentzian black holes with both electric and magnetic charge.  We should therefore think of the black hole in \eqref{eq:univBH} as a special case of the family of Euclidean black saddle solutions in \eqref{eq:univBS} obtained by sending $\beta_\tau \to \infty$. We illustrate this in Figure~\ref{fig:h2vsr2}.

The regularized on-shell action $I$ for the Euclidean black saddle solutions in \eqref{eq:univBS} was computed recently in \cite{BenettiGenolini:2019jdz}.\footnote{We show how to compute this on-shell action in detail in the context of the more general STU model in Section~\ref{sec:Supergravity}.} The result is that for any value of $Q$ one finds the simple answer
\begin{equation}\label{eq:univBSonshellA}
	I = -\frac{\pi}{4 g^2 G_N} (\mfrk{g}-1)\,,
\end{equation}
and so it is independent of $Q$.  This is exactly analagous to how the topologically twisted index is independent of $\beta$.  Moreover, using the holographic relation $\frac{1}{4g^2G_N} = \frac{\sqrt{2}}{3}N^{3/2}$, we find that for every value of $\beta_\tau$ in \eqref{eq:betaunivBS} the on-shell action of the black saddle solutions agrees with the topologically twisted index of the dual ABJM theory \eqref{eq:univlogZ}. As expected from the supersymmetric localization result the on-shell action is independent of the size of the $S^1$. We can therefore take the limit $\beta_\tau \to \infty $ and find that the on-shell action is related to the black hole entropy in \eqref{eq:univsbhsec2} as $I = -S_{\rm BH}$. In conclusion, we have shown that the supergravity dual of the universal twist of the ABJM theory on $S^1\times \Sigma_{\mathfrak{g}}$, with $\mathfrak{g}>1$ and $S^1$ with finite size $\beta$, is the Euclidean supergravity solution in \eqref{eq:univBS}.

There is yet another subtlety regarding the holographic description of the universal twist of the ABJM theory. The Lorentzian black hole solution in \eqref{eq:univBH} is free of naked singularities only for $\mathfrak{g}>1$~\cite{Caldarelli:1998hg}. Therefore, despite the fact that the universal twist of ABJM on $S^1 \times \sigg$ is well-defined for any genus $\mathfrak{g}$ there is no obvious candidate supergravity dual solution.  Yet again, the key to resolving this subtlety is to look for Euclidean black saddle solutions. A careful analysis of the Euclidean BPS equations, discussed in Section~\ref{sec:Supergravity}, shows that the solutions in \eqref{eq:univBS} can be generalized to arbitrary genus while still preserving supersymmetry.  The result is the following supersymmetric Euclidean background:
\begin{equation}\begin{aligned}
	ds^2 &= U(r) d\tau^2 + \frac{dr^2}{U(r)} + r^2 ds_{\sigg}^2~, \\
	U(r) &= \left(\sqrt{2} g r + \frac{\kappa}{2\sqrt{2} g r} \right)^2 - \frac{Q^2}{8 r^2}~, \\
	F^I &= -\frac{\kappa}{4\xi g}\,V_{\sigg} + \frac{Q}{4r^2} d\tau\wedge dr~, \\
	z^\alpha &=\tilde{z}^{\alpha}= 0~,
\label{eq:romansgen}
\end{aligned}\end{equation}
where $\kappa$ is the curvature of the Riemann surface with $\kappa=-1$ for $\mathfrak{g}>1$, $\kappa=0$ for $\mathfrak{g}=1$, and $\kappa=1$ for $\mathfrak{g}=0$. The metric function $U(r)$ has its outermost root at
\begin{equation}\label{eq:r+genBSR}
	r_+ = \frac{\sqrt{-\kappa + g |Q|}}{2g}~.
\end{equation}
This root has to be real in order for the spacetime to cap off at a real value of the coordinate $r$.  Additionally, we need to ensure that $r_+ > 0$ in order to avoid a curvature singularity.  Thus, to have a smooth solution we have to demand that
\begin{equation}
	g |Q| > \kappa~.
\label{eq:qgg}
\end{equation}
This means that $g |Q| >0$ for $\mathfrak{g} = 1$ and $g |Q| > 1$ for $\mathfrak{g} = 0$. It is clear from \eqref{eq:rpmunivBS} that there are no constraints on the charge $Q$ for $\mathfrak{g}>1$.  As $r \to r_+$, the metric in \eqref{eq:romansgen} asymptotes to $\mathbb{R}^2 \times \sigg$, and the periodicity of the $\tau$ coordinate is fixed to
\begin{equation}\label{eq:betaBSg01}
	\beta_\tau = \frac{\pi\sqrt{-\kappa + g |Q|}}{g^2 |Q|}~,
\end{equation}
in order to avoid any conical singularities.  Thus we once again find that for any finite electric charge $Q$, the solution has a finite periodicity for the $\tau$ coordinate that matches directly onto the size of the $S^1$ in the dual field theory.  In the limit $Q \to 0$ one finds that $\beta_\tau \to \infty$ and thus the $S^1$ decompactifies. However, due to the constraint in \eqref{eq:qgg}, this limit results in a smooth supergravity solution only for $\mathfrak{g}>1$ and this is precisely the limit in which one recovers the black hole solution in \eqref{eq:univBH}. Therefore we learn that there are black saddle solutions, like the ones in \eqref{eq:romansgen} with $\mathfrak{g}=0,1$, for which the $\beta_\tau \to \infty$ limit is not regular and they cannot be interpreted as regular black holes in Lorentzian signature.

We show in detail in Section~\ref{sec:Supergravity} that the on-shell action of the black saddle solution in \eqref{eq:romansgen} is given by \eqref{eq:univBSonshellA} for any value of the genus $\mathfrak{g}$. This supergravity result in turn agrees with the supersymmetric localization result in \eqref{eq:univlogZ}. Thus we conclude that the black saddle solutions in \eqref{eq:romansgen} are the holographic dual description of the universal twist of the ABJM theory on $\Sigma_{\mathfrak{g}}$.

While the black saddle solutions in \eqref{eq:romansgen} with $\beta_\tau$ in \eqref{eq:betaBSg01} and $Q$ obeying the bound \eqref{eq:qgg} are regular for all values of the genus $\mathfrak{g}$ they may not be the leading saddle point contribution to the supergravity path integral. In the saddle point approximation the supergravity partition function takes the form $Z_{\rm grav} = e^{-I}$ and it is clear from \eqref{eq:univBSonshellA} that for $\mathfrak{g}>1$ the black saddle solution is dominant in the small $G_N$, or equivalently large $N$, semi-classical approximation. For $\mathfrak{g}=0$ we find that $Z_{\rm grav} \sim e^{-N^{3/2}}$ which implies that the black saddle solution is subdominant in the path integral. The case $\mathfrak{g}=1$ is somewhat puzzling since then the on-shell action of the regular black saddle solution vanishes and thus  $Z_{\rm grav} \sim 1$. Superficially, this is in harmony with the vanishing of the topologically twisted index in \eqref{eq:univlogZ}. However, the result in \eqref{eq:univlogZ} is derived assuming $N^{3/2}$ scaling of the free energy in the supersymmetric localization matrix model, which means that there is a priori no reason to expect a match for this $\mathfrak{g}=1$ case. It will be most interesting to understand better this case both in field theory and in supergravity.

Finally, let us emphasize another important lesson from the black saddle solutions in \eqref{eq:romansgen}. In the topologically twisted ABJM theory the radius $\beta$ of the $S^1$ is a free parameter while the electric charge associated to the $U(1)$ R-symmetry, $\mathfrak{q}_R$, is not fixed since its associated chemical potential is fixed by supersymmetry \eqref{eq:cftconst}. The smooth black saddle solutions in \eqref{eq:romansgen}, on the other hand, uniquely determine $q_R$ (which is proportional to $Q$) in terms of $\beta$ as in \eqref{eq:betaBSg01}. Therefore, in situations where $Z_{\rm grav} \sim e^{N^{3/2}}$, we should think of the smooth black saddles as the dominant contribution of the gravitational path integral in the large $N$, or small $G_N$, limit. Presumably other values of $\mathfrak{q}_R$ for a fixed value of $\beta$ are manifested in supergravity by different, subdominant saddle points. As discussed above in the $\beta \to \infty $ limit for $\mathfrak{g}>1$ we find that $Q=0$, and thus $q_R=0$, and the dominant gravitational saddle point becomes the Lorentzian magnetic black hole solution in \eqref{eq:univBH}. The fact that $q_R$ is uniquely determined by requiring regularity of the black saddle solution is also a general feature of the more general black saddles we discuss in the rest of the paper.

After this initial foray into the physics of the supersymmetric black saddle solutions in the context of minimal $\mathcal{N}=2$ gauged supergravity we now proceed with studying similar solutions in the STU model. These more general black saddle solutions have many of the features of the solution in \eqref{eq:romansgen} but have different values for magnetic and electric parameters in the gauge fields and also non-trivial scalar profiles. As we show in detail below, this rich set of new parameters in turn allows us to construct the supergravity dual of the topologically twisted index of the ABJM theory in \eqref{eq:cftpart2} for general values of the complex fugacities and magnetic charges, without the need to invoke the $\mathcal{I}$-extremization procedure.

\section{Supergravity and holography}
\label{sec:Supergravity}

In this section we present the BPS equations for Euclidean black saddle solutions in the STU model and solve these BPS equations perturbatively around the UV and IR regions of the spacetime.  We use these solutions in tandem with a careful treatment of holographic renormalization to compute the on-shell action for the black saddles.  We then establish a precise holographic match between the black saddle on-shell action and the topologically twisted index of the dual planar ABJM theory.

\subsection{Euclidean BPS conditions in the STU model}
\label{subsec:EuclidSTU}

Our goal is to study Euclidean black saddle solutions in the STU model of $\mathcal{N}=2$ gauged supergravity \cite{Behrndt:1996hu,Duff:1999gh,Cvetic:1999xp}.  The STU model arises as a consistent truncation of the four-dimensional maximal $SO(8)$ gauged supergravity of \cite{deWit:1982bul}. Every solution of the STU model, including the black saddles discussed below, can be uplifted to eleven-dimensional supergravity using the results in \cite{Azizi:2016noi}. It is sufficient to restrict our study to this truncated supergravity theory since it contains precisely the bosonic fields needed to implement holographically the partial topological twist of the ABJM theory discussed in Section~\ref{sec:fieldtheorypartfunc}. In this section, we present the relevant details of the STU model, explain how to Euclideanize the supergravity theory, and then give the Euclidean BPS conditions for the black saddle solutions of interest.  Our conventions are given in Appendix~\ref{app:conventions}, and additional details on $\mathcal{N}=2$ supergravity are presented in Appendix~\ref{app:n2sugra}.

The STU model comprises an $\mathcal{N}=2$ gravity multiplet
\begin{equation}
	\left( g_{\mu\nu}~,~ \psi_\mu^i~,~ A_\mu^0\right)~,
\end{equation}
where $g_{\mu\nu}$ is the metric, $\psi_\mu^i$ is an $SU(2)$ doublet of gravitini, and $A_\mu^0$ is the graviphoton, as well as three $\mathcal{N}=2$ vector multiplets
\begin{equation}
	\left( A_\mu^\alpha~,~ \lambda^{\alpha}_i~,~ z^\alpha\right)~,
\end{equation}
where $A_\mu^\alpha$ is a $U(1)$ vector field, $\lambda^{\alpha}_i$ is an $SU(2)$ doublet of gaugini, $z^\alpha$ is a complex scalar, and the index $\alpha$ runs over $\alpha = 1, 2, 3$.  It is convenient to put all four vector fields on the same footing by denoting them as $A_\mu^I$, with $I = 0,\ldots,3$.  The scalars $z^\alpha$ each parameterize a copy of the Poincar\'e disk and so they must each have modulus less than one, i.e.
\begin{equation}
\label{eq:modulus}
	|z^\alpha| < 1~.
\end{equation}

To understand the interactions between the fields in the theory, we first need to specify the holomorphic sections $X^I$ in terms of the physical scalars $z^\alpha$ by
\begin{equation}\begin{aligned}
	X^0 &= \frac{1}{2\sqrt{2}}(1- z^1)(1- z^2)(1-z^3)~, &\quad& X^1 = \frac{1}{2\sqrt{2}}(1- z^1)(1+z^2)(1+z^3)~, \\
	X^2 &= \frac{1}{2\sqrt{2}}(1 + z^1)(1- z^2)(1+z^3)~, &\quad& X^3 = \frac{1}{2\sqrt{2}}(1+ z^1)(1+z^2)(1-z^3)~.
\end{aligned}\end{equation}
The prepotential $F$ of the theory is then given by
\begin{equation}\label{eq:prepot}
	F = -2 i \sqrt{X^0 X^1 X^2 X^3}~,
\end{equation}
with derivatives denoted by $F_I = \frac{\partial F}{\partial X^I}$, $F_{IJ} = \frac{\partial^2 F}{\partial X^I\partial X^J}$, etc.  The K\"ahler potential $\mathcal{K}$ and the K\"ahler metric $g_{\alpha \bar{\beta}}$ are correspondingly given by
\begin{equation}
	\mathcal{K} = -\sum_{\alpha=1}^3 \log\left[ (1- |z^\alpha|^2)\right]~, \qquad g_{\alpha \bar{\beta}} = \frac{\delta_{\alpha \bar{\beta}}}{(1-|z^\alpha|^2)^2}~.
\label{eq:kg}
\end{equation}
Additionally, the kinetic mixing matrix $\mathcal{N}_{IJ}$ for the vector fields is given by
\begin{equation}
	\mathcal{N}_{IJ} = \bar{F}_{IJ} + i \frac{N_{IK}X^K N_{JL}X^L}{N_{NM} X^N X^M} = \mathcal{R}_{IJ} + i \mathcal{I}_{IJ}~,
\label{eq:rimat}
\end{equation}
where we have defined $N_{IJ} \equiv 2 \,\text{Im}\,F_{IJ}$, and $\mathcal{R}_{IJ}$ and $\mathcal{I}_{IJ}$ are, respectively, the real and imaginary parts of $\mathcal{N}_{IJ}$.

The gauging of the supergravity theory is determined by the vector of three moment maps, $\vec{P}_I$, which for the STU model are given by
\begin{equation}
	P_I^1 = P_I^2 = 0~, \quad P_I^3 = -1~.
\end{equation}
The corresponding potential $\mathcal{P}$ for the scalars associated with this gauging is
\begin{equation}
	\mathcal{P}=  6 - \sum_{\alpha = 1}^3 \frac{4}{1- |z^\alpha|^2}~.
\label{eq:zpot}
\end{equation}
Alternatively, in $\mathcal{N}=1$ language, we can define a superpotential
\begin{equation}\label{eq:superpotdef}
	\mathcal{V} = 2 (z^1 z^2 z^3 - 1)~,
\end{equation}
from which we can generate the scalar potential $\mathcal{P}$:
\begin{equation}
	\mathcal{P} = \frac{1}{2} e^{\mathcal{K}}\left( g^{\alpha \bar{\beta}}\nabla_\alpha \mathcal{V} \nabla_{\bar{\beta}}\bar{\mathcal{V}} - 3 \mathcal{V}\bar{\mathcal{V}}\right)~,
\end{equation}
where $g^{\alpha \bar{\beta}}$ is the inverse of the metric in \eqref{eq:kg} and the K\"ahler-covariant derivative acts on the superpotential as $\nabla_\alpha \mathcal{V} = \partial_\alpha \mathcal{V} + (\partial_\alpha \mathcal{K})\mathcal{V}$.  

With all of these ingredients established, we can now present the bosonic part of the action for the STU model in $\mathcal{N}=2$ gauged supergravity:
\begin{equation}\label{eq:Slor}
	S = \frac{1}{8\pi G_N}\int d^4x\,\sqrt{-g}\,\left[ \frac{1}{2}R + \frac{1}{4} \mathcal{I}_{IJ} F^I_{\mu\nu}F^{J\mu\nu} - \frac{i}{4} \mathcal{R}_{IJ} F^I_{\mu\nu} \tilde{F}^{J\mu\nu} - g_{\alpha \bar{\beta}} \nabla^\mu z^\alpha \nabla_\mu \bar{z}^{\bar{\beta}} - g^2 \mathcal{P}\right]~,
\end{equation}
where $g$ is the gauging parameter.\footnote{We also use $g$ to denote the determinant of the four-dimensional metric. We hope that the meaning intended is clear from the context. } When we set all scalars and gauge fields to zero, the theory has a maximally-supersymmetric AdS$_4$ vacuum with size $L$, related to the gauging parameter $g$ by
\begin{equation}\label{eq:gLrelation}
	g = \frac{1}{\sqrt{2} L}~.
\end{equation}
The supersymmetry variations of this model are\footnote{See Appendix~\ref{app:n2sugra} for more details.} 
\begin{equation}\begin{aligned}
	\delta \psi_\mu^i &= \mathcal{D}_\mu \epsilon^i + \frac{i}{2} g e^{\mathcal{K}/2} P_I^3 X^I \gamma_\mu \varepsilon^{ik}(\sigma_3)\ind{_k^j} \epsilon_j + \frac{1}{4} e^{\mathcal{K}/2} \mathcal{I}_{IJ} X^I F^{-J}_{ab}\gamma^{ab} \gamma_\mu \varepsilon^{ij}\epsilon_j~, \\
	\mathcal{D}_\mu \epsilon^i &= \left(\partial_\mu + \frac{1}{4}\omega_\mu^{ab} \gamma_{ab} + \frac{i}{2}\mathcal{A}_\mu \right)\epsilon^i + \frac{i}{2} g P^3_I A_\mu^I (\sigma_3)\ind{_j^i}\epsilon^j~, \\ 
	\delta \lambda_i^\alpha &= - \frac{1}{2}g^{\alpha \bar{\beta}} f_{\bar{\beta}}^I \mathcal{I}_{IJ} F^{-J}_{\mu\nu} \gamma^{\mu\nu} \varepsilon_{ij}\epsilon^j + \gamma^\mu \nabla_\mu z^\alpha \epsilon_i - i g g^{\alpha \bar{\beta}} f_{\bar{\beta}}^I P_I^3 (\sigma_3)\ind{_i^k} \varepsilon_{kj}\epsilon^j~,
\label{eq:susyl}
\end{aligned}\end{equation}
where the K\"ahler connection $\mathcal{A}_\mu$ is
\begin{equation}
	\mathcal{A}_\mu = - \frac{i}{2}\left(\partial_\alpha \mathcal{K} \partial_\mu z^\alpha - \partial_{\bar{\alpha}} \mathcal{K} \partial_\mu \bar{z}^{\bar{\alpha}}\right)~,
\end{equation}
the $f_\alpha^I$ denote K\"ahler-covariant derivatives of the symplectic sections
\begin{equation}
	f_{\alpha}^I = e^{\mathcal{K}/2}\nabla_\alpha X^I = e^{\mathcal{K}/2}\left(\partial_\alpha X^I + (\partial_\alpha \mathcal{K}) X^I\right)~,
\end{equation}
and the supersymmetry parameters $\epsilon^i$ form an $SU(2)$ doublet of Weyl spinors.

At this point, we could in principle try to set the supersymmetry variations in \eqref{eq:susyl} to zero and solve the corresponding BPS conditions.  However, the theory we have presented so far is Lorentzian in nature, while our goal is to construct Euclidean supergravity solutions.  We therefore need to perform a suitable analytic continuation of our theory to Euclidean signature.  Crucially, as emphasized in \cite{Freedman:2013oja}, this Euclideanization forces us to treat the fermionic fields $\psi_\mu^i$, $\lambda_i^\alpha$, and $\epsilon^i$ as independent from their would-be conjugates $\psi_{i \mu}$, $\lambda^{i\bar{\alpha}}$, and $\epsilon_i$, since they do not fall into conjugate representations of the Euclidean isometry group $SO(4)$.  Furthermore, since bosons and fermions transform into one another via the action of the supersymmetry generators, we must also allow the scalars $z^\alpha$ and $\bar{z}^\alpha$ to now be \emph{independent} complex scalars, as well as allowing for the gauge fields $A_\mu^I$ to be complex.  In order to emphasize their new-found independence, we will follow the notation of~\citep{Freedman:2013oja} and denote the conjugate scalars by $\tilde{z}^\alpha$ instead of $\bar{z}^\alpha$.  We also note that this Euclideanization procedure does not act on the K\"ahler manifold parameterized by the scalars, and so the scalars $z^\alpha$ and $\tilde{z}^\alpha$ should all still take values on the Poincar\'e disk and correspondingly have modulus less than one.

With this subtlety accounted for, we now need to implement the Euclidean continuation via a Wick-rotation of the tangent space time direction:
\begin{equation}
	x^0 \to -i x^4~.
\end{equation}
Correspondingly, we must also Wick-rotate the time component of any vector fields, e.g. $A^I_0 \to i A^I_4$.  Note that this also means we must Wick-rotate the action accordingly, due to the integration over the time coordinate.  The end result of this procedure is that the Euclidean action for the bosonic fields in the STU model is
\begin{equation}
	S = \frac{1}{8\pi G_N}\int d^4x\,\sqrt{g}\,\left[ - \frac{1}{2}R - \frac{1}{4} \mathcal{I}_{IJ} F^I_{\mu\nu}F^{J\mu\nu} + \frac{i}{4} \mathcal{R}_{IJ} F^I_{\mu\nu} \tilde{F}^{J\mu\nu} + g_{\alpha \bar{\beta}} \nabla^\mu z^\alpha \nabla_\mu \tilde{z}^{\bar{\beta}} + g^2 \mathcal{P}\right]~,
\label{eq:SEucl}
\end{equation}
where we take the explicit forms of the matrices $\mathcal{I}_{IJ}$, $\mathcal{R}_{IJ}$ from \eqref{eq:rimat}, $g_{\alpha \bar{\beta}}$ from (\ref{eq:kg}), and $\mathcal{P}$ from (\ref{eq:zpot}), and replace all instances of $\bar{z}^\alpha$ with $\tilde{z}^\alpha$.  Similarly, we can also use the same supersymmetry variations in \eqref{eq:susyl} from the Lorentzian theory and Euclideanize each term.  Importantly, we need to also additionally include the conjugate variations to the ones in \eqref{eq:susyl} since they are no longer conjugates in Euclidean signature.

The procedure outlined above, for Wick-rotating a Lorentzian supergravity theory in order to find a Euclidean one, is well-established in many different holographic settings, and it has been used to derive several non-trivial tests of the AdS/CFT correspondence~\cite{Freedman:2013oja,Bobev:2013cja,Bobev:2016nua,Bobev:2018wbt,Bobev:2018ugk,Bobev:2019bvq}.  One could in principle follow \cite{deWit:2017cle} and use a different approach to construct a Euclidean supergravity theory via an off-shell time-like reduction of five-dimensional supergravity. This procedure leads to a few differences: all complex fields get split into two independent fields that must satisfy stringent reality conditions, and the prepotential $F$ in \eqref{eq:prepot} that determines vector multiplet couplings is now split into two separate functions $F_\pm$ that are a priori unrelated to one another.  Since we have a specific holographic setting in mind where the Lorentzian and Euclidean formulations of the dual ABJM theory have the same field content, we would eventually have to impose relations between these fields and the two prepotentials that would bring us back to the theory we obtained by a na\"ive Wick-rotation using our approach above.

Now that we have established our Euclidean STU model, we can present the ansatz for the black saddle solutions of interest:
\begin{equation}\begin{aligned}\label{eq:ansatz}
	ds^2 &= e^{2 f_1(r)} d\tau^2 + e^{2 f_2(r)} dr^2 + e^{2 f_3(r)} ds_{\sigg}^2~, \\
	A^I &= e^I(r) d\tau + p^I \omega_{\sigg}~, \\
	z^\alpha &= z^\alpha(r)~, \\
	\tilde{z}^{{\alpha}} &= \tilde{z}^{{\alpha}}(r)~.
\end{aligned}\end{equation}
Here $f_1$, $f_2$, $f_3$, $e^I$, $z^\alpha$, and $\tilde{z}^{{\alpha}}$ are all functions only of the radial coordinate $r$, $p^I$ are constants, and $ds^2_{\sigg}$ denotes the constant curvature metric on a smooth compact Riemann surface $\sigg$ of genus $\mg$.\footnote{Generalizations of this ansatz to more general metrics on the Riemann surface were studied in \cite{Bobev:2020jlb}, following \cite{Anderson:2011cz}. It is also possible to include point-like singularities on the Riemann surface using the results of \cite{Bobev:2019ore}.}  As in Section~\ref{sec:BlackSaddles} we denote the normalized curvature of the Riemann surface by $\kappa$ with $\kappa = 1$, $\kappa =0$, and $\kappa = -1$ for genus $\mathfrak{g} = 0$, $\mathfrak{g}=1$, and $\mathfrak{g} > 1$, respectively.  The volume form on the Riemann surface is $V_{\sigg}$ while  $\omega_{\sigg}$ is the local one-form potential such that $d \omega_{\sigg} = V_{\sigg}$.

The constants $p^I$ are precisely the magnetic charges of the solution,
\begin{equation}
	p^I = \frac{1}{\text{Vol}[\sigg]} \int_{\sigg} F^I~,
\end{equation}
while the electric charges take a somewhat more complicated form
\begin{equation}\label{eq:qi}
	q_I = \frac{1}{\text{Vol}[\sigg]} \int_{\sigg} G_I = e^{-f_1 - f_2+ 2 f_3}\mathcal{I}_{IJ}  \partial_r e^{J} - i \mathcal{R}_{IJ} p^J~,
\end{equation}
where the dual field strength $G_{I\mu\nu}$ is defined in Euclidean signature by, see \eqref{eq:FtildedefApp}, 
\begin{equation}
	\tilde{G}_{I\mu\nu} = -16 \pi G_N \frac{\delta S}{\delta F^{I\mu\nu}}~.
\end{equation}
Equivalently, the magnetic charges $p^I$ and electric charges $q_I$ can be obtained as the conserved charges that correspond to the Bianchi and Maxwell equations, respectively.

We have chosen a democratic basis for the charges $q_I$ where the index $I$ takes values $I = 0,1,2,3$.  However, in order to make better contact with the field theory, it will be useful to define a new basis of charges $q_R, q_F^\alpha$ as
\begin{equation}\begin{aligned}
	q_R &= \frac{1}{2} \left(q_0 + q_1 + q_2 + q_3\right)~, \\
	q_F^1 &= \frac{1}{2}\left(q_0 + q_1 - q_2 - q_3\right)~, \\
	q_F^2 &= \frac{1}{2}\left(q_0 - q_1 + q_2 - q_3\right)~, \\
	q_F^3 &= \frac{1}{2}\left(q_0 - q_1 - q_2 + q_3\right)~.
\label{eq:newbasis}
\end{aligned}\end{equation}
Similarly, we can define $p_R$ and $p_F^\alpha$ as the same linear combinations, but of the magnetic charges $p^I$, as well as $e_R$ and $e_F^\alpha$ as the same linear combinations of the functions $e^I$.  As discussed in Section~\ref{sec:fieldtheorypartfunc}, the utility of these definitions comes from the fact that the dual ABJM theory has a $U(1)_R$ symmetry as well as a $U(1)^3_F$ flavor symmetry.  The combinations of charges presented in \eqref{eq:newbasis} are precisely the ones dual to the charges of the CFT states under these symmetries.

We are interested in $\frac{1}{4}$-BPS solutions that preserve two of the eight supercharges of the theory.  Correspondingly, we impose the following projectors on the spinors:
\begin{equation}
	\gamma_{23} \epsilon^i = - i \xi (\sigma_3)\ind{_j^i}\epsilon^j~, \qquad 	\gamma_{23} \epsilon_i = i \xi (\sigma_3)\ind{_i^j} \epsilon_ j~,
\label{eq:projectxi}
\end{equation}
and
\begin{equation}
	\gamma_4 \epsilon^i = i M(r) \varepsilon^{ij}\epsilon_j~, \qquad  \gamma_4 \epsilon_i = i \wt{M}(r) \varepsilon_{ij}\epsilon^j~,
\label{eq:project}
\end{equation}
where $\xi = \pm 1$, while $M$ and $\wt{M}$ are both functions of the radial coordinate that satisfy the constraint $M \wt{M} = 1$.  Importantly, since the spinors $\epsilon^i$ and $\epsilon_i$ are no longer related by charge conjugation when we go to Euclidean signature, the projectors $M$ and $\wt{M}$ do not have to be complex conjugates of one another.  We also define the symplectic vector $g_I$ by
\begin{equation}
	g_I = \xi g P_I^3 = -\xi g(1,1,1,1)~,
\end{equation}
in order to simplify some of the subsequent formulae.  The end result of our analysis of the BPS conditions in Appendix~\ref{app:BPS} is that the magnetic charges are subject to the constraint
\begin{equation}\label{eq:toptwistcond}
	g_I p^I = \kappa~,
\end{equation}
which is the gravity analogue of the topological twist condition \eqref{eq:toptwistcft} in the dual CFT, while the metric functions $f_i$, the Wilson lines $e^I$, and the scalars $z^\alpha$, $\tilde{z}^\alpha$ obey the following differential equations:
\begin{equation}\label{eq:BPS}\begin{aligned}
	\frac{\partial f_1}{\partial r} &= \wt{M} e^{\mathcal{K}/2+ f_2}\left( i g_I X^I - e^{-2 f_3}(i q_I X^I - p^I F_I)\right) + e^{f_2 - f_1} g_I e^I \\
	&= M e^{\mathcal{K}/2+f_2} \left( -i g_I \bar{X}^I - e^{-2 f_3}( i q_I \bar{X}^I - p^I \bar{F}_I) \right) - e^{f_2 - f_1} g_I e^I~, \\
	\frac{\partial f_3}{\partial r} &= \wt{M} e^{\mathcal{K}/2+ f_2}\left( i g_I X^I + e^{-2 f_3}(i q_I X^I - p^I F_I)\right) \\
	&= M e^{\mathcal{K}/2+f_2} \left( -i g_I \bar{X}^I + e^{-2 f_3}( i q_I \bar{X}^I - p^I \bar{F}_I) \right) ~, \\
	\frac{\partial z^\alpha}{\partial r} &= - M e^{\mathcal{K}/2+f_2} g^{\alpha \bar{\beta}} \nabla_{\bar{\beta}} \left( -i g_I \bar{X}^I + e^{-2 f_3}( i q_I \bar{X}^I - p^I \bar{F}_I) \right) ~, \\
	\frac{\partial \tilde{z}^{\bar{\alpha}}}{\partial r} &= - \wt{M} e^{\mathcal{K}/2 + f_2} g^{\bar{\alpha} \beta} \nabla_\beta \left( i g_I X^I + e^{-2 f_3}(i q_I X^I - p^I F_I)\right)~.
\end{aligned}\end{equation}
Note that we are assuming that the metric and the charges $q_I$ and $p^I$ are all real.  We could in principle relax this reality condition by including complex conjugation on all charges and metric functions in every second equation in (\ref{eq:BPS}). However, the interpretation of these complex geometries is not clear. In particular if the metric is complex there is no notion of radial flow direction from the conformal boundary into the bulk and no clear regularity condition in the IR region.  In order to construct explicit solutions of these equations and study their holographic interpretation we must impose regularity in the IR region and employ the usual Fefferman-Graham expansion in the UV region.  These conditions necessitate the assumption of a real metric in our holographic analysis.

The most important feature of the Euclidean BPS conditions (\ref{eq:BPS}) is that, unlike the analogous equations in Lorentzian signature, they come in pairs of independent equations that are not related by complex conjugation.  This effectively doubles the number of equations we need to solve in order to find Euclidean black saddles. The BPS conditions are first-order in derivatives with respect to $r$, but they are highly non-linear due to the non-linear dependence of the prepotential on the scalar fields.  These two facts combine to make the equations very difficult to solve in full generality.  However, in order to compute the on-shell action for the solutions we are interested in, it suffices to solve these equations perturbatively around the boundaries of the spacetime.  In the rest of this section, we present these perturbative solutions and use them to compute the on-shell action and find agreement with the supersymmetric localization result in \eqref{eq:cftpart2}. To make this holographic match rigorous one has to construct bona fide solutions of the full system of non-linear equations. We will present some of these explicit Euclidean black saddle solutions in Section~\ref{sec:solutions}.

\subsection{UV expansion}
\label{sec:UV}

In order to get a grasp on the space of solutions to our BPS equations, it is informative to perform an asymptotic expansion.
We start here with a UV expansion, which will be especially useful for applying holographic renormalization and establishing a correspondence between bulk fields and field theory operators.  We are interested in solutions that are locally asymptotically AdS$_4$ with boundary $S^1 \times \Sigma_\mathfrak{g}$.
The UV asymptotics of such solutions are captured by a Fefferman-Graham expansion of the form
\begin{equation}\begin{aligned}\label{eq:metricUV}
	ds^2 &= e^{2 f_1} d\tau^2 + L^2 d\rho^2 + e^{ 2 f_3} ds_{\Sigma_{\mathfrak{g}}}^2~.
\end{aligned}\end{equation}
Note that we now use $\rho$ as a radial coordinate to signify that we are choosing the gauge $e^{f_2}=L$ in the ansatz \eqref{eq:ansatz}, with $\rho \to \infty$ corresponding to the location of the UV boundary.  Consistency of this Fefferman-Graham gauge with the BPS conditions and the equations of motion demands that the metric functions and bulk fields have asymptotic expansions of the form
\begin{equation}\label{eq:UVexpans}\begin{aligned}
	e^{f_1} &= e^\rho \left(a_0 + a_1 e^{-\rho} + \ldots\right)~, \\
	e^{f_3} &= e^\rho \left(b_0 + b_1 e^{-\rho} + \ldots\right)~, \\
	M &= M_0 + M_1 e^{-\rho} + \ldots~, \\
	\widetilde{M} &= \wt{M}_0 + \wt{M}_1 e^{-\rho} + \ldots~, \\
	z^\alpha &= e^{-\rho}\left(z^\alpha_0 + z^\alpha_1 e^{-\rho} + \ldots \right)~, \\
	\tilde{z}^\alpha &= e^{-\rho}\left(\tilde{z}^\alpha_0 + \tilde{z}^\alpha_1 e^{-\rho} + \ldots \right)~, \\
	e^I &= e^I_0 + e^I_1 e^{-\rho} +\ldots~,
\end{aligned}\end{equation}
for some set of constants $\{ a_n, b_n, M_n, \wt{M}_n, z_n^\alpha, \tilde{z}^\alpha_n, e_n^I\}$ that should be determined by solving the BPS conditions.
Note that the constants $a_0$ and $b_0$ can be chosen to take any value via a rescaling of the coordinate $\tau$ and a shift in $\rho$.
We use this freedom to set $b_0=L$ in the following, but keep $a_0$ a free constant.

Armed with this asymptotic expansion for the fields, we can now insert these into the BPS conditions (\ref{eq:BPS}) and solve the equations order-by-order in powers of $e^{\rho}$.  When doing so, we assume arbitrary magnetic and electric charges subject to the topological twist condition (\ref{eq:toptwistcond}).  At leading order, i.e. by collecting the highest power of $e^\rho$ and dropping all subleading terms, we find the following conditions: 
\begin{equation}\label{eq:UVzerothorder}
	g_Ie^I_{0} = 0~, \quad M_0 = -i \xi~, \quad \wt{M}_0 = i \xi~,
\end{equation}
with no constraints on $z^\alpha_0$, $\tilde{z}^\alpha_0$, nor the remaining combinations of $e^{I}_{0}$.
At first subleading order, we find that
\begin{equation}\label{eq:UV}\begin{aligned}
	a_1 &= b_1 = 0~, \\
	M_1 &= \wt{M}_1 = 0~, \\
	z^\alpha_1 &= - \frac{\tilde{z}^1_0 \tilde{z}^2_0 \tilde{z}^3_0}{\tilde{z}^\alpha_0} - \xi g \left(p^\alpha_F + q^\alpha_F\right)~,\\
	\tilde{z}^\alpha_1 &= - \frac{{z}^1_0{z}^2_0 {z}^3_0}{{z}^\alpha_0} - \xi g \left(p^\alpha_F - q^\alpha_F\right)~, \\
	e_1^I &= \sqrt{2} g a_0 q_I~.
\end{aligned}\end{equation}
Higher orders in this UV expansion of the black saddle solution can be computed systematically using \eqref{eq:UVexpans}. We will not need their explicit form for the calculation of the on-shell action. However, it is important to emphasize that these higher-order coefficients in the UV expansion are completely fixed in terms of the first two terms in the expansion \eqref{eq:UVexpans}. This is a general feature of the Fefferman-Graham expansion and is compatible with the dual QFT where the dynamics are uniquely determined by the sources and vacuum expectation values (VEVs).

One important feature to note in the subleading expansion (\ref{eq:UV}) is that the leading terms $z_0^\alpha$, $\tilde{z}_0^\alpha$ in the scalar fields are unconstrained, while the subleading pieces $z_1^\alpha$, $\tilde{z}_1^\alpha$ are determined by both the leading pieces and the magnetic and electric charges. As discussed in more detail in Section~\ref{sec:match}, this plays an important role when performing holographic renormalization for the black saddle solutions.

\subsection{IR expansion}
\label{sec:IR}

As discussed in the previous subsection, we are interested in Euclidean spacetimes with UV boundary $S^1 \times \Sigma_{\mg}$.  Then, as we go from the asymptotic boundary into the bulk, we want the spacetime to eventually cap off smoothly so as to avoid developing any singularities.  We refer to the location of this smooth cap-off in the bulk as the IR region of the solution.  In this section, we focus on the asymptotic solution to the BPS conditions in this IR region.

First we must ask what kind of asymptotic IR geometries are both smooth and also compatible with the BPS conditions and the equations of motion.  After carefully analyzing these equations, we found that there are two distinct possibilities: the solution can either approach $\mathbb{H}^2 \times \sigg$ or $\mathbb{R}^2 \times \sigg$.  The solutions with $\mathbb{H}^2 \times \sigg$ IR geometries are simply the Euclidean analogues of the extremal Lorentzian black hole solutions with near-horizon AdS$_2\times\Sigma_\mg$ geometry studied in \cite{Cacciatori:2009iz,Benini:2015eyy}.  We will not focus on these black hole solutions, as our goal is to establish that they are more naturally realized as a particular limit of Euclidean black saddle geometries.  We will therefore be interested in solutions where the $S^1$ shrinks to zero size  in the IR region such that we find a smooth $\mathbb{R}^2$ plane.

We denote the location of the IR to be at some radial coordinate $r = r_0$.  By requiring that $e^{f_1} \to 0$ in the IR, and also that $e^{f_1} f_1'$, $e^{f_2}$, and $e^{f_3}$ are all non-vanishing in the IR, the metric has an asymptotic expansion in the IR of the form
\begin{equation}
	ds^2 \to \left(e^{f_1} f_1'\right)^2|_\text{IR} (r - r_0)^2 d\tau^2 + e^{2f_2}|_\text{IR} dr^2 + e^{2 f_3}|_\text{IR} ds_{\Sigma_{\mathfrak{g}}}^2 + \ldots~,
\label{eq:irasymp}
\end{equation}
where the subscript IR denotes evaluation at $r=r_0$, and the dots indicate terms that are higher-order in $(r-r_0)$.  Neglecting these subleading terms, (\ref{eq:irasymp}) is simply a metric on $\mathbb{R}^2 \times \Sigma_{\mathfrak{g}}$, albeit written with a slightly unconventional set of polar coordinates.  Thus, the conditions above on the metric functions guarantee that the IR has the desired asymptotic $\mathbb{R}^2 \times \sigg$ geometry.  In order to avoid conical singularities in the metric \eqref{eq:irasymp}, the periodicity $\beta_\tau$ of the coordinate $\tau$ must be given by
\begin{equation}\label{eq:beta}
	\beta_\tau = \frac{2\pi e^{f_2 - f_1}}{f_1'} \bigg{|}_\text{IR}~.
\end{equation}

Our goal now is to determine what values the metric functions $e^{f_1}f_1'$, $f_2$ and $f_3$ and the fields $M$, $\wt{M}$, $z^\alpha$, $\tilde{z}^\alpha$, and $e^I$ all take in the IR by solving the BPS conditions perturbatively around the IR cap-off.  As in the previous section, we will do this assuming arbitrary magnetic and electric charges subject to the topological twist condition (\ref{eq:toptwistcond}).  In the IR region we have $e^{f_1} \to 0$ and this leads to various divergent and vanishing terms in the BPS equations. To remedy this we need to take one of the functions $M$ or $\wt{M}$ in the projectors \eqref{eq:project} to approach zero in the IR.  This in turn leads to two distinct branches of solutions to our BPS equations:
\begin{equation}\begin{aligned}
	&\text{Branch }1: \quad M \to 0 \text{ as } r \to r_0~, \\
	&\text{Branch }2: \quad \wt{M} \to 0 \text{ as } r \to r_0~.
\end{aligned}\end{equation}
If we now implement this in the BPS equations and solve them to leading order in powers of $(r-r_0)$, we find the following constraints on the IR values of the metric functions and fields:
\begin{equation}\label{eq:IR1}\begin{aligned}
\text{Branch } 1: \quad	g_I e^I|_\text{IR} &= - e^{-f_2+f_1}f_1'|_\text{IR} = -\frac{2\pi}{\beta_\tau}~, \\
	\left.e^{2f_3}\right|_\text{IR}&=-\frac{ i p^I F_I + q_I X^I}{g_I X^I}\bigg{|}_\text{IR}~, \\
	0 &= g^{\bar{\alpha} \beta} \nabla_\beta \left( i g_I X^I + e^{-2 f_3}(i q_I X^I - p^I F_I)\right)\bigg{|}_\text{IR}~, \\
\text{Branch } 2: \quad	g_I e^I|_\text{IR} &= e^{-f_2+f_1}f_1'|_\text{IR} = \frac{2\pi}{\beta_\tau}~, \\
	\left.e^{2f_3}\right|_\text{IR}&=\frac{ i p^I \bar{F}_I + q_I \bar{X}^I}{g_I\bar{X}^I}\bigg{|}_\text{IR}~, \\
	0 &= g^{\alpha \bar{\beta}} \nabla_{\bar{\beta}} \left( -i g_I \bar{X}^I + e^{-2 f_3}( i q_I \bar{X}^I - p^I \bar{F}_I) \right)\bigg{|}_\text{IR} ~.
\end{aligned}\end{equation}
The first relation for each branch puts a constraint on the periodicity $\beta_\tau$ in terms of the Wilson line parameters $e^I$.  For generic Euclidean solutions to the BPS equations, $\beta_\tau$ will be finite. However, in the $\beta_\tau \to \infty$ limit the cap-off in the IR is no longer controlled by the $\mathbb{R}^2$ metric in \eqref{eq:irasymp} but rather takes the form of two-dimensional hyperbolic space $\mathbb{H}^2$. These $\mathbb{H}^2 \times \Sigma_{\mathfrak{g}}$ IR solutions can be analytically continued to Lorentzian signature where they become precisely the AdS$_2\times \Sigma_{\mathfrak{g}}$ near-horizon geometries of the supersymmetric extremal black holes studied in \cite{Cacciatori:2009iz,Benini:2015eyy,Benini:2016rke}. For these black hole solutions the IR relation takes the form  $g_I e^I|_{\mathbb{H}^2 \times \sigg} = 0$.  This constraint has been noted before for BPS black holes in the STU model~\cite{Bobev:2018uxk} and it is clear from our analysis that it should be thought of as a limit of the more general relation in \eqref{eq:IR1}. 

The remaining relations in (\ref{eq:IR1}) on each branch give constraints on the size of the Riemann surface, which is determined by the metric function $f_3$, as well as the IR values of the scalars, purely in terms of the electric and magnetic charges.  Importantly, though, the constraints on Branch 1 are holomorphic and depend only on the scalars $z^\alpha$, while the constraints on Branch 2 are anti-holomorphic and depend only on the scalars $\tilde{z}^\alpha$.  Moreover, since there are four of these constraints on each branch, this means that on Branch 1, the function $f_3$ and the scalars $z^\alpha$ have their IR values determined completely by the charges, while the IR values of the remaining scalars $\tilde{z}^\alpha$ are free.  On Branch 2, it is the other way around: the $\tilde{z}^\alpha$ are constrained while the $z^\alpha$ are free.  

The crucial takeaway from this analysis is that either $z^\alpha$ or $\tilde z^\alpha$ are unconstrained in the IR.  For Lorentzian extremal black holes, these scalars are complex conjugates and thus all scalars are completely determined by the magnetic and electric charges. This is sometimes referred to as an attractor mechanism for asymptotically AdS$_4$ black holes \cite{Cacciatori:2009iz,DallAgata:2010ejj,Hristov:2010ri}. In our Euclidean setup however, $\tilde z^\alpha$ is independent from $z^\alpha$ and depending on the branch, only one of the two becomes fixed in terms of the charges.  It is precisely this freedom that allows for a large family of new Euclidean solutions with no corresponding Lorentzian analogue; only in the limit where $\tilde{z}^\alpha \to \bar{z}^\alpha$ do we recover the subset of solutions with nice Wick-rotated Lorentzian geometries that were studied in \cite{Cacciatori:2009iz}.

The IR BPS conditions can be studied to an arbitrarily high order in the $(r-r_0)$ expansion. This explicit calculation will not be needed to compute the on-shell action.  However, it is important to note that in the higher-order expansion, we find that all derivatives of the fields and metric functions are determined by the leading order expansion.  Thus, the Euclidean black saddle solutions are determined entirely by ten parameters: the magnetic charges $p^I$ (subject to the topological twist constraint (\ref{eq:toptwistcond})), the electric charges $q_I$, and either $z^\alpha|_\text{IR}$ or $\tilde{z}^\alpha|_\text{IR}$, depending on which branch of solutions to the BPS equations we work with.

\subsection{Holographic renormalization}
\label{sec:holorenorm}

We are now interested in evaluating the action \eqref{eq:SEucl} on-shell for solutions to the BPS equations.  As a first step, let us implement both our ansatz \eqref{eq:ansatz} and the equations of motion, without making use of the BPS equations yet.  We can use the equations of motion to trade all explicit dependence on the scalar fields for the metric functions and the gauge fields instead.  The end result is that the bulk action \eqref{eq:SEucl} on our ansatz simply becomes
\begin{equation}\begin{aligned}
	S_\text{bulk} &= \frac{\text{Vol}(\Sigma_{\mathfrak{g}})\beta_\tau}{8\pi G_N} \int dr\, \left[ e^{f_1 - f_2 + 2 f_3} f_1' - q_I e^I \right]'~,
\label{eq:sbulk}
\end{aligned}\end{equation}
where the prime denotes a derivative with respect to $r$, $\beta_\tau$ is the periodicity of the coordinate $\tau$, and $\text{Vol}(\Sigma_{\mathfrak{g}})$ is the volume of the Riemann surface, which depends on the genus as follows:
\begin{equation}\label{eq:VolSigmaeta}
	\text{Vol}(\Sigma_{\mathfrak{g}}) = 2 \pi \eta~, \quad \eta \equiv \begin{dcases} 2 |\mathfrak{g} - 1| & \mathfrak{g}\neq 1~, \\
	1 & \mathfrak{g} =1 ~.
	\end{dcases}
\end{equation}
The full details of how we derive this on-shell action are presented in Appendix~\ref{app:moreholorenorm}.  The steps involved are tedious but conceptually straightforward, so we will not dwell on them here.

An important and non-trivial feature of the on-shell action (\ref{eq:sbulk}) is that the integrand is a total derivative on all solutions of the equations of motion captured by the ansatz in \eqref{eq:ansatz}.  In similar studies of Euclidean BPS solutions dual to ABJM~\cite{Freedman:2013oja,Bobev:2018wbt} the on-shell action was computed only after constructing explicit solutions, either analytically or numerically.  Moreover, to derive this total-derivative form of the on-shell action we did not use supersymmetry. This is to be contrasted with the results in \cite{Genolini:2016ecx,BenettiGenolini:2019jdz}, where it was shown that the on-shell action for BPS solutions to minimal supergravity receives contributions only from the fixed points of a Killing vector associated to a Killing spinor of the background. 

Since the integrand in (\ref{eq:sbulk}) is a total derivative, the on-shell bulk Euclidean action reduces to a boundary term.  Therefore there are two contributions to this on-shell action: a UV contribution at the conformal boundary at $r \to \infty$, and an IR contribution from the cap-off at $r = r_0$.  That is,
\begin{equation}
	S_\text{bulk} = S_\text{UV} - S_\text{IR}~,
\end{equation}
where we have defined
\begin{equation}\label{eq:SUVSIRexpl}\begin{aligned}
	S_\text{UV} &\equiv \frac{\text{Vol}(\Sigma_{\mathfrak{g}})\beta_\tau}{8\pi G_N} \left( e^{f_1-f_2 + 2f_3}f_1' - q_I e^I\right)\bigg|_\text{UV}~, \\
	S_\text{IR} &\equiv \frac{\text{Vol}(\Sigma_{\mathfrak{g}})\beta_\tau}{8\pi G_N} \left( e^{f_1-f_2 + 2f_3}f_1' - q_I e^I\right)\bigg|_\text{IR}~.
\end{aligned}\end{equation}
Therefore the calculation of the on-shell action for the black saddle solutions of interest boils down to simply evaluating the particular combinations in \eqref{eq:SUVSIRexpl} in the UV and the IR.  We have already determined the conditions on the supergravity fields in these two asymptotic regions in Sections~\ref{sec:UV} and~\ref{sec:IR}, and we now use these in order to determine their corresponding contributions to the on-shell action.

\subsubsection*{The IR contribution}

First, we focus on the IR contribution.  Using the relation (\ref{eq:beta}) between the periodicity $\beta_\tau$ and the IR values of the metric functions, we first rewrite the IR piece of the on-shell action more compactly as
\begin{equation}\label{eq:SIR}
	S_\text{IR} = \frac{\text{Vol}(\Sigma_{\mathfrak{g}})}{4G_N} \left(e^{2 f_3} - \frac{\beta_\tau}{2\pi}\, q_I e^I \right)\bigg|_\text{IR}~.
\end{equation}
From here, our goal is to use the IR expansion of the BPS equations (\ref{eq:IR1}) to express this quantity in terms of the IR values of the scalars as much as possible.  In particular, we would like to express the IR action in terms of the scalars $z^\alpha$ on Branch 1 and the scalars $\tilde{z}^\alpha$ on Branch 2, since these are the scalar fields that get fixed entirely in terms of the charges.  In doing so, it will be convenient to first repackage these scalar fields into new quantities $\hat{u}^I$, defined by
\begin{equation}\begin{aligned}\label{eq:hatu}
	&\text{Branch }1: \quad \hat{u}^I = \frac{2\pi X^I}{\sum_I X^I}\bigg{|}_\text{IR}~, \\
	&\text{Branch }2: \quad \hat{u}^I = \frac{2\pi \bar{X}^I}{\sum_I \bar{X}^I}\bigg{|}_\text{IR}~.
\end{aligned}\end{equation}
That is, depending on the branch of BPS solutions we are interested in, the $\hat{u}^I$ are related to either the holomorphic or anti-holomorphic symplectic sections.  In particular, the $\hat{u}^I$ encode information about the scalars that are fixed by the charges in the IR.  We can also define $\hat{u}^\alpha_F$, with $\alpha = 1,2,3$, as the same objects, but in a new basis akin to the new basis for the charges in (\ref{eq:newbasis}):
\begin{equation}\begin{aligned}
	\hat{u}_F^1 &= \frac{1}{2}\left(\hat{u}^0 + \hat{u}^1 - \hat{u}^2 - \hat{u}^3\right)~, \\
	\hat{u}_F^2 &= \frac{1}{2}\left(\hat{u}^0 - \hat{u}^1 + \hat{u}^2 - \hat{u}^3\right)~, \\
	\hat{u}_F^3 &= \frac{1}{2}\left(\hat{u}^0 - \hat{u}^1 - \hat{u}^2 + \hat{u}^3\right)~.
\end{aligned}\end{equation}
The utility of the definitions of $\hat{u}^{I}$ in \eqref{eq:hatu} will become clear in Section~\ref{sec:match} when we establish a detailed holographic dictionary between quantities in the  ABJM theory and parameters of the Euclidean black saddle solutions.  In particular, the $\hat{u}^I$ are the supergravity quantities dual to the complex fugacities $u^I$ in the ABJM theory \eqref{eq:uIdefQFT}. In the analysis of~\cite{Benini:2015eyy,Benini:2016rke} a similar definition of the $\hat{u}^{I}$ was used. However we note that in \cite{Benini:2015eyy,Benini:2016rke} the supergravity values of $\hat{u}^{I}$ were computed in the near-horizon AdS$_2\times \Sigma_{\mathfrak{g}}$ region of a Lorentzian black hole. It was argued in \cite{Benini:2015eyy,Benini:2016rke} that these values correspond to complex field theory fugacities $u^I$ for which the topologically twisted index is extremized. In our setting we define $\hat{u}^{I}$ in \eqref{eq:hatu} for general Euclidean black saddle solutions with an $\mathbb{R}^2\times \Sigma_{\mathfrak{g}}$ IR region. Correspondingly, as we show in Section~\ref{sec:match}, we are able to recover the ABJM topologically twisted index for general values of the complex fugacities $u^{I}$. Put differently, the $\hat{u}^{I}$ in \eqref{eq:hatu} extend the results in \cite{Benini:2015eyy,Benini:2016rke} for general values of $\beta_\tau$ and reduce to theirs in the $\beta_\tau \to \infty$ limit.

With these definitions in mind, and using the IR BPS constraints in \eqref{eq:IR1}, we find that the IR contribution to the on-shell action is given by
\begin{equation}
S_\text{IR} = \frac{\text{Vol}(\sigg)}{8\pi G_N}\left[ \frac{1}{\xi g} \sqrt{\hat{u}^0 \hat{u}^1 \hat{u}^2 \hat{u}^3} \sum_{I=0}^3 \frac{p^I}{\hat{u}^I} + \sum_{\alpha = 1}^3 q_F^\alpha\left( \pm \frac{\hat{u}_F^\alpha}{\xi g} - \beta_\tau e_F^\alpha|_\text{IR}\right)\right]~,
\label{eq:sirans}
\end{equation}
where the $+/-$ signs correspond to Branch $1$ and Branch $2$, respectively. Interestingly, the BPS conditions on $e^I$ and $e^{2 f_3}$ conspire to eliminate the dependence of $S_\text{IR}$ on $q_R$.  What we are left with is a relatively simple expression that depends on the magnetic charges, the flavor electric charges, the IR values of the scalars that are fixed in terms of these charges, and the IR values of the flavor Wilson lines.  Importantly, the scalars not fixed by supersymmetry ($\tilde{z}^\alpha$ on Branch 1, and $z^\alpha$ on Branch 2) do not appear anywhere in \eqref{eq:sirans}.

\subsubsection*{The UV contribution}

Now, we proceed with the evaluation of the UV contribution to $S_\text{bulk}$.  We take the UV boundary to be at some radial coordinate $\rho_b$  that we will eventually send to infinity. This procedure results in a divergent expression for $S_\text{UV}$.  As is standard in holographic contexts, we must regularize this UV action by adding the appropriate boundary counterterms; see \cite{Skenderis:2002wp} for a review of the holographic renormalization procedure.  The counterterm action is the sum of three distinct terms
\begin{equation}
	S_\text{CT} = S_\text{GH} + S_{\mathcal{R}} + S_\text{SUSY}~,
\label{eq:sct1}
\end{equation}
with
\begin{equation}\begin{aligned}
	S_\text{GH} &\equiv -\frac{1}{8\pi G_N}\int d^3x\,\sqrt{h}\,K~, \\
	S_{\mathcal{R}} &\equiv \frac{L}{16\pi G_N} \int d^3x\,\sqrt{h}\, \mathcal{R}~, \\
	S_\text{SUSY} &\equiv  \frac{1}{8\pi G_N L} \int d^3x\,\sqrt{h}\,e^{\mathcal{K}/2} \sqrt{\mathcal{V} \tilde{\mathcal{V}}}~,
\label{eq:sct2}
\end{aligned}\end{equation}
where $h$ is the determinant of the induced metric, $h_{ab}$, on the conformal boundary, $K$ is the trace of the extrinsic curvature, $\mathcal{R}$ the Ricci scalar of $h_{ab}$, and $\mathcal{V}$ and $\tilde{\mathcal{V}}$ are the superpotentials in \eqref{eq:superpotdef}, given by
\begin{equation}
	\mathcal{V} = 2(z^1 z^2 z^3 - 1)~, \quad \tilde{\mathcal{V}} = 2(\tilde{z}^1 \tilde{z}^2 \tilde{z}^3 -1 )~.
\end{equation}
It is worth emphasizing once again that these superpotentials are no longer complex conjugates of one another in the Euclidean theory; they should be treated as \emph{distinct} functions of the scalars $z^\alpha$ and $\tilde{z}^\alpha$.

The first term, $S_\text{GH}$, in (\ref{eq:sct1}) is the familiar Gibbons-Hawking boundary term, which is needed not only for a well-posed variational principle but also to remove the usual divergences of the Einstein-Hilbert action.  The second term, $S_\mathcal{R}$, is another standard counterterm in holographic renormalization; it is the term associated with the curvature of the boundary surface.  Both of these counterterms are divergent and end up contributing no finite pieces to the regularized action in the limit where we send $\rho_b \to \infty$.  

The third term, $S_\text{SUSY}$, in (\ref{eq:sct1}) is perhaps less familiar.  It depends both on the metric and on the scalar fields at the boundary.  If we expand it out in successive powers of the scalar fields, it takes the form
\begin{equation}
	S_\text{SUSY} = \frac{1}{4\pi G_N L}\int d^3x\,\sqrt{h}\,\left(1 + \frac{1}{2}\left(z^1 \tilde{z}^1 + z^2 \tilde{z}^2 + z^3 \tilde{z}^3\right) - \frac{1}{2}\left(z^1 z^2 z^3 + \tilde{z}^1 \tilde{z}^2 \tilde{z}^3\right) + \ldots \right)~,
\label{eq:ssusyexp}
\end{equation}
where the dots indicate terms that are at least quartic in the scalar fields.  In the limit $\rho_b \to \infty$, the constant term and the quadratic terms above are both divergent, while the cubic terms remain finite. All higher-order terms vanish in this limit, due to the $e^{-\rho}$ fall-off behavior of the scalar fields in the UV expansion (\ref{eq:UVexpans}).  The divergent terms in (\ref{eq:ssusyexp}) are precisely the ones we would obtain by doing the usual near-boundary analysis in holographic renormalization.  The finite and subleading terms, on the other hand, cannot be determined in this way; instead, they are determined by demanding that the counterterms respect the supersymmetry of the theory.  The particular combination that shows up in (\ref{eq:ssusyexp}) is the one demanded by supersymmetry, as demonstrated in~\cite{Freedman:2013oja,Freedman:2016yue}.  

The counterterms in \eqref{eq:sct2} are all of the ``infinite'' counterterms, i.e. the ones necessary to cancel all divergences in the bulk action. One should in principle allow also for arbitrary finite counterterms which are covariant in the supergravity fields. These terms should take the schematic form\footnote{We  do not include parity-breaking finite counterterms like Chern-Simons terms for the flavor gauge fields. These counterterms should be compatible with supersymmetry but affect only the imaginary part of the on-shell action, see for instance \cite{Closset:2012vg,Closset:2012vp}.}
\begin{equation}
	\int d^3x\,\sqrt{h}\, f\left(z^\alpha, \tilde{z}^\alpha, A^I_\mu\right) \mathcal{R}~,
\end{equation}
for some function $f$ that depends on the scalars and vector fields.  However, from the fall-off conditions satisfied by the fields in the UV expansion (\ref{eq:UVexpans}), one can show that the only such counterterms that are covariant and also remain finite and non-zero when we send $\rho_b \to \infty$ are the ones where $f$ is \emph{linear} in the scalar fields.  This means that the only possible finite counterterms we can add to the STU model at hand are
\begin{equation}
	S_{\text{finite}} = \frac{L}{16\pi G_N}\int d^3x\,\sqrt{h}\,\sum_{\alpha=1}^3 \left(c^\alpha z^\alpha + \tilde{c}^\alpha \tilde{z}^\alpha\right)\mathcal{R}~,
\label{eq:sfin}
\end{equation}
for some constants $c^\alpha$ and $\tilde{c}^\alpha$.  These constants should in principle be fixed by demanding that the counterterm action is compatible with supersymmetry, which requires doing a careful study of how the bulk supersymmetry is realized on the boundary of AdS$_4$.  A partial analysis of this question has been done in~\cite{Freedman:2016yue}, but no full systematic analysis along the lines of~\cite{Belyaev:2007bg,Grumiller:2009dx} has thus far been attempted in four-dimensional gauged supergravity.  In view of this, we keep $c^\alpha$ and $\tilde{c}^\alpha$ as arbitrary constants for the moment and will fix them later using symmetry arguments and holographic Ward identities.

The regularized UV contribution to the action, with the addition of these counterterms, is given by
\begin{equation}
	S_\text{UV}^{\text{(reg)}} = S_\text{UV} + S_\text{CT} + S_\text{finite}~.
\end{equation}
Writing all terms out explicitly and integrating over the conformal boundary coordinates, this regularized action takes the explicit form
\begin{equation}\label{eq:SregUVgeneral}\begin{aligned}
	S_\text{UV}^\text{(reg)} = \frac{\text{Vol}(\sigg)\beta_\tau}{8\pi G_N} \bigg{[} &- 2 e^{f_1 - f_2 + 2 f_3} f_3' - q_I e^I + \frac{\kappa e^{f_1}}{\sqrt{2} g} + \sqrt{2} g e^{f_1 + 2 f_3}e^{\mathcal{K}/2} \sqrt{\mathcal{V} \tilde{\mathcal{V}}} \\
	&  + \frac{\kappa e^{f_1}}{\sqrt{2}g}\sum_{\alpha=1}^3 \left(c^\alpha z^\alpha + \tilde{c}^\alpha \tilde{z}^\alpha\right) \bigg{]}_\text{UV}~,
\end{aligned}\end{equation}
with all fields evaluated at the boundary cut-off region located at $\rho = \rho_b$.  Importantly, this result applies for any arbitrary solution to the bulk equations of motion of the form in \eqref{eq:ansatz}. 

We now focus on evaluating \eqref{eq:SregUVgeneral} on the Euclidean black saddle solutions we are interested in.  To this end we use the UV expansion of the BPS equations (\ref{eq:UVexpans}) to eliminate all metric functions in \eqref{eq:SregUVgeneral} in favor of the scalar fields, the charges, and the Wilson lines.  The result of this procedure is that the regularized UV contribution to the on-shell action is finite, and given simply by
\begin{equation}\begin{aligned}
	S_\text{UV}^{\text{(reg)}} = \frac{\text{Vol}(\sigg)\beta_\tau}{8\pi G_N} \sum_{\alpha = 1}^3 \bigg{[}&\frac{\xi a_0}{\sqrt{2}}\left( p^\alpha_F (z_0^\alpha + \tilde{z}_0^\alpha) - q^\alpha_F (z_0^\alpha - \tilde{z}_0^\alpha)\right)  - q^\alpha_F e^{\alpha}_{F,0} \\
	&+ \frac{\kappa a_0}{\sqrt{2}g}( c^\alpha z^\alpha_0 + \tilde{c}^\alpha \tilde{z}^\alpha_0 )\bigg{]}~,
\label{eq:suv}
\end{aligned}\end{equation}
where we have sent $\rho_b \to \infty$ and dropped all terms that go as $\mathcal{O}(e^{-\rho})$.  Note that this expression, like the IR contribution (\ref{eq:sirans}), depends only on the flavor charges $p^\alpha_F$ and $q_F^\alpha$, the flavor Wilson lines $e_F^\alpha$, and the scalars.  Although there is a priori no reason to exclude the possibility that this UV piece of the action depends on $q_R$ and $e_R$, the BPS conditions conspire to cancel all such dependence from the result (\ref{eq:suv}).  Additionally, this regularized action only depends on the \emph{leading}-order pieces in the Wilson lines and the scalars in the UV; as we mentioned earlier, the higher-order terms drop out as we take the limit $\rho_b \to \infty$, and thus the higher-order terms in the UV expansion \eqref{eq:UVexpans} do not affect the calculation of the on-shell action.

\subsubsection*{Alternative quantization}

To summarize, we have so far shown that the bulk on-shell action of our theory has only two contributions: one from the IR  and another from the UV region (supplemented with the appropriate boundary counterterms).  The full on-shell action for any solution to the BPS equations is hence given by
\begin{equation}
	S_\text{on-shell} = S_\text{UV}^\text{(reg)} - S_\text{IR}~,
\end{equation}
where $S_\text{IR}$ is given in (\ref{eq:sirans}), and $S_\text{UV}^\text{(reg)}$ is given in (\ref{eq:suv}).  However, we cannot yet identify the partition function of our theory with this on-shell action; we must first tackle an additional subtlety related to how to properly quantize the scalar fields in the theory.

We have been somewhat cavalier in our notation and referred to all of the fields $z^\alpha$ and $\tilde{z}^\alpha$ as scalar fields.  However, this is not quite correct; the full $\mathcal{N}=8$ supergravity theory (of which our theory is a truncation) contains both scalars and pseudoscalars, which are dual to scalar operators of dimension one and pseudoscalar operators of dimension two, respectively, in the ABJM theory. If we track down how our fields are built out of the full $\mathcal{N}=8$ theory, we find that the linear combinations $z^\alpha + \tilde{z}^\alpha$ transform as scalars, while $z^\alpha - \tilde{z}^\alpha$ transform as pseudoscalars.  This distinction is important, because, as shown in~\cite{Breitenlohner:1982jf,Klebanov:1999tb}, supersymmetry requires that in Lorentzian signature, the real and imaginary parts of the scalars are quantized differently.  The corresponding statement in our Euclidean theory is that all pseudoscalars are quantized with regular boundary conditions, while the scalars obey \emph{alternative quantization}, i.e. they satisfy the alternate boundary conditions prescribed in~\cite{Klebanov:1999tb}.

With standard boundary conditions, one takes the asymptotics of the scalar field
\begin{equation}
	z^\alpha= z^\alpha_0 e^{-\rho}+z^\alpha_1 e^{-2\rho}+\mathcal{O}\left(e^{-3\rho}\right)~,
\end{equation}
and interprets the leading piece $z^\alpha_0$ as determining the source and the subleading piece $z_1^\alpha$ the VEV in the dual field theory.  The standard on-shell action is then a function of the sources $z^\alpha_0$ and $\tilde z^\alpha_0$ and it should match the partition function of the dual field theory.

For alternative quantization, though, the role of the source and the VEV are interchanged, in the sense that one must take the source to be proportional to the canonical conjugate to $z^\alpha_0$ and the VEV to the canonical conjugate to $z^\alpha_1$.  Correspondingly, the on-shell action must be Legendre-transformed with respect to the scalars $z^\alpha + \tilde{z}^\alpha$ in order for it to be a function of only the sources in the dual field theory.  This Legendre-transformed action, which we denote by $J$, is the bulk dual of the generating functional for correlation functions in the dual CFT.

In order to implement the Legendre transform, we first compute the canonical conjugates $\Pi^\alpha$ and $\tilde{\Pi}^\alpha$ of $z^\alpha$ and $\tilde z^\alpha$, respectively, to be
\begin{equation}\begin{aligned}
	\Pi^\alpha &= \frac{\delta S_\text{on-shell}}{\delta (\partial_r z^\alpha)} = -\frac{a_0 e^\rho}{8 \sqrt{2}\pi g G_N} \left( \tilde{z}^\alpha_1 + \frac{z^1_0 z^2_0 z^3_0}{z^\alpha_0}\right) + \frac{\kappa a_0 e^\rho}{8\sqrt{2} \pi g G_N} c^\alpha +\ldots~, \\
	\tilde{\Pi}^\alpha &= \frac{\delta S_\text{on-shell}}{\delta (\partial_r \tilde{z}^\alpha)} = -\frac{a_0 e^\rho}{8\sqrt{2}\pi g G_N} \left( {z}^\alpha_1 + \frac{\tilde{z}^1_0 \tilde{z}^2_0 \tilde{z}^3_0}{\tilde{z}^\alpha_0}\right) + \frac{\kappa a_0 e^\rho}{8\sqrt{2}\pi g G_N} \tilde{c}^\alpha + \ldots~,
\end{aligned}\end{equation}
where the dots indicate terms that are finite as $\rho \to \infty$ and do not contribute to the Legendre transform. The second term in between brackets results from the boundary term $S_\text{SUSY}$ in \eqref{eq:ssusyexp}, required in order to preserve supersymmetry. The BPS constraints impose the relations \eqref{eq:UV} between the subleading pieces $z^\alpha_1$  and the leading pieces $z^\alpha_0$, which in turn leads to
\begin{equation}\begin{aligned}
	\Pi^\alpha &= \frac{\xi a_0 e^\rho}{8\sqrt{2}\pi G_N} \left( p^\alpha_F - q^\alpha_F\right) + \frac{\kappa a_0 e^\rho}{8\sqrt{2}\pi g G_N} c^\alpha + \ldots ~, \\
	\tilde{\Pi}^\alpha &= \frac{\xi a_0 e^\rho}{8\sqrt{2}\pi G_N} \left( p^\alpha_F + q^\alpha_F\right) + \frac{\kappa a_0 e^\rho}{8\sqrt{2}\pi g G_N} \tilde{c}^\alpha + \ldots ~.
\end{aligned}\end{equation}
The pseudoscalars $z^\alpha - \tilde z^\alpha$ obey standard quantization, while the scalars $z^\alpha + \tilde z^\alpha$ obey alternative quantization.  This leads to the following quantities being proportional to sources for pseudoscalars and scalars, respectively:
\begin{equation}\label{eq:sources}\begin{aligned}
	\lim_{\rho \to \infty} e^\rho(z^\alpha - \tilde{z}^\alpha) &= z^\alpha_0 - \tilde{z}^\alpha_0~, \\
	\lim_{\rho \to \infty} e^{-\rho}(\Pi^\alpha + \tilde{\Pi}^\alpha) &= \frac{\sqrt{2} \xi a_0 p^\alpha_F}{8\pi G_N} + \frac{\kappa  a_0}{8\sqrt{2}\pi g G_N}( c^\alpha + \tilde{c}^\alpha )~.
\end{aligned}\end{equation}
The Legendre-transformed on-shell action then takes the form
\begin{equation}\begin{aligned}\label{eq:J0}
	J &= S_\text{on-shell} - \frac{1}{2}\sum_{\alpha=1}^3 \int_{\text{UV}} d^3x\, (z^\alpha + \tilde{z}^\alpha)(\Pi^\alpha + \tilde{\Pi}^\alpha)~,
\end{aligned}\end{equation}
and is a function of the sources \eqref{eq:sources}, rather than $z^\alpha_0$ and $\tilde z^\alpha_0$.

Using \eqref{eq:sirans} and \eqref{eq:suv} we find that the full Legendre-transformed on-shell action is
\begin{equation}\label{eq:J}\begin{aligned}
	 J = \frac{\text{Vol}(\sigg)}{8\pi G_N}\bigg{[}&-\frac{1}{\xi g} \sqrt{\hat{u}^0 \hat{u}^1 \hat{u}^2 \hat{u}^3} \sum_{I=0}^3 \frac{p^I}{\hat{u}^I} \\
	 & - \sum_{\alpha = 1}^3 q_F^\alpha\left( \pm \frac{\hat{u}_F^\alpha}{\xi g} + \beta_\tau \left( e_F^\alpha|_\text{IR} - e_F^\alpha|_\text{UV}\right) + \frac{\xi a_0 \beta_\tau}{\sqrt{2}} \left(z_0^\alpha - \tilde{z}_0^\alpha\right)\right) \\
	 & + \sum_{\alpha=1}^3 \frac{\kappa a_0 \beta_\tau}{\sqrt{2} g}\left(c^\alpha - \tilde{c}^\alpha\right)\left(z_0^\alpha - \tilde{z}_0^\alpha\right)\bigg{]}~,
\end{aligned}\end{equation}
where $\pm$ corresponds to Branch $1$ or Branch $2$.  As discussed earlier $J$, and not $S_\text{on-shell}$, should be related to the gravitational partition function, which in turn should match the partition function of the dual CFT.  More precisely we should identify
\begin{equation}
	\log Z_\text{grav} = - J~,
\end{equation}
in the semi-classical saddle-point approximation to the gravity path integral.
%

\subsubsection*{Fixing the finite counterterms}

Before we compare the gravitational partition to the one in the dual field theory, we must revisit the finite counterterms (\ref{eq:sfin}) that we included in our computation.  As we discussed earlier, we could in principle constrain the coefficients $c^\alpha$ and $\tilde{c}^\alpha$ via an elaborate analysis in supergravity by tracking how the bulk symmetries and supersymmetry are realized at the UV boundary and which finite counterterms respect these symmetries.  However, there is a much simpler way to fix $c^\alpha$ and $\tilde{c}^\alpha$; we can employ holography and demand that the bulk supergravity theory agrees with the dual field theory, both in its symmetries and in all observables.

First, we note that the finite counterterms (\ref{eq:sfin}) can be rewritten as
\begin{equation}
	S_\text{finite} = \frac{L}{32 \pi G_N} \int d^3x\,\sqrt{h}\,\sum_{\alpha =1 }^3\bigg{(}(c^\alpha + \tilde{c}^\alpha)(z^\alpha + \tilde{z}^\alpha) + (c^\alpha - \tilde{c}^\alpha)(z^\alpha - \tilde{z}^\alpha)\bigg{)} \mathcal{R}~.
\end{equation}
The combinations $z^\alpha + \tilde{z}^\alpha$ and $z^\alpha - \tilde{z}^\alpha$ transform as scalars and pseudoscalars, respectively.  However, since the dual $U(N)_{-k} \times U(N)_k$ ABJM theory has equal and opposite Chern-Simons levels for each $U(N)$ gauge group, the theory is parity invariant.  Since pseudoscalars are odd under parity transformations, no bare pseudoscalars can appear in the counterterm action. This symmetry argument immediately leads to the constraint
\begin{equation}\label{eq:c+ctilde=0}
	c^\alpha - \tilde{c}^\alpha = 0~,
\end{equation}
for all values of the index $\alpha$.

Another constraint on the finite counterterms comes from supersymmetry.  As discussed in Section~\ref{sec:fieldtheorypartfunc}, when we place ABJM on $S^1 \times \sigg$ with a topological twist, we are free to turn on background vector multiplets for the $U(1)_F^3$ symmetry that act as sources for the dynamical fields in the theory.  In Lorentzian signature, these three vector multiplets $(A_{F,a}^\alpha, \lambda_F^\alpha, \sigma_F^\alpha, D_F^\alpha)$ consist of a vector field $A_{F,a}^\alpha$, a complex fermion $\lambda_F^\alpha$, and two real scalars $\sigma_F^\alpha$ and $D_F^\alpha$, where $\alpha = 1,2,3$.  In Euclidean signature the same field content is present, but the bosonic fields become complex and the fermions $\lambda^\alpha_F$ and $\lambda_F^{\dagger,\alpha}$ are treated as independent fermions.  Importantly, $D_F^\alpha$ is an auxiliary field whose value is fixed by supersymmetric Ward identities.  For a large class of three-dimensional $\mathcal{N}=2$ gauge theories, including the ABJM theory of interest for us, the Ward identities on $S^1 \times \sigg$ with a partial topological twist imply that~\cite{Benini:2015noa}
\begin{equation}\label{eq:DFWard}
	D_F^\alpha = i\mathfrak{p}^\alpha_F~,
\end{equation}
where $\mathfrak{p}^\alpha_F$ is the flavor magnetic flux across the Riemann surface.  In holography, the values of $\sigma_F^\alpha$ and $D_F^\alpha$ can be read off from the source relations in (\ref{eq:sources}). Ignoring an overall normalization constant we therefore find the relations
\begin{equation}\label{eq:sigmaFDFholo}\begin{aligned}
	\sigma_F^\alpha &\propto z_0^\alpha - \tilde{z}_0^\alpha~, \\
	D_F^\alpha &\propto \frac{\sqrt{2} \xi a_0 p^\alpha_F}{8\pi G_N} + \frac{\kappa  a_0}{8\sqrt{2}\pi g G_N}( c^\alpha + \tilde{c}^\alpha )~.
\end{aligned}\end{equation}  
The supergravity magnetic fluxes $p_F^\alpha$ are proportional to the field theory fluxes $\mathfrak{p}_F^\alpha$.  Therefore the only way to reproduce the Ward identity in \eqref{eq:DFWard} in the supergravity dual description is to set 
\begin{equation}
\label{eq:c-ctilde=0}
	c^\alpha + \tilde{c}^\alpha = 0~,
\end{equation} in \eqref{eq:sigmaFDFholo} and thus arrive $D_F^\alpha = i \mathfrak{p}_F^\alpha \propto p_F^\alpha$.

By combining the above constraints \eqref{eq:c+ctilde=0} and \eqref{eq:c-ctilde=0} on the finite counterterms, we find that we must set $c^\alpha = \tilde{c}^\alpha = 0$ in order for the bulk gravity theory to be consistent with the dual field theory.  As a consistency check, we can use the fact that counterterms are fixed at the level of the supergravity theory itself and do not depend on a particular solution.  This means that the counterterms we use here for the black saddle solutions of the form \eqref{eq:ansatz} must be compatible with the counterterms that have been used in previous successful holographic studies of the STU model.  In particular, it was shown in~\cite{Freedman:2013oja} that when placing ABJM on $S^3$ and studying its supergravity dual, the finite counterterms in \eqref{eq:sfin} must vanish in order to both match the field theory partition function as well as reproduce the supersymmetric Ward identity for the field theory sources.  More specifically, the combination $c^\alpha + \tilde{c}^\alpha$ affects the source relations in~\cite{Freedman:2013oja}, while the combination $c^\alpha - \tilde{c}^\alpha$ affects their result for the partition function, and thus both of these counterterms should be set to zero.  Since in \cite{Freedman:2013oja} they study the same Euclidean supergravity model we therefore conclude that we need to set
\begin{equation}
	c^\alpha = \tilde{c}^\alpha = 0~,
\end{equation}
in \eqref{eq:sfin}, which is exactly what we concluded from \eqref{eq:c+ctilde=0} and \eqref{eq:c-ctilde=0} using symmetry arguments.  By setting these finite counterterms to zero in \eqref{eq:J} we arrive at our end result for the semiclassical partition function for the black saddles of our supergravity theory 
\begin{equation}\label{eq:JJ}\begin{aligned}
	 \log Z_\text{grav} = \frac{\text{Vol}(\sigg)}{8\pi G_N}&\bigg{[}\frac{1}{\xi g} \sqrt{\hat{u}^0 \hat{u}^1 \hat{u}^2 \hat{u}^3} \sum_{I=0}^3 \frac{p^I}{\hat{u}^I} \\
	 & + \sum_{\alpha = 1}^3 q_F^\alpha\left( \pm \frac{\hat{u}_F^\alpha}{\xi g} + \beta_\tau \left( e_F^\alpha|_\text{IR} - e_F^\alpha|_\text{UV}\right) + \frac{\xi a_0 \beta_\tau}{\sqrt{2}} \left(z_0^\alpha - \tilde{z}_0^\alpha\right)\right)\bigg{]}~.
\end{aligned}\end{equation}

It is worth comparing our holographic renormalization analysis to the results in ~\cite{Halmagyi:2017hmw,Cabo-Bizet:2017xdr} where this question was studied for Lorentzian black hole solution of the STU model. In particular, it is argued in these references that one has to choose finite counterterms such that any source for dimension-one operators in the dual field theory must vanish.  In our context, this statement translates into choosing the finite counterterms in \eqref{eq:sfin} such that we obtain $D_F^\alpha = 0$, since $D_F^\alpha$ source dimension-one scalar operators in the field theory.  Such a prescription is in direct conflict with the ABJM Ward identities in \eqref{eq:DFWard} that clearly impose $D_F^\alpha$ to be non-zero and proportional to the magnetic fluxes along the Riemann surface.  Moreover, the arguments of~\cite{Cabo-Bizet:2017xdr} are in direct tension with the precise holographic computations of sources found in~\cite{Freedman:2013oja} for the holographic dual of the ABJM theory on $S^3$ and in~\cite{Bobev:2018wbt} for a superpotential mass-deformation of the ABJM theory on $S^3$. 

As we discuss in detail below one can show that the second line in \eqref{eq:JJ} vanishes on all regular solutions of the BPS equations.  Such a cancellation is not obvious a priori but is indeed crucial to establish a detailed agreement between \eqref{eq:JJ} and the topologically twisted index of the ABJM theory. This is to be contrasted with the analysis in \cite{Cabo-Bizet:2017xdr} where it is stated that only the terms proportional to $\beta_{\tau}$ in \eqref{eq:JJ} vanish on regular solutions. We suspect that this discrepancy arises from the fact that the authors of \cite{Cabo-Bizet:2017xdr} work with regular Lorentzian solutions which necessarily have $\beta_\tau = \infty$. Keeping the parameter $\beta_\tau$ finite is crucial for establishing the result in \eqref{eq:JJ} which in turn allows for a proper treatment of the $\beta_\tau \to \infty$ limit.

\subsection{The holographic match} 
\label{sec:match}

In this section, we establish a detailed holographic match between the topologically twisted index of the ABJM theory in the planar limit \eqref{eq:cftpart2} and the partition function of the black saddle solutions in the supergravity theory \eqref{eq:JJ}. To achieve this we need to spell out a number of details on the holographic dictionary for the Euclidean black saddles. In particular, we need a holographic dictionary that maps the field theory parameters onto related gravitational quantities.  That is, we need to find supergravity quantities ($\hat{\beta}$, $\hat{\Delta}^I$, $\hat{\sigma}^I$), computed using the bulk Euclidean black saddle solutions, that precisely match the field theory parameters ($\beta$, $\Delta^I$, $\sigma^I$) that determine the topologically twisted index.

The parameter $\beta$ in \eqref{eq:metQFT} controls the size of the $S^1$ in the QFT.  This should be related to the periodicity $\beta_\tau$ of the $\tau$-coordinate in gravity, which we can read off from \eqref{eq:beta}.  To establish the precise relation we need to take into account a possible rescaling of the coordinate $\tau$ with respect to the one used in the field theory metric \eqref{eq:metQFT}.  This is taken care of by considering the UV expansion of the supergravity metric in \eqref{eq:metricUV} and \eqref{eq:UVexpans} which leads to the identification
\begin{equation}\label{eq:betahata0}
	\hat \beta = a_0 \beta_\tau~,
\end{equation}
where $a_0$ is the leading coefficient in the Fefferman-Graham expansion of the metric, see \eqref{eq:UVexpans}.

Let us proceed with the electric chemical potentials $\Delta^I$.  As is standard in holographic calculations, these should map onto the electric chemical potentials in supergravity, which are controlled by the Wilson lines $A^I_\tau$ in the black saddle solutions \eqref{eq:ansatz}.  Therefore, we expect a relation of the form
\begin{equation}\label{eq:hatdelta}
	\hat{\Delta}^I = \alpha \left(A^I_\tau |_\text{IR} - A^I_\tau |_\text{UV}\right) = \alpha \left(e^I|_\text{IR} - e^I|_\text{UV}\right)~,
\end{equation}
for some proportionality constant $\alpha$.  Note that this involves the difference between the UV and IR values of the gauge field in order to ensure gauge invariance of the chemical potential under constant shifts.  To determine the normalization $\alpha$, we recall from Section~\ref{sec:fieldtheorypartfunc} that the field theory chemical potentials satisfy the constraint \eqref{eq:cftconst}. The corresponding quantities in the gravity dual, $\hat{\Delta}^I$, on the other hand, satisfy
\begin{equation}\label{eq:Deltahatalpha}
	\sum_I \hat{\Delta}^I = \pm \frac{2\pi \alpha}{\beta_\tau \xi g}~,
\end{equation}
which can be obtained by using the UV and IR supersymmetry constraints (\ref{eq:UVzerothorder}) and (\ref{eq:IR1}). We therefore find that in order to have agreement between the field theory and supergravity parameters we need to set $\alpha = \pm \beta_\tau \xi g$, where the sign corresponds to Branch $1$ and Branch $2$ respectively.  This normalization is also precisely what we would obtain by applying the holographic dictionary for chemical potentials proposed in~\cite{Silva:2006xv,Dias:2007dj}: the overall factor of $\beta_\tau$ is crucial in order to consistently define a BPS limit of the quantum statistical relation and the factor of $g$ arises simply by accounting for our normalization of the electric flavor charges in (\ref{eq:qi}).  Therefore the gravitational dual to the electric chemical potentials, with the proper normalization, are given by
\begin{equation}
	\hat{\Delta}^I = \pm \beta_\tau \xi g \left(e^I|_\text{IR} - e^I|_\text{UV}\right)~, \qquad \sum_I \hat{\Delta}^I = 2\pi~.
\label{eq:deltaconstr}
\end{equation}
We note that this supergravity analysis reproduces the field theory constraint on the parameters $\Delta^I$ which leads to $\Delta_R= \pi$, i.e. we have $n=1$ in \eqref{eq:toptwistcft}.\footnote{We could have also chosen $n=-1$ by changing the sign of the parameter $\alpha$ in \eqref{eq:Deltahatalpha}.}  It would be interesting to understand how to construct black saddle solutions for other values of the integer $n$. It is tempting to speculate that these solutions may arise by relaxing the IR regularity condition \eqref{eq:IR1} which ensures that the metric caps off smoothly as $\mathbb{R}^2\times \Sigma_{\mathfrak{g}}$ and instead allow for an $\mathbb{R}^2/{\mathbb{Z}_{|n|}}$ orbifold singularity.

Now let us discuss the real mass parameters $\sigma^I$ which are sources for dimension-two operators in the ABJM theory.  As shown in Section~\ref{sec:holorenorm}, in supergravity these parameters can be read off from the leading-order term in the Fefferman-Graham expansion for the pseudoscalars $z^\alpha - \tilde{z}^\alpha$.  This leads to the relation (\ref{eq:sigmaFDFholo}), which implies that the gravitational counterpart to the real masses is given by
\begin{equation}
	\hat{\sigma}_F^\alpha \sim z_0^\alpha - \tilde{z}_0^\alpha~,
\end{equation}
up to some undetermined constant of proportionality. To determine this proportionality constant it is convenient to work in the democratic basis $\hat{\sigma}^I$ using the same basis-change relation as in (\ref{eq:basis}).  This in turn yields
\begin{equation}
	\hat{\sigma}^I \sim \lim_{\rho\to\infty} e^\rho(X^I - \bar{X}^I)~,
\end{equation}
We can now use the results in \cite{Freedman:2013oja}, where the relation between the real masses and the bulk scalars is made explicit in the case of supergravity solutions with an $S^3$ boundary. Since we are dealing with scalar source and operators, the same relation should apply for the black saddle solutions with an $S^1 \times \sigg$ boundary.  Transforming the results of \cite{Freedman:2013oja} to our conventions leads to the compact expression
\begin{equation}
\hat{\sigma}^I = \lim_{\rho\rightarrow\infty} \pm i g e^{\rho}\left(\bar X^I- X^I\right)~,
\end{equation}
where $+$ and $-$ correspond to Branch $1$ and Branch $2$ respectively.  In terms of the leading order values of the supergravity scalars in the UV expansion \eqref{eq:UVexpans}, we find
\begin{equation}\label{eq:hatsigma}\begin{aligned}
	\hat{\sigma}^0 &= \pm\frac{i g}{2\sqrt{2}}\left( (z_0^1 - \tilde{z}_0^1) + (z_0^2 - \tilde{z}_0^2) + (z_0^3 - \tilde{z}_0^3) \right)~, \\
	\hat{\sigma}^1 &= \pm\frac{i g}{2\sqrt{2}}\left( (z_0^1 - \tilde{z}_0^1) - (z_0^2 - \tilde{z}_0^2) - (z_0^3 - \tilde{z}_0^3) \right)~,
\end{aligned}\end{equation}
plus cyclic permutations for $\hat{\sigma}^2$ and $\hat{\sigma}^3$. These supergravity quantities satisfy $\sum_I \hat{\sigma}^I = 0$, which agrees precisely with the field theory constraint \eqref{eq:cftconst}.

Finally, let us recall the judicious repackaging made in \eqref{eq:hatu}, where we defined the quantities $\hat{u}^I$ to encode information about the IR values of the scalars.  In particular, on Branch $1$ we have
\begin{equation}\label{eq:uhat}\begin{aligned}
	\hat{u}^I &= \frac{ 2 \pi X^I}{X^0 + X^1 + X^2 + X^3}\bigg{|}_\text{IR}~,
\end{aligned}\end{equation}
while on Branch 2 we find
\begin{equation}\begin{aligned}
	\hat{u}^I &= \frac{ 2 \pi \bar{X}^I}{\bar{X}^0 + \bar{X}^1 + \bar{X}^2 + \bar{X}^3}\bigg{|}_\text{IR}~.
\end{aligned}\end{equation}
By construction, these obviously satisfy the condition $\sum_I \hat{u}^I = 2\pi$.\footnote{We are again free to choose a convention in which $\sum_I \hat{u}^I = -2\pi$ by putting a minus sign in the definition of $\hat{u}^I$ in \eqref{eq:hatu}. This corresponds to taking $n=-1$ in \eqref{eq:toptwistcft}.} It is also convenient to define the corresponding quantities $\hat{u}_F^\alpha$, $\hat{\Delta}_F^\alpha$, and $\hat{\sigma}_F^\alpha$ in the flavor basis.  That is:
\begin{equation}\label{eq:udeltasigma}\begin{aligned}
	\hat{u}_F^1 &= \frac{1}{2}\left(\hat{u}^0 + \hat{u}^1 - \hat{u}^2 - \hat{u}^3 \right)~, \\
	\hat{\Delta}_F^1 &= \frac{1}{2}\left(\hat{\Delta}^0 + \hat{\Delta}^1 - \hat{\Delta}^2 - \hat{\Delta}^3 \right) = \pm \beta_\tau \xi g \left(e^1_F|_\text{IR} - e^1_F|_\text{UV}\right)~, \\
	\hat{\sigma}_F^1 &= \frac{1}{2}\left(\hat{\sigma}^0 + \hat{\sigma}^1 - \hat{\sigma}^2 - \hat{\sigma}^3 \right) = \pm \frac{i g}{\sqrt{2}}(z_0^1 - \tilde{z}_0^1)~,
	\end{aligned}\end{equation}
plus cyclic permutations, where $+$ and $-$ correspond to Branches 1 and 2, respectively.  Note that the 9 flavor basis objects in \eqref{eq:udeltasigma} are unconstrained, unlike the 12 objects in the democratic basis, \eqref{eq:deltaconstr}, \eqref{eq:hatsigma}, and \eqref{eq:uhat},   that satisfy 3 constraints.

With all of these definitions at hand, we can express the gravitational partition function (\ref{eq:JJ}) for the Euclidean black saddle solutions as:
\begin{equation}\label{eq:Zgravcomp}
	\log Z_\text{grav} = \frac{\text{Vol}(\sigg)}{8\pi G_N}\bigg{[}\frac{1}{\xi g}\sqrt{\hat{u}^0 \hat{u}^1 \hat{u}^2 \hat{u}^3} \sum_I \frac{p^I}{\hat{u}^I} \pm \sum_\alpha \frac{ q_F^\alpha}{\xi g}\left( \hat{u}_F^\alpha - \hat{\Delta}_F^\alpha - i \hat \beta \hat{\sigma}_F^\alpha\right)\bigg{]}~.
\end{equation}
The corresponding expression for the partition function of the dual CFT is given in (\ref{eq:cftpart2}), which we repeat here for clarity:
\begin{equation}
	\log Z_\text{CFT} = \frac{\sqrt{2} N^{3/2}}{3} \sqrt{u^0 u^1 u^2 u^3} \sum_I \frac{\mathfrak{p}^I}{u^I}~.
\label{eq:matchft}
\end{equation}
As described above, the quantities $\hat{\beta}$, $\hat{\Delta}^I$, and $\hat{\sigma}^I$ should be precisely the bulk quantities that match the CFT parameters $\beta$, $\Delta^I$, and $\sigma^I$.  Upon inspection of the gravitational and field theory partition functions, though, this map is not enough to establish the agreement between the two partition functions.  We have to include also a prescription to include the $\hat{u}^I$ terms, which are related to the values that the scalars take at the IR cap-off, in our holographic dictionary.  With this in mind we therefore propose the following holographic dictionary:
\begin{equation}
\begin{aligned}
	\hat \beta		&\longleftrightarrow	\beta~, \\
	\hat \Delta^I 	&\longleftrightarrow	\Delta^I~, 		\\
	\hat \sigma^I 	&\longleftrightarrow	\sigma^I~, 		\\
	\hat u^I 		&\longleftrightarrow	u^I~.	
\end{aligned}
\end{equation}
In order for this dictionary to be true, the supergravity quantities should conspire to reproduce the field theory identity $u^I=\Delta^I+i\beta \sigma^I$ in \eqref{eq:uIdefQFT}.  Therefore, we arrive at the following relation between gravitational quantities:
\begin{equation}\label{eq:UVIR}
	\boxed{\hat{u}^I = \hat{\Delta}^I + i \hat{\beta} \hat{\sigma}^I}~.
\end{equation}
This is a remarkable and highly non-trivial relation between parameters determined by the UV and IR expansion of the supergravity solutions. The $\hat{u}^I$ are a function only of the IR values of the scalars, which are in turn related to the electric and magnetic charges.  The quantities $\hat{\Delta}^I$ and $\hat{\sigma}^I$, on the other hand, require knowing the UV expansion for the scalars $z^\alpha$, $\tilde{z}^\alpha$, as well as the full flow solution for the Wilson lines $e^I$. 

As we show in detail in Section \ref{sec:solutions}, the UV/IR relation in \eqref{eq:UVIR} holds for all analytic and numerical Euclidean black saddle solutions we have found.   We therefore assume it is true on all possible solutions of the form \eqref{eq:ansatz} and proceed to analyze its consequences here.  Using \eqref{eq:UVIR} in \eqref{eq:Zgravcomp} we immediately find that the gravitational partition function reads
\begin{equation}
	\log Z_\text{grav} = \frac{\eta}{4 \xi g G_N}\sqrt{\hat{u}^0 \hat{u}^1 \hat{u}^2 \hat{u}^3} \sum_I \frac{p^I}{\hat{u}^I}~,
\label{eq:matchg}
\end{equation}
where we have used \eqref{eq:VolSigmaeta} for the volume of the Riemann surface.  The usual AdS$_4$/CFT$_3$ dictionary for the ABJM theory provides a relation between the rank of the gauge group $N$ and the four-dimensional supergravity couplings, see for instance \cite{Freedman:2013oja},
\begin{equation}
	\frac{\sqrt{2} N^{3/2}}{3} \longleftrightarrow \frac{1}{4 g^2 G_N}~.
\label{eq:dict}
\end{equation}
To find complete agreement between the supergravity and field theory results in \eqref{eq:matchg} and \eqref{eq:matchft} we need to fix the constant of proportionality between the magnetic charges on the two side of the correspondence, $p^I$ and $\mathfrak{p}^I$. This can be done by comparing the topological twist condition in field theory \eqref{eq:cftconst} with the supergravity BPS constraint in \eqref{eq:toptwistcond}. Spelling out these relations in detail we find
\begin{equation}
	\eta \xi g \sum_I p^I = \eta\xi g \times \left(-\frac{\kappa}{\xi g}\right) = - \eta \kappa = 2 (\mathfrak{g} - 1) = \sum_I \mathfrak{p}^I~.
\end{equation}
From this we find the following map between the magnetic charges
\begin{equation}\label{eq:pImap}
	\eta \xi g p^I \longleftrightarrow \mathfrak{p}^I ~.
\end{equation}
We note that this relation is in agreement with the results in~\cite{Benini:2016rke}.

Using \eqref{eq:pImap} and \eqref{eq:dict} in \eqref{eq:matchg} and comparing with the CFT result in \eqref{eq:matchft} we find a perfect agreement between the two partition functions:
\begin{equation}\label{eq:holoworks}
	\boxed{\log Z_\text{CFT}\left(\beta,\mathfrak{p}^I,\Delta^I,\sigma^I\right) = \log Z_\text{grav}\left(\hat \beta,p^I,\hat\Delta^I,\hat\sigma^I\right)}~.
\end{equation}
We emphasize that the agreement between the supergravity and CFT partition functions holds for arbitrary values of the parameters. In particular we do not need to fix the supergravity parameters such that we have a Lorentzian black hole solution and therefore we eschew the extremization of the topologically twisted index employed in~\cite{Benini:2015eyy,Benini:2016rke}.

As a further check of our proposed holographic dictionary, let us consider the definition of electric charges for the flavor symmetries on both sides of the correspondence. In the field theory, one defines the $\mathfrak{q}_F^\alpha$ by
\begin{equation}
\mathfrak{q}_F^\alpha \equiv \langle J_F^\alpha \rangle = -i \frac{\partial \log Z_\text{CFT}}{\partial \Delta_F^\alpha}~.
\label{eq:dzddelta_cft}
\end{equation}
On the gravity side, we can use \eqref{eq:Zgravcomp} to explicitly compute 
\begin{equation}\label{eq:Zqsugra}
\frac{\partial \log Z_\text{grav}}{\partial \hat{\Delta}_F^\alpha} = \mp \frac{\eta}{4\xi g G_N}\,q_F^\alpha~,
\end{equation}
where $-$ and $+$ correspond to Branch 1 and Branch 2, respectively.  Note that this supergravity result is deceptively simple. The left-hand side of \eqref{eq:Zqsugra} leads to a complicated function of $\hat u^I$ (which are in turn related to the IR values of the scalars) and $p^I$.  The IR values of the scalars are fixed in terms of the electric and magnetic charges via the IR constraint \eqref{eq:IR1}.  Using this constraint, we find that we can eliminate the scalars entirely in terms of the charges and we obtain the simple expression on the right-hand side of \eqref{eq:Zqsugra}. We thus find a dictionary between the charges $\mathfrak{q}_I$ in the field theory and the $q_I$ in the bulk:
\begin{equation}
	\mathfrak{q}_I \longleftrightarrow \pm \frac{i \eta}{4 \xi g G_N} q_I~.
\end{equation}
For black saddles that admit an analytic continuation to a Lorentzian black hole, the charges $q_I$ in Euclidean signature are Wick-rotated back to Lorentzian signature by sending $q_I^\text{(Eucl)} \to - i q_I^\text{(Lor)}$.  We therefore find that $\mathfrak{q}_I \leftrightarrow \pm \frac{\eta}{4\xi g G_N} q_I^\text{(Lor)}$ for black holes, which agrees exactly with the results in~\cite{Benini:2016rke}.

\section{Explicit solutions}
\label{sec:solutions}

In the previous section, we established a precise agreement between the partition function of the Euclidean black saddles and the topologically twisted index of the planar ABJM theory. This match hinges upon the UV/IR correspondence, \eqref{eq:UVIR}, a precise relationship between the values of the scalars and Wilson lines at the IR cap-off and at the UV conformal boundary. The validity of this UV/IR correspondence cannot be established by the perturbative UV and IR analysis discussed in the previous section and one needs to construct a full non-linear supergravity solution of the BPS equations \eqref{eq:BPS}.

To demonstrate the validity of the UV/IR correspondence \eqref{eq:UVIR} in this section we present explicit constructions of several classes of black saddle solutions by utilizing both analytic and numerical techniques. For all such solutions we find that the UV/IR relation in \eqref{eq:UVIR} is obeyed.

\subsection{Euclidean Romans solutions}
\label{subsec:EucRom}

The simplest class of black saddle solutions are the Euclidean generalizations of the Romans black hole solution~\cite{Romans:1991nq}, as we previously introduced in Section~\ref{subsec:grav}.  These solutions are obtained by turning off all scalars and setting all $U(1)$ gauge fields equal.  The most general supersymmetric black saddle solution to the BPS equations (\ref{eq:BPS}) that we can find in this truncation is given in \eqref{eq:romansgen}. Note that $\xi = \pm 1$ is the same quantity that shows up in the spinor projector \eqref{eq:projectxi}. The functions $M$ and $\wt{M}$ that appear in the other spinor projectors \eqref{eq:project} for this solution are given by
\begin{equation}
	M = \frac{1}{\wt{M}} = -i \xi \sqrt{\frac{4 g^2 r^2 +\kappa - \xi g Q}{4 g^2 r^2 +\kappa + \xi g Q}}~.
\label{eq:euclromsproj}
\end{equation}
This is hence a one-parameter family of $\frac{1}{4}$-BPS black saddles labelled by the electric charge $Q$. 

The metric function $U(r)$ has its outermost root at $r_+$ in \eqref{eq:r+genBSR} and the periodicity $\beta_\tau$ of the $\tau$-coordinate is given by \eqref{eq:qgg} which ensures that the solution caps off smoothly as $\mathbb{R}^2 \times \sigg$ with no conical singularities.  Regularity of the full solution as $r \to r_+$ also requires that the electric charge is bounded as in \eqref{eq:betaBSg01}. This in turn means that $g |Q| >0$ for Riemann surfaces with genus $\mathfrak{g} = 1$ and $g |Q| > 1$ for Riemann surfaces with genus $\mathfrak{g} = 0$, while imposing no constraints on the charge for higher-genus Riemann surfaces.  Importantly, this implies that there is no regular solution for $Q \to 0$ when $\mathfrak{g} = 0,1$.

Of course, in order for the Euclidean solution (\ref{eq:romansgen}) to have an on-shell action that matches the CFT partition function, we need to prove that the solutions satisfies the UV/IR correspondence (\ref{eq:UVIR}).  The analysis of this correspondence is fairly straightforward for this simple black saddle solution and we go through it in detail below.

First, we must analyze the different branches of solutions that are allowed.  As discussed in the analysis of the IR BPS conditions in Section~\ref{sec:IR}, either $M$ or $\wt{M}$ must vanish in the IR for consistency of the BPS equations.  For the Euclidean solutions at hand, this means that either $M(r_+) = 0$ or $\wt{M}(r_+) = 0$, which corresponds to Branch 1 or Branch 2, respectively.  If we take the explicit projector expressions in (\ref{eq:euclromsproj}) and evaluate them at $r_+$, we find that
\begin{equation}
	M(r_+) = -i \xi \sqrt{\frac{|Q| - \xi Q}{|Q| + \xi Q}}~, \qquad \wt{M}(r_+) = i \xi \sqrt{\frac{|Q| + \xi Q}{|Q| - \xi Q}}~.
\end{equation}
Clearly, $M(r_+) = 0$ if $\xi = \text{sgn}(Q)$, while $M(r_+) = 0$ if $\xi = - \text{sgn}(Q)$.  That is, the choice of how $\xi$ correlates with the sign of the charge $Q$ tells us exactly which branch of solutions we are looking at.  As detailed in Section~\ref{sec:match}, these branch choices also affect how we match supergravity quantities to their counterparts in the CFT.  If we run through the definitions in that section and apply them here, we find that the two different branches of solutions give us the following relations:
\begin{equation}\begin{aligned}
	&\textbf{Branch 1:} \quad & &\textbf{Branch 2:} \\
	&\xi = \text{sgn}(Q) && \xi = - \text{sgn}(Q) \\
	&\hat{u}^I = \frac{2\pi X^I}{\sum_J X^J}\bigg{|}_{\text{IR}} & & \hat{u}^I = \frac{2\pi \bar{X}^I}{\sum_J \bar{X}^J}\bigg{|}_{\text{IR}}\\
	&\hat{\Delta}^I = \beta_\tau g\,\text{sgn}(Q) \left(e^I|_\text{IR} - e^I|_\text{UV}\right)\quad\quad && \hat{\Delta}^I =\beta_\tau g\,\text{sgn}(Q) \left(e^I|_\text{IR} - e^I|_\text{UV}\right) \\
	&\hat{\sigma}_F^\alpha = \frac{i g\, \text{sgn}(Q)}{\sqrt{2}}\left(z_0^\alpha - \tilde{z}_0^\alpha\right) && \hat{\sigma}_F^\alpha =\frac{i g\, \text{sgn}(Q)}{\sqrt{2}}\left(\tilde{z}_0^\alpha - z_0^\alpha\right) \\
	&\hat{\beta} = \beta_\tau && \hat{\beta} = \beta_\tau
\end{aligned}\end{equation}
where we recall that $e^I \equiv A^I_\tau$ are the Wilson lines wrapping the $\tau$ direction, while $z_0^\alpha$ and $\tilde{z}_0^\alpha$ are the leading-order terms in the UV expansions of the scalars.  Note that the definition of $\hat{\Delta}^I$ does not change between branches because our choice of $\xi$ does not affect the Wilson lines in the solution, and so the dual CFT source should be unaffected by the choice of branch.  What does change, though, is whether we think of the scalars $z^\alpha$ or $\tilde{z}^\alpha$ as being fixed by the charges in the IR.  Additionally, we chose a coordinate frame in which there is no rescaling of the $\tau$ coordinate when comparing the UV and the IR, so $\hat{\beta}$ is precisely the $\tau$-periodicity $\beta_\tau$.

Equipped with this understanding of the two different branches of our holographic dictionary, we compute these quantities in detail. Since the scalars all vanish, $z^\alpha = \tilde{z}^\alpha = 0$, the gravitational counterpart $\hat{\sigma}^I$ to the field theory real masses all vanish:
\begin{equation}
	\hat{\sigma}^I = 0~.
\end{equation}
Additionally, since the scalars vanish, the symplectic sections are given by $X^I = \bar{X}^I = \frac{1}{2\sqrt{2}}$, and so the gravitational counterparts $\hat{u}^I$ to the complex fugacities are given by

\begin{equation}\begin{aligned}
	\hat{u}^I &= \frac{\pi}{2}~,
\end{aligned}\end{equation}
irrespective of the branch. The last piece in the puzzle are the bulk chemical potentials $\hat{\Delta}^I$.  The Wilson lines are given by
\begin{equation}
	e^I = \frac{Q}{4r}~,
\end{equation}
and thus they vanish in the UV. Therefore the chemical potentials are determined by evaluating the Wilson lines in the IR:
\begin{equation}
	\hat{\Delta}^I = \beta_\tau g\,\text{sgn}(Q) \frac{Q}{r_+} = \frac{\pi}{2}\, \text{sgn}(Q) \frac{Q}{|Q|} = \frac{\pi}{2}~,
\end{equation}
assuming that $Q$ is non-zero, or equivalently that the size of the $S^1$ is finite.  Putting this all together, we find that the UV/IR correspondence is satisfied, namely:
\begin{equation}
	\hat{u}^I = \hat{\Delta}^I + i \hat{\beta} \hat{\sigma}^I = \frac{\pi}{2}~,
\end{equation}
for all values of the index $I$ and for both branches of the solution.  Therefore we have indeed established the equivalence of the supergravity and field theory partition functions as in \eqref{eq:holoworks} for the universal black saddle solution in \eqref{eq:romansgen}. When we choose $\mathfrak{g}>1$ and tune $\beta \to \infty$ by taking the $Q \to 0 $ limit we recover the universal magnetic black hole solution. The on-shell action of this black hole is then related to its entropy and in turn is equal to the topologically twisted index in the dual CFT \cite{Benini:2015eyy,Azzurli:2017kxo}. For general values of the parameters $\mathfrak{g}$ and $\beta_{\tau}$ however the black saddle solution has no interpretation as a black hole with a given entropy. Nevertheless, as discussed in Section~\ref{subsec:grav} there is a nice agreement between the black saddle on-shell action and the topologically twisted index.

\subsection{Universal solutions with scalars}
\label{subsec:scalarsltns}

We now move on to more elaborate Euclidean black saddle solutions.  In particular, we construct solutions with non-trivial scalar profiles and verify that the UV/IR relationship \eqref{eq:UVIR} is in general satisfied.  However, solving the BPS equations with all six scalar fields turned on proves to be very difficult in general.  One strategy to simplify the problem is to turn on only a few of the scalars and find solutions within a correspondingly smaller parameter space.  This simplification has to be implemented with care since the IR boundary conditions (\ref{eq:IR1}) fix the values of the scalars in the IR in terms of the electric and magnetic charges, and for general values of the charges it is not consistent to turn any of the scalars off. 

The IR boundary conditions become very simple when we take the universal alignment of charges, just as we did for the Euclidean Romans solutions:
\begin{equation}
	p^I = -\frac{\kappa}{4 \xi g}~, \quad q_I = \frac{Q}{4}~,
\label{eq:univcharge}
\end{equation}
such that $Q$ is the total electric charge and the topological twist condition is satisfied.  This is equivalent to turning off all flavor magnetic charges $p_F^\alpha$ and electric charges $q_F^\alpha$.  Within this universal charge sector, the IR BPS conditions (\ref{eq:IR1}) inform us that the scalars must obey the following constraints on Branch 1:
\begin{equation}\begin{aligned}
	0 &= \left[\left(1 + (z^1)^2\right)z^2 z^3 + z^1 \left( (z^2)^2 + (z^3)^2\right)\right]_\text{IR}~, \\
	0 &= \left[\left(1 + (z^2)^2\right)z^3 z^1 + z^2 \left( (z^3)^2 + (z^1)^2\right)\right]_\text{IR}~, \\
	0 &= \left[\left(1 + (z^3)^2\right)z^1 z^2 + z^3 \left( (z^1)^2 + (z^2)^2\right)\right]_\text{IR}~.
\label{eq:irz}
\end{aligned}\end{equation}
Similarly, on Branch 2 we obtain the same expression, except with $z^\alpha \rightarrow \tilde{z}^\alpha$.    These equations allow solutions with some of the scalars turned off in the IR.  Moreover, the structure of the BPS equations are such that turning off the scalars in the IR ensures that they will not flow along the whole solution.  Thus, by carefully analyzing the various ways in which (\ref{eq:irz}) can be satisfied, we can find consistent truncations of the BPS equations with some of the scalars turned off, thus simplifying the differential equations we have to solve and allowing for analytic solutions.  We will detail some of these analytic solutions below.

\subsubsection*{Two-scalar solution}

First, let us note that the IR boundary conditions (\ref{eq:irz}) are all simultaneously solved on both Branch 1 and Branch 2 if we choose to set
\begin{equation}
	z^2 = z^3 = \tilde{z}^2 = \tilde{z}^3 = 0~,
\end{equation}
at the IR cap-off.  Moreover, it is consistent within the universal charge sector (\ref{eq:univcharge}) to turn these four scalars off along the entire flow.  We will allow for the remaining two scalars $z^1$ and $\tilde{z}^1$ to have non-trivial profiles.  By turning on both of these scalars, the K\"ahler potential becomes non-trivial and thus the scalars end up back-reacting on the metric.  Using the BPS equations, one can show that the most general form of this backreacted two-scalar solution is given by
\begin{equation}\begin{aligned}
	ds^2 &= U(r) d\tau^2 + \frac{dr^2}{U(r)} + r^2 ds_{\sigg}^2~, \\
	U(r) &= \left(\sqrt{2} g \sqrt{r^2 - c_1 \tilde{c}_1} + \frac{\kappa}{2\sqrt{2} g \sqrt{r^2 - c_1 \tilde{c}_1}} \right)^2 - \frac{Q^2}{8 (r^2 - c_1 \tilde{c}_1) }~, \\
	A^I &= -\frac{\kappa}{4\xi g}\,\omega_{\sigg} + e^I d\tau~, \\
	e^0 &= e^1 = \frac{Q r}{4(r^2 - c_1 \tilde{c}_1)} - \frac{ (c_1 - \tilde{c}_1)\kappa + (c_1 + \tilde{c}_1) \xi g Q}{8 \xi g (r^2 - c_1 \tilde{c}_1)}~, \\
	e^2 &= e^3 = \frac{Q r}{4(r^2 - c_1 \tilde{c}_1)} +\frac{ (c_1 - \tilde{c}_1)\kappa + (c_1 + \tilde{c}_1) \xi g Q}{8 \xi g (r^2 - c_1 \tilde{c}_1)}~, \\
	z^1 &= \frac{c_1}{r}~, \quad \tilde{z}^1 = \frac{\tilde{c}_1}{r}~, \quad z^2 = z^3 = \tilde{z}^2 = \tilde{z}^3 = 0~,
\label{eq:twoscalar}
\end{aligned}\end{equation}
where $\kappa$ is the curvature of the Riemann surface $\sigg$, $\omega_{\sigg}$ is the local one-form potential such that $d \omega_{\sigg} = V_{\sigg}$, $\xi = \pm 1$, and $c_1$ and $\tilde{c}_1$ are arbitrary complex parameters that characterize how the scalars $z^1$ and $\tilde{z}^1$ flow from the UV to the IR.  Additionally, the functions $M$ and $\wt{M}$ in the projector relation (\ref{eq:project}) are given by
\begin{equation}
	M = \wt{M}^{-1} = -i \xi \sqrt{\frac{4 g^2 \left(r^2 - c_1 \tilde{c}_1\right) +\kappa - \xi g Q}{4 g^2 \left(r^2 - c_1 \tilde{c}_1\right) +\kappa + \xi g Q}}~.
\label{eq:twoscalarm}
\end{equation}
Note that if we set $c_1 = \tilde{c}_1 = 0$, we recover the Euclidean Romans solution discussed in Section~\ref{subsec:EucRom}.

This two-scalar solution solves the BPS equations for general values of $c_1$ and $\tilde{c}_1$, but we will now restrict ourselves to values such that $c_1 \tilde{c}_1 \in \mathbb{R}$, in order to ensure that the metric is real.  With this in mind, the outermost root of the metric function $U(r)$ is located at
\begin{equation}
	r_+ = \frac{\sqrt{-\kappa + 4 g^2 c_1 \tilde{c}_1 + g |Q|}}{2g}~.
\end{equation}
The scalars $z^1$ and $\tilde{z}^1$ have monotonically increasing modulus as $r \to r_+$, and so the condition that the scalars take values on the Poincar\'e disk can be enforced by choosing $c_1$ and $\tilde{c}_1$ such that
\begin{equation}
	\frac{|c_1|}{r_+} < 1~, \quad \frac{|\tilde{c}_1|}{r_+} < 1~.
\end{equation}
Demanding the absence of conical singularities at $r_+$ informs us that the periodicity $\beta_\tau$ of the $\tau$-coordinate is
\begin{equation}
	\beta_\tau = \frac{\pi}{g^2 |Q|} \frac{ g |Q| - \kappa}{\sqrt{g |Q| - \kappa + 4 g^2 c_1 \tilde{c}_1}}~.
\end{equation} 
In order to ensure regularity of the solution at this IR cap-off, we must impose the inequalities
\begin{equation}\begin{aligned}
	g |Q| &\geq  \kappa - 4 g^2 c_1 \tilde{c}_1~, \\
	g |Q| &> \kappa~,
\end{aligned}\end{equation}
on the magnitude of the electric charge $Q$.  Just as with the Euclidean Romans solutions in Section~\ref{subsec:EucRom}, this imposes $g|Q| > 0$ for genus $\mathfrak{g}=1$ Riemann surfaces and $g|Q| > 1$ for genus $\mathfrak{g} = 0$ Riemann surfaces, which means that we cannot take the $Q \to 0$ limit for these lower-genus solutions.  Additionally, depending on the value of $c_1 \tilde{c}_1$, it may or may not be possible to send $Q \to 0$ for solutions with $\mathfrak{g} > 1$.  This is in contrast to the Euclidean Romans solution where there were no constraints on $Q$ for such higher-genus solutions.

Now that we have presented the two-scalar solution, we want to check that it satisfies the UV/IR relation (\ref{eq:UVIR}).  To do so, we first evaluate the projectors $M$ and $\wt{M}$ at $r_+$ and determine which of the two vanish, thus informing us whether we have a solution on Branch 1 or Branch 2.  This analysis goes exactly as it did for the Euclidean Romans solution, and so we find that $\xi = \text{sgn}(Q)$ corresponds to Branch 1 while $\xi = - \text{sgn}(Q)$ corresponds to Branch 2.

Armed with this branch analysis, it is relatively straightforward to apply the holographic dictionary established in Section~\ref{sec:match} and compute the bulk quantities $(\hat{\beta},\hat{u}^I,\hat{\Delta}^I, \hat{\sigma}^I)$ that must satisfy the UV/IR relation (\ref{eq:UVIR}).  The only subtlety in this calculation is that it requires knowing the Fefferman-Graham expansion of the scalars in the UV as described in Section~\ref{sec:UV}.  To determine this, we first recall that the coordinate $r$ is related to the Fefferman-Graham coordinate $\rho$ by the condition that
\begin{equation}
	e^{2 f_2} dr^2 = L^2 d\rho^2 = \frac{1}{2g^2} d\rho^2~.
\end{equation}
For the two-scalar solution at hand, this leads to a series expansion of the form
\begin{equation}
	r = \frac{1}{\sqrt{2}g} e^{\rho}\left(1 + \frac{2 g^2 c_1 \tilde{c}_1-\kappa}{4} e^{-2\rho} + \frac{g^2 Q^2 - \kappa^2}{32}e^{-4\rho} + \ldots\right)~,
\end{equation}
which means that the UV expansion of the scalars takes the form
\begin{equation}\begin{aligned}
	z^1 &= \sqrt{2} g c_1 e^{-\rho}\left(1 - \frac{2 g^2 c_1 \tilde{c}_1-\kappa}{4} e^{-2\rho} + \ldots\right)~, \\
	\tilde{z}^1 &= \sqrt{2} g \tilde{c}_1 e^{-\rho}\left(1 - \frac{2 g^2 c_1 \tilde{c}_1-\kappa}{4} e^{-2\rho} + \ldots\right)~,
\end{aligned}\end{equation}
and thus the leading-order pieces that go into the real masses are
\begin{equation}
	z_0^1 = \sqrt{2} g c_1~, \quad \tilde{z}_0^1 = \sqrt{2} g \tilde{c}_1~.
\end{equation}
Using all of this, we find that the parameters $(\hat{\beta}, \hat{u}^I, \hat{\Delta}^I, \hat{\sigma}^I)$ for the two-scalar solution for both branches are given by
\begin{equation}\begin{aligned}
	&\textbf{Branch 1:} \\
	& \hat{\beta} = \frac{\pi}{g^2 |Q|} \frac{ g |Q| - \kappa}{\sqrt{g |Q| - \kappa + 4 g^2 c_1 \tilde{c}_1}}~, \\
	&\hat{u}^0 = \hat{u}^1 = \frac{\pi}{2} - \frac{ c_1 g \pi}{\sqrt{1+ 4g^2 c_1 \tilde{c}_1 +g |Q|}}~, \\
	&\hat{u}^2 = \hat{u}^3 = \frac{\pi}{2} + \frac{ c_1 g \pi}{\sqrt{1+ 4 g^2 c_1 \tilde{c}_1 + g |Q|}}~, \\
	&\hat{\Delta}^0 = \hat{\Delta}^1 = \frac{\pi}{2}\left(1 + \frac{(c_1 - \tilde{c}_1) - (c_1 + \tilde{c}_1) g |Q|}{|Q|\sqrt{1 + 4 g^2 c_1 \tilde{c}_1 + g |Q|}}\right)~, \\
	&\hat{\Delta}^2 = \hat{\Delta}^3 = \frac{\pi}{2}\left(1 - \frac{(c_1 - \tilde{c}_1) - (c_1 + \tilde{c}_1) g |Q|}{|Q|\sqrt{1 + 4 g^2 c_1 \tilde{c}_1 + g |Q|}}\right)~, \\
	&\hat{\sigma}^0 = \hat{\sigma}^1 = -\hat{\sigma}^2 = -\hat{\sigma}^3 = \frac{i g^2 (c_1 -\tilde{c}_1)}{2}~,
\end{aligned}\end{equation}
on Branch 1, and
\begin{equation}\begin{aligned}
	&\textbf{Branch 2:} \\
	& \hat{\beta} = \frac{\pi}{g^2 |Q|} \frac{ g |Q| - \kappa}{\sqrt{g |Q| - \kappa + 4 g^2 c_1 \tilde{c}_1}}~, \\
	&\hat{u}^0 = \hat{u}^1 = \frac{\pi}{2} - \frac{ \tilde{c}_1 g \pi}{\sqrt{1+ 4g^2 c_1 \tilde{c}_1 +  g |Q|}}~, \\
	&\hat{u}^2 = \hat{u}^3 = \frac{\pi}{2} + \frac{ c_1 g \pi}{\sqrt{1+ 4 g^2 c_1 \tilde{c}_1 + g |Q|}}~, \\
	&\hat{\Delta}^0 = \hat{\Delta}^1 = \frac{\pi}{2}\left(1 + \frac{(\tilde{c}_1 - {c}_1) - (\tilde{c}_1 + {c}_1) g |Q|}{|Q|\sqrt{1 + 4 g^2 c_1 \tilde{c}_1 + g |Q|}}\right)~, \\
	&\hat{\Delta}^2 = \hat{\Delta}^3 = \frac{\pi}{2}\left(1 - \frac{(\tilde{c}_1 - {c}_1) - (\tilde{c}_1 + {c}_1) g |Q|}{|Q|\sqrt{1 + 4 g^2 c_1 \tilde{c}_1 + g |Q|}}\right)~, \\
	&\hat{\sigma}^0 = \hat{\sigma}^1 = -\hat{\sigma}^2 = -\hat{\sigma}^3 = \frac{i g^2 (\tilde{c}_1 - c_1)}{2} ~. 
\end{aligned}\end{equation}
on Branch 2.  Note that we can go from one branch to the other simply by swapping $z^1$ and $\tilde{z}^1$, or equivalently by exchanging $c_1$ and $\tilde{c}_1$.  Putting all of these relations together, we find that
\begin{equation}
	\hat{u}^I = \hat{\Delta}^I + i \hat{\beta} \hat{\sigma}^I~,
\end{equation}
for all values of $I$, thus recovering the UV/IR relationship for both branches.

To summarize: the two-scalar solution (\ref{eq:twoscalar}) has three independent parameters: $c_1$, $\tilde{c}_1$, and $Q$.  The supergravity quantities $\hat{\beta}$, $\hat{u}^I$, $\hat{\Delta}^I$, and $\hat{\sigma}^I$ satisfy the UV/IR relationship, as required to fully establish a match with the dual field theory.  As discussed in Section~\ref{sec:match}, the gravitational partition function should be thought of as a function of $\hat{\beta}$, $p^I$, $\hat{\Delta}^I$, and $\hat{\sigma}^I$, and these are exactly dual to the field theory sources $\beta$, $\mathfrak{p}^I$, $\Delta^I$, and $\sigma^I$.  In the flavor basis, these sources are more compactly expressed by
\begin{equation}\begin{aligned}
	&\hat{\beta} = \frac{\pi}{g^2 |Q|} \frac{1 + g |Q|}{\sqrt{1 + 4 g^2 c_1 \tilde{c}_1 + g |Q|}}~, \\
	&\hat{\Delta}_F^1 = \frac{\pi\left((c_1 - \tilde{c}_1) - (c_1 + \tilde{c}_1) g |Q|\right) }{|Q|\sqrt{1 + 4 g^2 c_1 \tilde{c}_1 + g |Q|}}~, \\
	&\hat{\sigma}_F^1 = i g^2 (c_1 -\tilde{c}_1)~, \\
	&p_F^\alpha = \hat{\Delta}_F^2 = \hat{\Delta}_F^3 =\hat{\sigma}_F^2 = \hat{\sigma}_F^3 = 0~, \\
\end{aligned}\end{equation}
for Branch 1, while swapping $c_1 \leftrightarrow \tilde{c}_1$ gives us the sources for Branch 2.  The dual CFT interpretation of our solution is therefore that we have a finite value of $\beta$ and we turn on a background chemical potential $\Delta_F^1$ and a real mass $\sigma_F^1$ while leaving all other sources turned off.  There are hence three independent parameters that determine the ensemble of the dual field theory, and so it is in one-to-one correspondence with our gravitational ensemble.

As we have discussed in detail in Section~\ref{sec:BlackSaddles}, in the limit $Q \to 0$ the periodicity $\beta_\tau \to \infty$ and so the $S^1$ decompactifies.  This causes the solution to develop an infinite throat and so the IR geometry asymptotes to $\mathbb{H}^2 \times \sigg$ instead of $\mathbb{R}^2 \times \sigg$.  Na\"ively, it seems that we can do this while keeping the parameters $c_1$ and $\tilde{c}_1$ both turned on.  However, the IR boundary conditions become much more stringent in this limit, and we are no longer free to choose arbitrary values for both $z^1$ and $\tilde{z}^1$.  Instead, we are forced to pick one of them to vanish, depending on the branch of the solution.  A Lorentzian interpretation of the solution also requires $z^1$ and $\tilde{z}^1$ to be complex conjugates, and thus we are forced to set $z^1 = \tilde{z}^1 = 0$ alongside sending $Q \to 0$ in order to find a smooth black hole solution.  Thus we find that the Romans black hole (\ref{eq:univBH}) is the only black hole solution accessible within the full three-parameter space of the two-scalar solution (\ref{eq:twoscalar}).

\subsubsection*{Three-scalar solution}

Another way to satisfy the IR boundary conditions (\ref{eq:irz}) in the universal charge sector \eqref{eq:univcharge} is to set $z^1 = z^2 = \tilde{z}^3 = 0$ on Branch 1, or $\tilde{z}^1 = \tilde{z}^2 = z^3 = 0$ on Branch 2.  This allows us to consider a truncation of the BPS equations where only three of the scalars are turned on.  Moreover, this choice of scalars ensures that both the K\"ahler potential (\ref{eq:kg}) and the superpotential (\ref{eq:superpotdef}) are trivial along the entire solution and so the scalars do not backreact on the geometry.  This indicates that there should exist black saddle solutions with three scalar fields turned on that have precisely the same form of the metric as the Euclidean Romans solution (\ref{eq:romansgen}).  However, the scalars will source non-trivial terms in the Maxwell equation (\ref{eq:qi}) and so the solution will have non-trivial profiles for the Wilson lines $e^I$. 

We will now restrict ourselves to Branch 1 and set $z^1 = z^2 = \tilde{z}^3 = 0$ while letting $\tilde{z}^1$, $\tilde{z}^2$, and $z^3$ have non-trivial profiles.\footnote{All of the results in this section can be generalized to Branch 2 simply by exchanging $z^\alpha \leftrightarrow \tilde{z}^\alpha$ and flipping the sign of $\xi$.}  The most general three-scalar solution to the BPS equations in this truncation is given by
\begin{equation}\begin{aligned}
	ds^2 &= U(r) d\tau^2 + \frac{dr^2}{U(r)} + r^2 ds_{\sigg}^2~, \\
	U(r) &= \left(\sqrt{2} g r + \frac{\kappa}{2\sqrt{2} g r} \right)^2 - \frac{Q^2}{8 r^2}~, \\
	A^I &= - \frac{\kappa}{4 \xi g} \omega_{\sigg} + e^I d\tau~, \\
	 & \tilde{z}^1 = \frac{\tilde{c}_1}{r}~, \quad \tilde{z}^2 = \frac{\tilde{c}_2}{r}~, \quad z^1 = z^2 = \tilde{z}^3 = 0~, \\
	& z^3 = \frac{c_3}{r} + \frac{ \tilde{c}_1 \tilde{c}_2}{r^2 (\kappa + \xi g Q)}\left(\kappa - \xi g Q+ \frac{4 \kappa \xi g r \tan^{-1}\left(\frac{2 \xi g r}{\sqrt{\kappa + \xi g Q}}\right)}{\sqrt{\kappa + \xi g Q }}\right)~,
\end{aligned}\end{equation}
where $\tilde{c}_1$, $\tilde{c}_2$, and $c_3$ are arbitrary complex numbers that parameterize the scalar flows of the solution.  The Wilson lines $e^I$ in this solution can be compactly expressed in the basis $(e_R, e_F^\alpha)$ defined in (\ref{eq:basis}), wherein they take the following forms:
\begin{equation}\begin{aligned}
	e_R &= \frac{Q}{2r}~, \\
	e_F^1 &= \frac{\tilde{c}_1 (\kappa - \xi g Q)}{4 \xi g r^2}~, \\
	e_F^2 &= \frac{\tilde{c}_2 (\kappa - \xi g Q)}{4 \xi g r^2}~, \\
	e_F^3 &= -\frac{c_3 (\kappa + \xi g Q)}{4 \xi g r^2} - \frac{\tilde{c}_1 \tilde{c}_2 (\kappa^2 +4 \kappa g^2 r^2 - g^2 Q^2)}{2 \xi g r^3 (\kappa + \xi g Q)} \\
	&\quad - \frac{\tilde{c}_1 \tilde{c}_2 \kappa (\kappa + \xi g Q + 4 g^2 r^2) \tan^{-1}\left(\frac{2 \xi g r}{\sqrt{\kappa + \xi g Q}}\right)}{r^2 (\kappa + \xi g Q)^{3/2}}~.
\end{aligned}\end{equation}
Since the geometry is unchanged from the Euclidean Romans solution, the location $r_+$ of the IR cap-off and the periodicity $\beta_\tau$ are unchanged and given by:
\begin{equation}
	r_+ = \frac{\sqrt{-\kappa + g |Q|}}{2g}~, \quad \beta_\tau = \frac{\pi\sqrt{-\kappa + g |Q|}}{g^2 |Q|}~.
\end{equation}
Additionally, the projectors $M$ and $\wt{M}$ also retain the same form as in~\eqref{eq:euclromsproj}.  This means that we must set $\xi = \text{sgn}(Q)$, in order to ensure that $M \to 0$ in the IR.  Equivalently, this condition also ensures that the scalar field $z^3$ remains finite as $r \to r_+$.  Note also that since the complex parameters $\tilde{c}_1$, $\tilde{c}_2$, and $c_3$ do not show up in the metric, we do not need to impose any reality conditions on them.  We do however have to choose them such that $\tilde{z}^1$, $\tilde{z}^2$, and $z^3$ all have modulus less than one at $r = r_+$, which in turn guarantees that they have modulus less than one along the entire flow.

The coordinate $r$ is related to the Fefferman-Graham coordinate $\rho$ via a series expansion of the form
\begin{equation}
	r = \frac{1}{\sqrt{2}g} e^{\rho}\left(1 - \frac{\kappa}{4} e^{-2\rho} + \frac{(g^2 Q^2 - \kappa^2)}{32} e^{-4\rho} + \ldots\right)~,
\end{equation}
which we can use to show that the leading-order pieces in the scalars that go into the real masses are
\begin{equation}\begin{aligned}
	z_0^3 = \sqrt{2} g c_3 + \frac{2\sqrt{2}\pi \kappa g^2 \tilde{c}_1 \tilde{c}_2 }{(\kappa + g |Q|)^{3/2}}~, \quad \tilde{z}_0^1 = \sqrt{2} g \tilde{c}_1~, \quad \tilde{z}_0^2 = \sqrt{2} g \tilde{c}_2~.
\end{aligned}\end{equation}
Putting all of this together, we find that the gravitational duals to the CFT sources are given by:
\begin{equation}\begin{aligned}\label{eq:3scalarsources}
	\hat{\beta} &= \frac{\pi\sqrt{-\kappa + g |Q|}}{g^2 |Q|}~, \\
	\hat{u}_F^1 &= \hat{u}_F^2 = 0~, \quad \hat{u}_F^3 =- \frac{2 \pi g c_3}{\sqrt{-\kappa + g |Q|}} +  \frac{4\pi g^2 \tilde{c}_1 \tilde{c}_2}{\kappa + g |Q|} \left( 1 - \frac{2\kappa\tan^{-1}\left(\sqrt{\frac{-\kappa + g |Q|}{\kappa + g |Q|}}\right)}{\sqrt{-\kappa^2 + g^2 Q^2}} \right)~, \\
	\hat{\Delta}^1_F &= -  \frac{\pi \tilde{c}_1 \sqrt{-\kappa + g |Q|}}{|Q|}~, \quad \hat{\Delta}^2_F = - \frac{\pi  \tilde{c}_2 \sqrt{-\kappa + g |Q|}}{|Q|}~, \\
	\hat{\Delta}_F^3 &=  -\frac{\pi c_3(\kappa + g |Q|)}{|Q|\sqrt{-\kappa + g|Q|}} \\
	&\quad + \frac{2 \pi g^2 \tilde{c}_1 \tilde{c}_2}{(\kappa + g |Q|)^{3/2}}\left(\frac{\kappa \sqrt{-\kappa + g |Q|}}{g |Q|} + 2 \sqrt{\kappa + g |Q|} - \frac{4 \kappa \tan^{-1}\left(\sqrt{\frac{-\kappa + g |Q|}{\kappa + g |Q|}}\right)}{\sqrt{-\kappa + g |Q|}}\right)~, \\
	\hat{\sigma}^1_F &= -i g^2 \tilde{c}_1~, \quad \hat{\sigma}^2_F = -i g^2 \tilde{c}_2~, \quad \hat{\sigma}^3_F = i g^2 c_3 + \frac{2 i \pi \kappa g^3 \tilde{c}_1 \tilde{c}_2}{(\kappa + g |Q|)^{3/2}}~.
\end{aligned}\end{equation}
After some careful algebra, one can show that these all precisely satisfy the UV/IR correspondence $\hat{u}^I = \hat{\Delta}^I + i \hat{\beta} \hat{\sigma}^I$.  This three-scalar solution is therefore a four-parameter family of solutions (where $Q$, $\tilde{c}_1$, $\tilde{c}_2$, and $c_3$ are the free parameters) with an on-shell action that captures the topologically twisted index of the planar ABJM theory for the choice of sources specified by \eqref{eq:3scalarsources}.

\subsection{Solutions with flavor charges}

The saddle solutions presented in the previous section all have a finite periodicity $\beta_\tau$ in the IR as well as non-trivial profiles for the Wilson lines $e^I$ and (some of) the scalars $z^\alpha$, $\zt^\alpha$.  This in turn means that the solutions have finite values for the gravitational quantities $\hat{\beta}$, $\hat{\Delta}^I$, and $\hat{\sigma}^I$ that are dual to the size $\beta$ of the $S^1$ and the sources $\Delta^I$ and $\sigma^I$ in the field theory.  Our goal now is to additionally incorporate flavor magnetic charges $p_F^\alpha$ in the bulk that are dual to the background fluxes $\mathfrak{p}_F^\alpha$ for the flavor symmetries in the field theory.  In this section we present a number of explicit black saddle solutions with flavor fluxes turned on and verify that they all satisfy the UV/IR correspondence (\ref{eq:UVIR}), thus demonstrating concretely the match (\ref{eq:holoworks}) between the supergravity partition function and the topologically twisted index in ensembles with non-zero magnetic fluxes.

\subsubsection*{One-scalar solutions with flavor charges}

The first class of saddles we focus on has one of the magnetic flavor fluxes turned on alongside a non-trivial profile for exactly one of the six scalars.  Turning on more scalars or more of the magnetic flavor fluxes is difficult in general, due in part to how the magnetic fluxes source new terms in the Maxwell equations (\ref{eq:qi}) that are highly non-linear in the scalar fields.  We therefore restrict ourselves to this smaller parameter space where we can obtain explicit solutions.

First, we set
\begin{equation}
	p_F^2 = p_F^3 = 0~,
\end{equation}
while letting $p_F^1$ take an arbitrary value.  With this choice of flavor magnetic fluxes, the IR BPS conditions (\ref{eq:IR1}) inform us that we cannot consistently turn off all scalars.  In particular, if we first focus on Branch 1, the IR value of $z^1$ is related to $p_F^1$ and so we must in general allow for $z^1$ to have some profile as it flows from the UV to the IR.  However, we can consistently turn off all other scalars along the flow, which allows us to find the following black saddle solution: 
\begin{equation}\begin{aligned}
	ds^2 &= U(r) d\tau^2 + \frac{dr^2}{U(r)} + r^2 ds_{\sigg}^2~, \\
	U(r) &= \left(\sqrt{2} g r + \frac{\kappa}{2\sqrt{2} g r} \right)^2 - \frac{Q^2}{8 r^2}~, \\
	A^I &=  p^I \omega_{\sigg} + e^I d\tau~, \\
	z^1 &= \frac{c_1}{r} + \frac{2 p_F^1 \tan^{-1}\left(\frac{2\xi g r}{\sqrt{\kappa + \xi g Q}}\right)}{r \sqrt{\kappa+ \xi g Q}}~, \\
	z^2 &= z^3 = \zt^1 = \zt^2 = \zt^3 = 0~,
\end{aligned}\end{equation}
where $c_1$ is a general complex number, $p_F^1$ and $Q$ are real-valued charges, the Wilson lines in the basis $(e_R,e_F^\alpha)$ are given by
\begin{equation}\begin{aligned}
	e_R &= \frac{Q}{2r}~, \quad e_F^2 = e_F^3 = 0~,\\
	e_F^1 &= -\frac{c_1 (\kappa + \xi g Q)}{4 \xi g r^2} - \frac{2 p_F^1(\kappa+ \xi g Q + 4 g^2 r^2)\tan^{-1}\left(\frac{2\xi g r}{\sqrt{\kappa + \xi g Q}}\right)}{4\xi g r^2 \sqrt{\kappa + \xi g Q}}~,
\end{aligned}\end{equation}
and the projectors $M$ and $\wt{M}$ are given by (\ref{eq:euclromsproj}).  This is a three-parameter solution labelled by the values of $c_1$, $p_F^1$, and $Q$.  We can compute the full set of electric charges $q_I$ of this solution using (\ref{eq:qi}), which leads us to
\begin{equation}
	q_F^1 = p_F^1~, \quad q_F^2 = q_F^3 = 0~.
\end{equation}
The bulk saddle solution should therefore be thought of as having a single dyonic flavor charge turned on, in addition to an arbitrary electric R-charge $Q$ and a magnetic R-charge fixed by the topological twist.  However, the gravitational path integral is a grand canonical partition function, which means that the ensemble is one where the electric chemical potentials are fixed, not the charges.  The electric charges are instead a property of the black saddle that we must compute after we have constructed a saddle solution with the specified chemical potentials.

Alternatively, one can also consider turning on $p_F^1$ and looking for solutions on Branch 2.  In this case the IR boundary conditions (\ref{eq:IR1}) instead relate the IR value of $\zt^1$ to $p_F^1$.  Turning off all other scalars, we find the following bulk solution:
\begin{equation}\begin{aligned}
	ds^2 &= U(r) d\tau^2 + \frac{dr^2}{U(r)} + r^2 ds_{\sigg}^2~, \\
	U(r) &= \left(\sqrt{2} g r + \frac{\kappa}{2\sqrt{2} g r} \right)^2 - \frac{Q^2}{8 r^2}~, \\
	A^I &=  p^I \omega_{\sigg} + e^I d\tau~, \\
	\tilde{z}^1 &= \frac{\tilde{c}_1}{r} + \frac{2 p_F^1 \tan^{-1}\left(\frac{2\xi g r}{\sqrt{\kappa - \xi g Q}}\right)}{r \sqrt{\kappa - \xi g Q}}~, \\
	z^1 &= z^2 = z^3 = \zt^2 = \zt^3 = 0~,
\end{aligned}\end{equation}
where $\tilde{c}_1$ is a complex number, $p_F^1$ and $Q$ are again real-valued charges, the projectors $M$ and $\wt{M}$ are again given by (\ref{eq:euclromsproj}), and the Wilson lines are specified by
\begin{equation}\begin{aligned}
	e_R &= \frac{Q}{2r}~, \quad e_F^2  = e_F^3 = 0~, \\
	e_F^1 &= \frac{\tilde{c}_1 (\kappa - \xi g Q)}{4 \xi g r^2} + \frac{2 p_F^1(\kappa - \xi g Q + 4 g^2 r^2)\tan^{-1}\left(\frac{2\xi g r}{\sqrt{\kappa - \xi g Q}}\right)}{4\xi g r^2 \sqrt{\kappa - \xi g Q}}~. \\
\end{aligned}\end{equation}
The electric charges are similar to the case above, but we now have a single anti-dyonic flavor charge:
\begin{equation}
	q_F^1 = -p_F^1~, \quad q_F^2 = q_F^3 = 0~.
\end{equation}
We again stress that this relation is only a property of the particular solutions we have constructed, not the path integral itself, since there is no way to fix the electric charges in the grand canonical ensemble of consideration here.

Now that we have established these solutions, we can run through the machinery of Section~\ref{sec:match} and compute the gravitational quantities dual to the sources in the field theory.  This computation follows the exact same steps as done in Section~\ref{subsec:scalarsltns} and the end result is that both of the one-scalar solutions above satisfy the UV/IR correspondence (\ref{eq:UVIR}).  These black saddles therefore serve as tests of the correspondence for ensembles with non-zero flavor magnetic fluxes turned on.  Additionally, while both solutions solve the BPS equations for general values of $p_F^1$, we additionally impose that this flux across the Riemann surface obeys the Dirac quantization condition. In our supergravity conventions, this condition takes the form
\begin{equation}
	\eta \xi g p_F^1 \in \mathbb{Z}~,
\label{eq:dirac}
\end{equation} 
or equivalently the dual field theory flux must obey $\mathfrak{p}_F^1 \in \mathbb{Z}$.

\subsubsection*{Three-scalar solution with flavor charges}

The solutions described above involve turning off as many of the scalars as possible when one of the flavor magnetic fluxes is turned on.  However, as discussed in Section~\ref{sec:IR}, the IR boundary conditions only fix three of the six scalars in terms of the charges.  The remaining scalars ($\zt^\alpha$ on Branch 1 and $z^\alpha$ on Branch 2) are freely specified in the IR and generically flow along the radial direction.  It is therefore informative to construct solutions that have a magnetic flavor flux turned on alongside multiple flowing scalars.  

We again choose $p_F^1$ to be non-zero while setting $p_F^2 = p_F^3 = 0$, and we restrict our attention to Branch 1.  The IR BPS conditions (\ref{eq:IR1}) on Branch 1 inform us that we can consistently turn off $z^2$ and $z^3$ at the cost of giving $z^1$ a non-zero IR value sourced by $p_F^1$.  For the remaining scalars $\tilde{z}^\alpha$, we employ the strategy used in Section~\ref{subsec:scalarsltns} and set $\tilde{z}^1 = 0$ while keeping $\tilde{z}^2$ and $\tilde{z}^3$ general.\footnote{We can also similarly construct a three-scalar solution with non-zero flavor flux $p_F^1$ on Branch 2 by setting $z^1 = \tilde{z}^2 = \tilde{z}^3 = 0$ and letting the remaining scalars have profiles.}  This choice of scalars makes the K\"ahler potential (\ref{eq:kg}) and the superpotential (\ref{eq:superpotdef}) trivial along the entire flow thus greatly simplifying the BPS equations.

Within this setup, we find that the most general three-scalar black saddle solutions are as follows:
\begin{equation}\begin{aligned}
	ds^2 &= U(r) d\tau^2 + \frac{dr^2}{U(r)} + r^2 ds_{\sigg}^2~, \quad U(r) = \left(\sqrt{2} g r + \frac{\kappa}{2\sqrt{2} g r} \right)^2 - \frac{Q^2}{8 r^2}~, \\
	A^I &=  p^I \omega_{\sigg} + e^I d\tau~, \\
	z^1 &= \frac{c_1}{r} + \frac{\tilde{c}_2 \tilde{c}_3 (\kappa - \xi g Q)}{(\kappa + \xi g Q) r^2} + \frac{(4 \kappa \tilde{c}_2 \tilde{c}_3 \xi g + 2 p_F^1(\kappa + \xi g Q))\tan^{-1}\left(\frac{2\xi g r}{\sqrt{\kappa + \xi g Q}}\right)}{(\kappa + \xi g Q)^{3/2}r}~, \\
	\tilde{z}^2 &= \frac{\tilde{c}_2}{r}~, \quad \tilde{z}^3 = \frac{\tilde{c}_3}{r}~, \quad z^2 = z^3 = \zt^1 = 0~,
\end{aligned}\end{equation}
where $c_1$, $\tilde{c}_2$, and $\tilde{c}_3$ are complex numbers, $p_F^1$ and $Q$ are real-valued charges, the projectors $M$ and $\wt{M}$ are given by (\ref{eq:euclromsproj}), and the Wilson lines are
\begin{equation}\begin{aligned}
	e_R &= \frac{Q}{2r}~, \quad e_F^2 = \frac{\tilde{c}_2 (\kappa - \xi g Q)}{4 \xi g r^2}~, \quad e_F^3 = \frac{\tilde{c}_3 (\kappa - \xi g Q)}{4 \xi g r^2}~, \\
	e_F^1 &= \frac{p_F^1}{r} -\frac{c_1(\kappa + \xi g Q)}{4 \xi g r^2} - \frac{\tilde{c}_2 \tilde{c}_3 (\kappa - \xi g Q)}{2 \xi g r^3} \\
	& - \left(\frac{\kappa \tilde{c}_2 \tilde{c}_3}{r^2} + \frac{p_F^1 (\kappa + \xi g Q)}{2 \xi g r^2}\right)\left(\frac{2\xi g \sqrt{\kappa + \xi g Q} + (4 g^2 r^2 + \kappa + \xi g Q)\tan^{-1}\left(\frac{2\xi g r}{\sqrt{\kappa + \xi g Q}}\right)}{(\kappa + \xi g Q)^{3/2}}\right)\,.
\end{aligned}\end{equation}
The electric charges associated with this solution are 
\begin{equation}
	q_F^1 = p_F^1~, \quad q_F^2 = q_F^3 = 0~,
\end{equation}
and so these solutions once again correspond to turning on a dyonic flavor charge in the bulk.  We have verified that this solution satisfies the UV/IR correspondence (\ref{eq:UVIR}) for general choices of the parameters $c_1$, $\tilde{c}_2$, $\tilde{c}_3$, $p_F^1$ and $Q$, though we again have in mind that the magnetic fluxes satisfy the Dirac quantization condition (\ref{eq:dirac}).  This black saddle solution provides a highly non-trivial check of the UV/IR correspondence in the presence of magnetic flux, where the three independent real masses $\hat{\sigma}_F^\alpha$ and the three independent chemical potentials $\hat{\Delta}_F^\alpha$ are all non-zero and have non-linear dependence on the R-charge $Q$.

\subsection{Numerics}

So far we described several families of analytic black saddle solutions to the BPS equations by restricting the supergravity parameters that specify the solution.
Unfortunately a full analytical exploration of the solution space remains unfeasible. Therefore, in this section we resort to a numerical analysis to construct black saddle solution and further establish the holographic map spelled out in Section~\ref{sec:match}.

We are interested in constructing only regular black saddle solutions. Regularity is imposed in the IR region by requiring a smooth cap off at the origin near $r=r_0$.
It is most convenient to set up the numerics by specifying these regular boundary conditions at $r=r_0$ and integrating the BPS equations towards the UV region.
The constraints that imposed by regularity in the IR region are described in Section \ref{sec:IR}. Upon implementing these constraints, the solution space has 10  degrees of freedom that we are left to specify:
\begin{enumerate}
\item Four magnetic charges $p^I$ subject to one constraint, \eqref{eq:toptwistcond}, coming from the topological twist.
\item Four electric charges $q_I$.
\item For Branch $1$, three IR values $\zt^\alpha|_\text{IR}$; and for Branch $2$, three IR values $z^\alpha|_\text{IR}$.
\end{enumerate}
Furthermore, one needs to specify the genus of the Riemann surface which in turn fixes the parameter $\kappa=0,\pm 1$. 
Given a value for each of these 10 degrees of freedom and $\kappa$, a unique numerical solution might exist.
Still a solution is not guaranteed to exist, for some values of the degrees of freedom might be excluded on physical grounds.

Note that this setup of the solution space seems slightly different from the discussion in Section~\ref{sec:match} where instead of the electric charges and the IR values of the scalars the degrees of freedom were taken to be $\hat\beta,\hat \Delta^I,\hat \sigma^I$. This reorganization of the parameters specifying the solution is a consequence of shooting from the IR and therefore specifying IR boundary conditions. More specifically, one needs the dependence of all functions in the black saddle solution on the radial coordinate $r$ in order to determine the values of $\hat\beta,\hat \Delta^I,\hat \sigma^I$ from these IR boundary conditions alone. Importantly the parameters $\hat u^{I}$ are determined by the IR data of the solution, see \eqref{eq:hatu}. Therefore by constructing explicit numerical solutions we can confirm the UV/IR relation \eqref{eq:UVIR}.\footnote{It will be very interesting to understand if there is an alternative method to integrate the BPS equations and find a way to determine the parameters $\hat \Delta^I,\hat \sigma^I$ individually in terms of the data in the IR region.}

To set up the numerical integration, consider the following reduction of the BPS equations.
First, we take the difference between the first and third line of \eqref{eq:BPS} and the difference between the second and fourth line of \eqref{eq:BPS}, and solve the resulting equations for $M$ and $\wt M$. We are then left with eight equations for $f_1', f_3', z^{\alpha'}$ and $\zt^{\alpha'}$.
We also need four equations for ${e^I}'$, which can be obtained from \eqref{eq:qi}.
The metric function $f_2(r)$ keeps track of the gauge freedom to perform coordinate transformations on $r$.
We fix the gauge by picking
\begin{equation}
	e^{f_2(r)} = L = \frac{1}{\sqrt{2}g}~,
\end{equation}
and choose the IR region to be at $\rho=0$. We then take an IR expansion of the BPS equations of the form
\begin{equation}\label{eq:IRexpans}\begin{aligned}
	e^{f_1(r)} &= f_1'(0)e^{f_1(0)} r+\ldots~, \\
	e^{f_3(r)} &= e^{f_3(0)}+ f_3'(0)e^{f_3(0)}r+\ldots~,\\
	z^\alpha (r) &= z^\alpha(0) + {z^\alpha}{}'(0) r + \ldots~, \\
	\zt^\alpha (r) &= \zt^\alpha(0) + {\zt^\alpha}{}'(0) r + \ldots~, \\
	e^I (r) &= e^I(0) + {e^I}'(0) r+ \ldots~.
\end{aligned}\end{equation}
Rescaling of the $S^1$ coordinates allows us to set $f_1'(0)e^{f_1(0)}=1$ and we can use the gauge freedom for the potential to set $e_F^\alpha(0)=0$.
Performing an IR analysis as described in Section~\ref{sec:IR} fixes the remaining coefficients in the above expansion in terms of the magnetic and electric charges, except for $\zt^\alpha(0)$ on Branch $1$ or $z^\alpha(0)$ on Branch $2$. We find a perturbative IR solution up to order $r^2$ and set up boundary conditions as
\begin{equation}
	\phi_i(r=\epsilon) = \phi_i^\text{IR}(r=\epsilon)~,
\end{equation}
where $\phi_i$ is a vector containing all fields $\phi_i = \{f_1,f_3,z^\alpha,\zt^\alpha,e^I\}$ and $\phi_i^\text{IR}$ is the same vector containing the perturbative IR solution. The parameter
$\epsilon$ is taken to be small (order $10^{-10}$) and determines at what value of $r$ to start the numerical integration.
We then perform the numerical integration using the function $\texttt{NDSolve}$ in $\textrm{Mathematica}$.

This ten-dimensional solution space is too large for a comprehensive numerical analysis. For a concise presentation of our numerical solutions, we will restrict the solution space in several ways. First, let us focus on Branch $1$, such that we can choose  $\zt^\alpha(0)$, while $z^\alpha(0)$ is fixed in terms of the electric and magnetic charges. Second, we restrict the parameter space by imposing $SU(3)$ symmetry in the STU model, i.e. we take:
\begin{equation}\label{eq:SU3numerics}\begin{aligned}
	&p_F^1=p_F^2=p_F^3\equiv p_F~, \\
	&q_F^1=q_F^2=q_F^3\equiv q_F~, \\
	&\zt^1(0)=\zt^2(0)=\zt^3(0) \equiv \zt(0)~.
\end{aligned}\end{equation}
This leaves a four-dimensional solution space spanned by $p_F,q_F,q_R$ and $\zt(0)$, for any given genus $\mathfrak{g}$ of the Riemann surface.  Importantly, the restrictions in \eqref{eq:SU3numerics} ensure that $z^\alpha(r)=z(r)$, $\zt^\alpha(r)=\zt(r)$ and $e_F^\alpha(r) = e_F(r)$ along the entire flow.
The IR expansion informs us that $z(0)$ is related to the charges by
\begin{equation}
	q_F = -\frac{2 p_{R} z(0)^2+p_F \left(z(0)^4+2 z(0)^3+2 z(0)+1\right)}{(z(0)-1)^3 (z(0)+1)}~.
\end{equation}
It is worth discussing also the connection with Lorentzian black holes. The simplest and most explicit black hole solutions in this class are magnetic black holes \cite{Cacciatori:2009iz}. To obtain this class of solutions we further set $q_F=0$, such that
\begin{equation}\label{eq:cpF}
p_F = -\frac{2 p_R z(0)^2}{z(0)^4+2 z(0)^3+2 z(0)+1}~.
\end{equation}
The IR value of $e^{2f_3}$ now reads
\begin{equation}
e^{2f_3(0)} = \frac{1}{2\xi g} \left(q_R-\frac{p_R (z(0)-1) (z(0)+1)^3}{z(0)^4+2 z(0)^3+2 z(0)+1}\right)~.
\end{equation}
With these restrictions, the only Lorentzian regular solutions are magnetic black holes with $q_R=0$.
One can check that $e^{2f_3(0)}|_{q_R=0}>0$ only if $\xi p_F<0$.
Moreover, Lorentzian magnetic black holes have $z=\zt$.
We thus conclude that Lorentzian regular solutions exist for $q_F=q_R=0$, with $z(0)=\zt(0)$ determined by solving \eqref{eq:cpF}.
These solutions depend on the two free parameters $p_F$ and $\kappa$ which should obey the constraint $\xi p_F<0$. In Euclidean signature there exists a much larger family of regular solutions determined by the constants $p_F$, $\kappa$, $\zt(0)$ and $q_R$.
In particular, we can access the range $\xi p_F>0$, by making $q_R$ large enough such that $e^{2f_3(0)}>0$. This illustrates once again that black saddles represent a more general class of supergravity solutions which includes the known supersymmetric static black holes studied in \cite{Cacciatori:2009iz}.

Let us illustrate this with an explicit example.
A Lorentzian magnetic black hole is specified by the flavor magnetic charge $p_F$ and the curvature of the Riemann surface through $p_R = -\kappa/(2\xi g)$.
Keeping the curvature of the Riemann surface $\kappa$ free for now, we tune the flavor magnetic charge as $p_F = -p_R =\kappa/(2\xi g)$. In terms of the $p^I$, this implies:
\begin{equation}
p^0 = -p^1 = -p^2 = -p^3 = \frac{-\kappa}{2\xi g}~,
\end{equation}
which is consistent with the quantization condition $\eta \xi   g \, p^I \in \mathbb{Z}$. The constraint \eqref{eq:cpF} is then solved with $z(0) = \frac{1}{2} \left(\sqrt{2 \left(\sqrt{5}+1\right)}-1-\sqrt{5}\right)\approx 0.346$, where we picked the root with $|z(0)|<1$.
One can solve the BPS equations to find the following Lorentzian magnetic black hole solution
\begin{equation}
ds^2 = - e^{K(r)}V(r) dt^2 + e^{-K(r)}\frac{dr^2}{V(r)} + e^{-K(r)} r^2 d\Sigma_\mathfrak{g}^2 ~,
\end{equation}
with
\begin{equation}
e^{K(r)} = 2 \frac{(1-z(r)^3)^2}{(1-z(r)^2)^3}~, \quad V(r) = \left(g r+\frac{5\kappa}{4 g r}\right)^2~,
\end{equation}
and the scalar field takes the form
\begin{equation}
z(r) = \zt(r) = \frac{1}{2} \left(-1-\frac{2 g r}{\sqrt{-\kappa }}+\sqrt{\frac{\left(\sqrt{-\kappa }-2 g r\right) \left(3 \sqrt{-\kappa }+2 g r\right)}{\kappa }}\right)~.
\end{equation}
Note that $z(r)$ obeys
\begin{equation}
\frac{z(r)^2+z(r)+1}{z(r)} = -\frac{2 g}{\sqrt{-\kappa }} r~.
\end{equation}
Now let us consider the curvature of the Riemann surface.
The metric function $V(r)$ only has a root for $\kappa=-1$ and moreover $z(r)$ is real only for $\kappa=-1$.
Evidently, in Lorentzian signature this solution is only well-defined for $\kappa=-1$.
As expected, we see that the regular black hole lies in the range $\xi p_F=-1/(2g)<0$.

The space of regular solutions in Euclidean signature is quite distinct from Lorentzian signature.
To find the solutions for generic values of $\zt(0)$ and $q_R$ we resort to a numerical analysis.
We have checked numerically that Euclidean solutions exist in the following cases:
\begin{itemize}
	\item Case 1: $p_F=-\frac{1}{2\xi g}$, $z(0)=\zt(0)=\frac{1}{2} \left(\sqrt{2 \left(\sqrt{5}+1\right)}-1-\sqrt{5}\right)$, $\kappa=-1$ and $q_R=n/20$ for $n\in (1,\dots,100)$.
	For these parameters there is a regular Lorentzian black hole at $q_R=0$. 
	\item Case 2: $p_F=-\frac{1}{2\xi g}$, $z(0)=\frac{1}{2} \left(\sqrt{2 \left(\sqrt{5}+1\right)}-1-\sqrt{5}\right)$, $q_R=0$, $\kappa=-1$ and $\zt(0)=z(0)+\frac{n}{100}(1-z(0))$ for $n\in(1,\dots,99)$.	For these parameters there is a regular Lorentzian black hole at $\zt(0)=z(0)$. 
	\item Case 3: $p_F=\frac{1}{2\xi g}$, $z(0)=\zt(0)=\frac{1}{2} \left(\sqrt{2 \left(\sqrt{5}+1\right)}-1-\sqrt{5}\right)$, $\kappa=1$ and $q_R=\frac{3}{2}+\frac{n}{25}$,  for $n\in(1,\dots,100)$.
	For these parameters there is no smooth limit of $q_R$ to a regular Lorentzian black hole.
\end{itemize}
In the numerical construction we choose the normalization $L=1$ and $\xi=1$.
The value of $z(0)$ is found by solving \eqref{eq:cpF} and picking the root satisfying $|z(0)|<1$.
Case 1 serves to illustrate that the regular Lorentzian magnetic black hole can be extended in Euclidean signature by turning on the electric charge $q_R\neq0$.
Similarly, Case 2 shows that the regular Lorentzian magnetic black hole can be generalized in Euclidean signature by taking $z(r)\neq \zt(r)$.
Finally, Case 3 contains a family of Euclidean black saddles that do not allow for a smooth limit to a Lorentzian black hole.

\begin{figure}
	\begin{center}
	\includegraphics[width=0.6\textwidth]{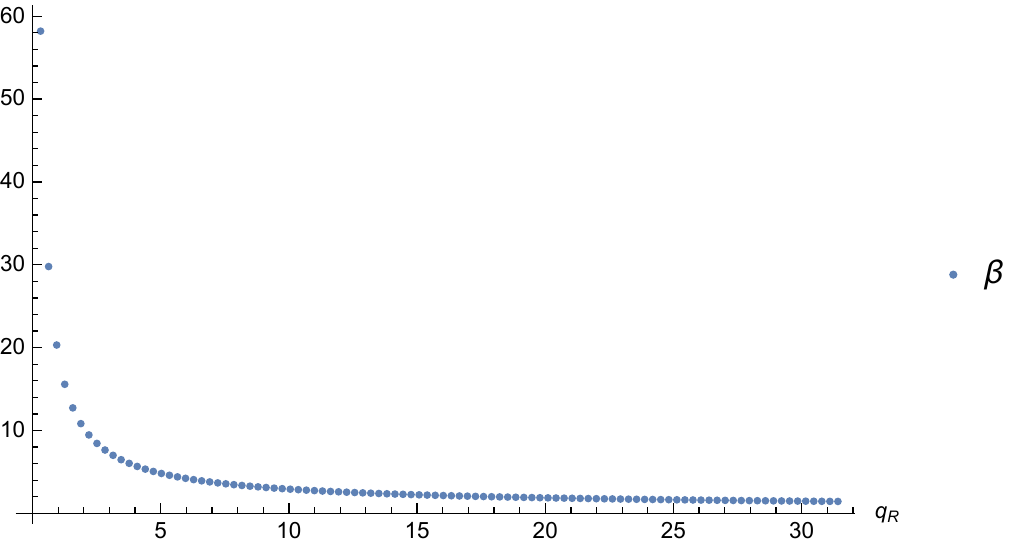}
	\caption{Numerical values of $\hat\beta$ for Case 1 as a function of $q_R$.}
	\label{fig:case_1_beta}
	\end{center}
\end{figure}

\begin{figure}
	\begin{center}
	\includegraphics[width=0.75\textwidth]{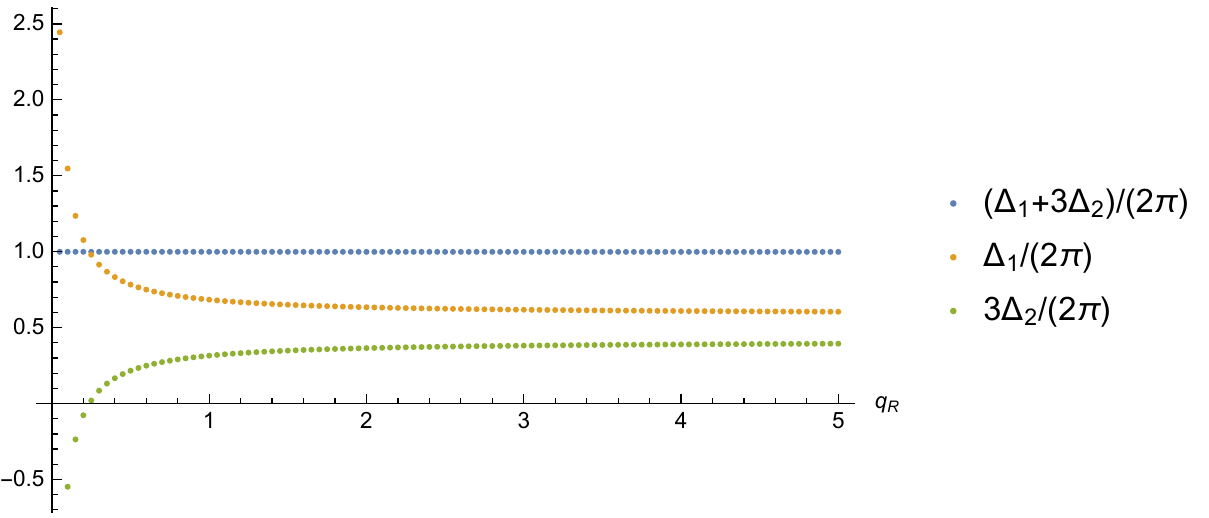}
	\caption{Numerical values of the chemical potentials $\hat\Delta_1$ and $\hat\Delta_2$ normalized by $2\pi$ for Case 1 as a function of $q_R$. Also the combination $\hat\Delta_1+3\hat\Delta_2$ is shown which is expected to be equal to $2\pi$.}
	\label{fig:case_1_chem}
	\end{center}
\end{figure}

\begin{figure}
	\begin{center}
	\includegraphics[width=0.75\textwidth]{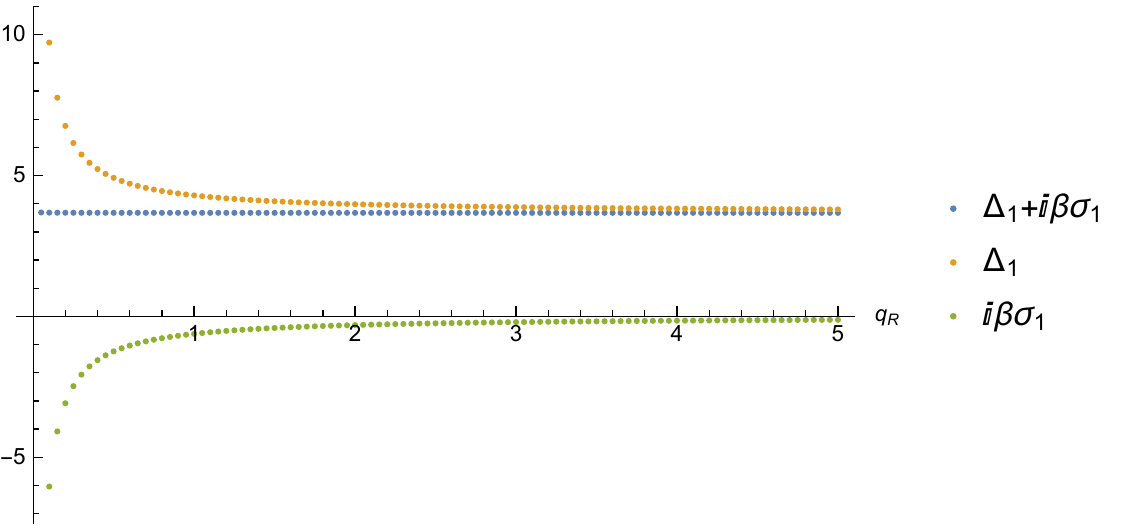}
	\caption{Numerical values of $\hat\Delta_1$, $i\hat\beta \hat\sigma_1$ and the combination $\hat\Delta_1+i\hat\beta\hat\sigma_1$ for Case 1 as a function of $q_R$.}
	\label{fig:case_1_UVIRrel}
	\end{center}
\end{figure}

\begin{figure}
	\begin{center}
	\includegraphics[width=0.75\textwidth]{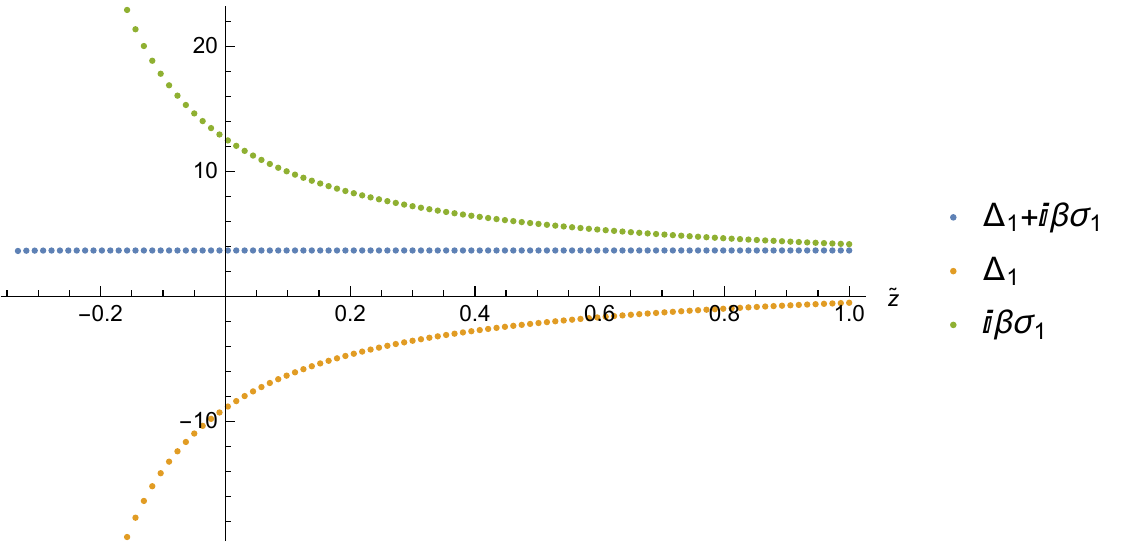}
	\caption{Numerical values of $\hat\Delta_1$, $i\hat\beta \hat\sigma_1$ and the combination $\hat\Delta_1+i\hat\beta\hat\sigma_1$ for Case 2 as a function the IR value $\zt(0)$.}
	\label{fig:case_2_UVIRrel}
	\end{center}
\end{figure}

\begin{figure}
	\begin{center}
	\includegraphics[width=0.75\textwidth]{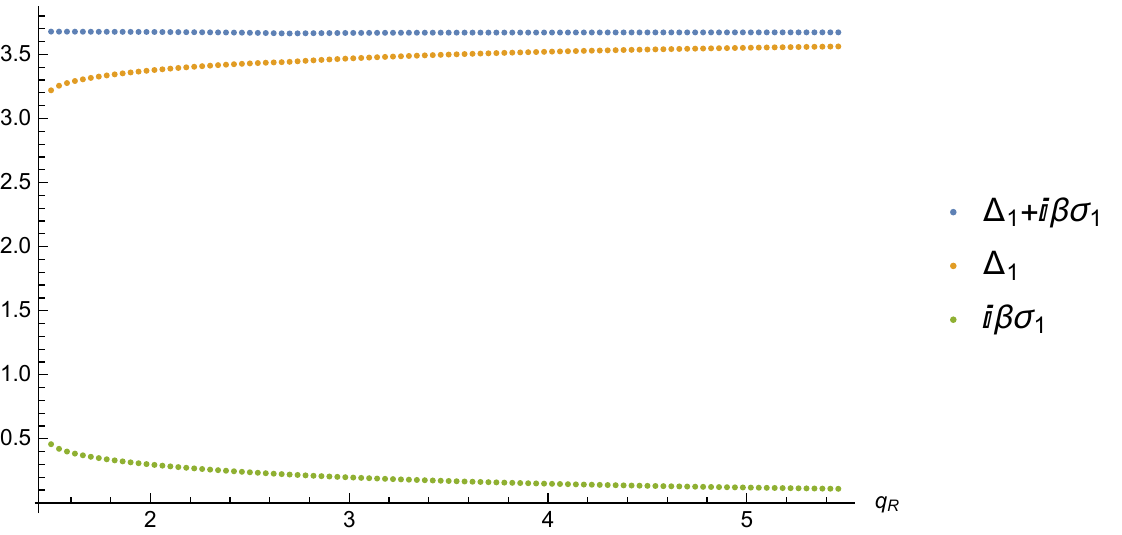}
	\caption{Numerical values of $\hat\Delta_1$, $i\hat\beta \hat\sigma_1$ and the combination $\hat\Delta_1+i\hat\beta\hat\sigma_1$ for Case 3 as a function of $q_R$.}
	\label{fig:case_3_UVIRrel}
	\end{center}
\end{figure}

The parameters $\hat \beta$, $\hat u^I$, $\hat\Delta^I$ and $\hat \sigma^I$ should be extracted from the numerical solutions using \eqref{eq:betahata0}, \eqref{eq:hatdelta}, \eqref{eq:hatsigma} and \eqref{eq:uhat}.
However, because we chose the IR to lie at $r=0$, the UV expansion of our fields differs from \eqref{eq:UVexpans} by a shift $r_\text{shift}$ in the radial coordinate.
This shift affects the UV behaviour of $z(r)$, $\zt(r)$, $e^{-f_1(r)}$ and $e^{-f_3(r)}$ by the same factor $e^{r_\text{shift}}$, so we can use the combination $e^{f_3(r)}z(r)$, $e^{f_3(r)}z$ and $e^{f_1(r)-f_3(r)}$ whose UV behaviour is independent of $r_\text{shift}$.
Additionally, we used the freedom in rescaling the $\tau$ coordinate to set $e^{f_1(0)}f_1'(0)=1$, such that $\beta_\tau=2\pi L$.
Taking this into account, we can extract the parameters as follows
\begin{equation}\begin{aligned}
\hat \beta &= 2\pi e^{f_1(r_\text{max})-f_3(r_\text{max})}~,\\
\hat{u}^1 &= \frac{\pi  (z(0)-1)^2}{2 \left(z(0)^2+z(0)+1\right)}~, & \hat{u}^2 &= \hat u^3= \hat u^4 =\frac{\pi  (z(0)+1)^2}{2 \left(z(0)^2+z(0)+1\right)}~, \\
\hat{\Delta}^I &= 2\pi \xi g \left(e^I(\epsilon) - e^I(r_\text{max})\right)~, \\
\hat{\sigma}_F^1 &= \frac{i}{4}e^{f_3(r_\text{max})}3\left(z(r_\text{max}) - \zt(r_\text{max}) \right)~, & \hat{\sigma}_F^2 &= -\frac{i}{4}e^{f_3(r_\text{max})}\left(z(r_\text{max}) - \zt(r_\text{max}) \right)~,
\end{aligned}\end{equation}
where $\epsilon$ and $r_\text{max}$ are the minimal and maximal numerical values of the radial coordinate $r$.
We choose $\epsilon=10^{-10}$ and we let $r_\text{max}$ be chosen dynamically by stopping the integration once $f_1'(r)-1<10^{-6}$, resulting in $5 \lesssim r_\text{max} \lesssim 15$ , depending on the chosen IR parameters.
Due to the $SU(3)$ symmetry, one finds $\hat u^2=\hat u^3=\hat u^4$ and $\hat \sigma^2=\hat \sigma^3=\hat \sigma^4$.
All cases have the same value for $z(0)=\frac{1}{2} \left(\sqrt{2 \left(\sqrt{5}+1\right)}-1-\sqrt{5}\right)$, such that
\begin{equation}
\hat u^1 = \frac{1}{10} \left(3 \sqrt{5}+5\right) \pi \approx 3.67824~, \qquad \hat u^2 = \hat u^3= \hat u^4 = \frac{1}{10} \left(5-\sqrt{5}\right) \pi \approx 0.868315~. 
\label{eq:u1value}
\end{equation}
Figure~\ref{fig:case_1_beta} shows the parameter $\hat \beta$ for Case 1 as a function of the parameter $q_R$.
We observe that as $q_R$ tends to zero, the parameter $\hat \beta$ goes to infinity.
This is precisely the limit where a supersymmetric Lorentzian black hole with $\hat \beta = \infty$ exists.
Figure~\ref{fig:case_1_chem} shows the chemical potentials for Case 1 as a function of the parameter $q_R$.
As expected, see \eqref{eq:deltaconstr}, the combination $\sum_I\Delta^I=\Delta^1+3\Delta^2 = 2\pi$.
Figure~\ref{fig:case_1_UVIRrel} displays the values of $\hat\Delta^1$, $i \hat \beta \hat \sigma^1$ and the combination $\hat{\Delta}^1+i\hat{\beta}\hat{\sigma}^1$ for Case 1 as a function of the parameter $q_R$.
While both $\hat\Delta^1$ and $i \hat \beta \hat \sigma^1$ have a non-trivial dependence on $q_R$, their sum is constant and equal to the value of $\hat u^1$ as in~\eqref{eq:u1value}.  Similar results can be obtained for $\hat u^2 = \hat u^3= \hat u^4$.
Figure~\ref{fig:case_2_UVIRrel} and \ref{fig:case_3_UVIRrel} similarly validate the UV/IR relation \eqref{eq:UVIR} for Case 2 and Case 3.

\begin{figure}
	\begin{center}
	\includegraphics[width=0.75\textwidth]{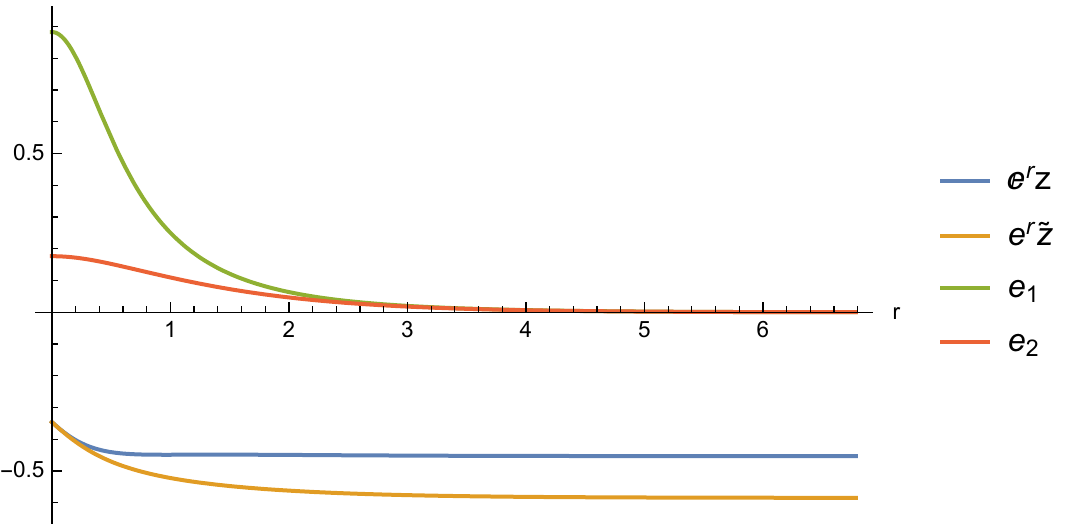}
	\caption{Numerical solutions of $e^r z(r)$, $e^r \zt(r)$, $e_1(r)-e_1(r_\text{end})$ and $e_2(r)-e_2(r_\text{end})$ as a function of $r$ for Case 1 with $q_R=\frac{5}{2}$.}
	\label{fig:case_random}
	\end{center}
\end{figure}

Figure \ref{fig:case_random} shows the profiles of the fields as a function of $r$ for Case 1 with $n=50$, corresponding to $q_R=\frac{5}{2}$.
The combination $e^r z$ and $e^r \zt$ is shown, which allows for extracting the parameters $z_0^\alpha$, ${\zt}_0^\alpha$, as defined in \eqref{eq:UVexpans}.
Note that even though $z$ and $\zt$ start off at the same IR value, their profiles and asymptotic values differ.

%
We emphasize that large families of Euclidean black saddle solutions exist, irrespective of whether there exists a limit of the parameters which allow for a smooth Lorentzian black hole.
This is exemplified in Case 3, where, keeping $p_F,\kappa$ and $q_F$ fixed, there is no limit that allows for a Wick rotation to a regular Lorentzian black hole.
Nevertheless, there is a 2-parameter family of Euclidean solutions spanned by $\zt(0)$ and $q_R$.

Note that reducing the BPS equations to the equations used for the numerical integration, we obtain solutions for $M$ and $\Mt$ as a function of the fields $\phi_i$.
Recall that $M$ and $\Mt$ satisfy the condition $M \Mt=1$, which we have not implemented explicitly in the numerical integration procedure.
We can thus perform a consistency check by evaluating the product of $M$ and $\Mt$ on the numerical solutions.
Indeed, we find that $M\Mt-1$ is of the order $10^{-10}$ for the solution presented in Figure \ref{fig:case_random}.

In this section we have numerically verified the existence of a large family of regular Euclidean black saddle solutions.
Moreover, we have checked the validity of the UV/IR relation \eqref{eq:UVIR} in all cases presented above.
This analysis underscores the importance of the black saddle solutions as generalizations of the Lorentzian black holes of \cite{Cacciatori:2009iz} and provides compelling evidence that \eqref{eq:UVIR} holds in general.

\section{Discussion}
\label{sec:discussions}

After the detailed presentation of the black saddle solutions, their on-shell action and holographic interpretation, it is time to take stock and discuss the interpretation of some of our results in the context of $\mathcal{I}$-extremization.  We will also discuss some possible generalizations and open questions.

\subsection{An ode to extremization}

As discussed in Section~\ref{sec:fieldtheorypartfunc}, the topologically twisted index $\mathcal{I} \equiv \log Z(u,\mathfrak{p})$ is well-defined for arbitrary values of the deformations parameters $u = \Delta+i\beta \sigma$ for a fixed choice of the $\Sigma_{\mathfrak{g}}$ and background magnetic fluxes $\mathfrak{p}$. It was argued in  \cite{Benini:2015eyy,Benini:2016rke} that there is a preferred choice of the parameters $u$: namely, the one that extremizes the index. This $\mathcal{I}$-extremization principle is analogous to similar extremization results for conformal anomalies and partition functions in other dimensions and setups \cite{Intriligator:2003jj,Jafferis:2010un,Benini:2012cz}. The physical interpretation of $\mathcal{I}$-extremization is also clear. The topologically twisted three-dimensional $\mathcal{N}=2$ SCFT on $S^1\times\Sigma_{\mathfrak{g}}$ is deformed by the parameters $(u,\mathfrak{p})$ and undergoes an RG flow that results in an effective one-dimensional quantum mechanics theory with two supercharges in the deep IR. If the dynamics of this one-dimensional theory are conformal, then $\mathcal{I}$-extremization is the principle that determines which of the available Abelian global symmetries is the unique superconformal R-symmetry.\footnote{We do not discuss here the subtle and important question whether such a superconformal quantum mechanical theory exists at finite $N$ and what are its detailed properties. The near-horizon AdS$_2$ region of the supersymmetric black holes discussed in this work suggests that there is some notion of superconformal invariance at least in the large $N$ limit.} Put differently, for a given choice of $\mathfrak{g}$ and $\mathfrak{p}$ one can use $\mathcal{I}$-extremization to determine which choice of $u$ will lead to conformal dynamics in the IR. Presumably for other choices of the parameters $u$ the IR quantum mechanics theory is gapped and thus there is no notion of a superconformal R-symmetry. It is important to recall also that the topologically twisted index is a partition function in the canonical ensemble with respect to the magnetic charges $\mathfrak{p}$ but grand canonical with respect to the electric charges, since we fix the fugacities $u$ instead of the charges $\mathfrak{q}$. In order to preserve supersymmetry the fugacity for the R-symmetry is fixed as in \eqref{eq:toptwistcft} and thus one cannot determine the electric R-charge $\mathfrak{q}_R$.

The supergravity manifestation of $\mathcal{I}$-extremization is also discussed in \cite{Benini:2015eyy,Benini:2016rke} and is easy to state. The supersymmetric Lorentzian static black hole solutions presented in \cite{Benini:2015eyy,Benini:2016rke} have an AdS$_2$ near-horizon geometry. The supergravity BPS equations uniquely determine the values of the three scalars $z^{\alpha}$ in the STU model in this near-horizon region. When this IR boundary condition is interpreted in the asymptotically AdS$_4$ region of the black hole, it amounts to fixing the field theory parameters $u$ precisely as dictated by $\mathcal{I}$-extremization. 

The black saddle solutions address two important subtleties in this dialogue between field theory and supergravity. To illustrate this, let us first focus on a setup where the flavor electric charges $q_F^{\alpha}$ vanish. This can be arranged in supergravity by using the relation in \eqref{eq:Zqsugra}. One can then choose to not fix the electric R-charge nor the asymptotic values of the supergravity scalar fields and thus avoid $\mathcal{I}$-extremization altogether. This in turn is reflected in a supergravity black saddle solution which caps-off smoothly as $\mathbb{R}^2\times \Sigma_{\mathfrak{g}}$ with finite $\beta$ without developing an AdS$_2$ throat. Alternatively one can invoke $\mathcal{I}$-extremization and determine the values of the fugacities $u$ as a function of the magnetic charges $p$. This should na\"ively correspond to selecting a black saddle solution with infinite $\beta$ which can be analytically continued to a Lorentzian black hole with an AdS$_2$ near-horizon region. 

However, this is not always possible; it could happen that such a regular black hole solution does not exist at all or exists only for a specific value of the electric R-charge $\mathfrak{q}_R$. In the cases when the black hole solution indeed exists, the topologically twisted index after $\mathcal{I}$-extremization was shown to reproduce the black hole entropy \cite{Benini:2015eyy,Benini:2016rke}. In all other situations the supergravity description of the topologically twisted index is in terms of the on-shell action of the Euclidean black saddle solutions that we have constructed in this work. The simple black saddle solution with $\mathfrak{g}>1$ presented in Section~\ref{subsec:grav} can act as an illustrative example. In Euclidean signature this solution is regular for all values of the electric R-charge $Q$. The Lorentzian black hole exists only in the limit $\beta\rightarrow \infty$ which corresponds to $Q\rightarrow 0$. Notice that even though $Q$ is a parameter which determines the black saddle solution, the topologically twisted index is independent of the value of $\mathfrak{q}_R$ and correspondingly the on-shell action in supergravity does not depend on the value of $Q$ \eqref{eq:univBSonshellA}. This is in line with the discussion in \cite{Benini:2016rke} where the authors argue that the black hole solutions is the dominant supergravity saddle for a specific value for $q_R$. For all other values of $q_R$ the black saddles should dominate the supergravity path integral. 
 
In the case where the flavor electric charges $q_F^{\alpha}$ do not vanish the relation between $\mathcal{I}$-extremization, black hole entropy, and the on-shell action of the black saddle solutions requires a minor modification. One must first perform a Legendre transformation and obtain the following quantity in the canonical ensemble with respect to electric charges:
\begin{equation}\label{eq:ILegconcl}
\mathcal{I}(p^I,q^I) \equiv \log Z_\text{grav}(\hat{u}^I,p^I) + \frac{\eta}{4 \xi gG}\,\hat u_F^\alpha q_F^\alpha~.
\end{equation}
Then $\mathcal{I}$-extremization amounts to extremizing the right hand side of \eqref{eq:ILegconcl} over all fugacities $u$ for a fixed set of electric and magnetic charges. For the topologically twisted index of the ABJM theory this extremization is performed in detail in Section 2 of~\cite{Bobev:2018uxk}, and the end result should be compared with the entropy of a sypersymmetric dyonic black hole with this collection of electric and magnetic charges \cite{Benini:2016rke}. To do this, however, one again has to uniquely fix the value of the electric charge $q_R$ for the R-symmetry in order to ensure that there exists a corresponding smooth black hole. For all other values of $q_R$ the topologically twisted index of the ABJM theory should be compared to the on-shell action of the black saddle solutions discussed in this work. Once again, if the $\mathcal{I}$-extremization procedure is not performed then the appropriate solutions to consider are the Euclidean black saddles.

The somewhat lengthy discussion above can be summarized as follows. If $Z_\text{grav}$ in \eqref{eq:Zgravcomp} is viewed as a holomorphic function of $\hat{u} = \hat{\Delta} +i \hat{\beta} \hat{\sigma}$ then the expression which determines the electric charges in \eqref{eq:Zqsugra} can be written as
\begin{equation}\label{eq:odeext}
\frac{\partial \log Z_\text{grav}}{\partial \hat{u}_F^\alpha} = \mp \frac{\eta}{4\xi g G_N}\,q_F^\alpha~.
\end{equation}
If one in addition imposes that $\hat{\Delta}$, $\hat{\sigma}$, and $q_F^\alpha$ are real then \eqref{eq:odeext} can be thought of as fixing completely the parameters $\hat{\Delta}$ and $\hat{\sigma}$ in terms of the electric and magnetic charges. This is the point of view advocated in \cite{Benini:2016rke}. The two subtleties discussed at length above are that $q_R$ is not fixed by \eqref{eq:odeext} and that the reality condition on $\hat{\Delta}$ and $\hat{\sigma}$ can be relaxed in Euclidean signature.  As we discussed above the dyonic Lorentzian black holes indeed impose a reality condition on $\hat{\Delta}$ and $\hat{\sigma}$ and fix uniquely the charge $q_R$ (if there is a regular black hole solution). In this sense the $\mathcal{I}$-extremization procedure in  \cite{Benini:2015eyy,Benini:2016rke} is an essential ingredient in the holographic dictionary for black holes. For more general situations, however, $\mathcal{I}$-extremization can be eschewed and the honor of holography is saved by the black saddle solutions discussed in this work.
 
\subsection{Generalizations and open questions}

Our work suggests a number of avenues for future research. Here we summarize some of them.

\begin{itemize}

\item In this work we studied black saddle solutions in the STU model of four-dimensional gauged supergravity which arises as a consistent truncation of eleven-dimensional supergravity. It is clear that there should be black saddle solutions in other four-dimensional gauged supergravity models that arise from string and M-theory. To look for such solutions it is natural to consider models with known field theory dual descriptions in which there are explicit constructions of supersymmetric dyonic black holes. Two specific examples in this context involve coupling the STU model to a hyper multiplet with two distinct gaugings which in turn admits uplifts to massive IIA \cite{Guarino:2017eag,Guarino:2017pkw,Azzurli:2017kxo,Hosseini:2017fjo,Benini:2017oxt} or eleven-dimensional supergravity \cite{Bobev:2018uxk,Bobev:2018wbt}. It will be very interesting to study supersymmetric black saddle solutions in these models and relate their on-shell actions to the topologically twisted index in the holographically dual QFT. It is also desirable to delineate the properties of supersymmetric black saddles in general matter-coupled 4d $\mathcal{N}=2$ gauged supergravity theories.

\item The Euclidean black saddle solutions we constructed should admit several generalizations. First, it will be interesting to consider solutions with more general metrics on $\Sigma_{\mathfrak{g}}$ along the lines of \cite{Anderson:2011cz,Bobev:2019ore,Bobev:2020jlb}. Second, it should be possible to find black saddles for which the geometry at asymptotic infinity is a more general three-manifold, rather than $S^1\times \Sigma_{\mathfrak{g}}$. Solutions of this type have been studied in \cite{Toldo:2017qsh} and it will be very interesting to generalize them, understand their relation to the Lorentzian rotating black hole solutions in \cite{Hristov:2018spe,Hristov:2019mqp}, and establish their dual field theory description. It is important also to study the relation between black saddles and the ``gravitational blocks'' discussed recently in \cite{Hosseini:2019iad,Choi:2019dfu}. We also note that it will be interesting to construct black saddle solutions which correspond to general values of the parameter $n \neq 1$ which specifies the R-symmetry chemical potential in \eqref{eq:toptwistcft} and understand their interpretation in the dual field theory.

\item In Section~\ref{sec:solutions} we presented a number of explicit examples of analytic black saddle solutions. However, we were not able to solve the full system of BPS equations and find the most general analytic black saddle solutions. It will be interesting to explore whether such explicit solutions can be constructed. A fruitful strategy could be to understand whether specific limits of the non-supersymmetric dyonic black hole solutions in \cite{Chow:2013gba} can lead to new Euclidean black saddle solutions.\footnote{This is similar in spirit to the procedure used in~\cite{Cabo-Bizet:2018ehj,Cassani:2019mms} to understand the BPS limit of the quantum statistical relation for AdS black holes by first imposing supersymmetry and then extremality on suitably Euclideanized versions of non-extremal black hole solutions.} It should also be possible to use the method outlined in \cite{Kim:2019feb} to construct approximate black saddle solutions.

\item Using the explicit formulae in \cite{Azizi:2016noi} it is possible to uplift our black saddle solutions to backgrounds of eleven-dimensional supergravity. It will be very interesting to do this explicitly and to investigate possible relations with the studies of $\mathcal{I}$-extremization in \cite{Couzens:2018wnk,Hosseini:2019ddy,Gauntlett:2019roi,Kim:2019umc,Gauntlett:2019pqg,vanBeest:2020vlv}. A similar question can be posed about the relation between the holographic realization of $F$-maximization in four-dimensional gauged supergravity \cite{Freedman:2013oja} and the Sasaki-Einstein volume minimization principle studied in \cite{Martelli:2005tp,Martelli:2006yb}.

\item In Section~\ref{sec:holorenorm} we derived a relatively simple result for the on-shell action of the black saddle solutions. This was ultimately possible due to the fact that the on-shell action can be rewritten as a total derivative which in turn implies that it receives contributions only from the UV and IR regions of the geometry. This is reminiscent of the recent results on the ``localization of the on-shell action'' in minimal gauged supergravity \cite{Genolini:2016ecx,BenettiGenolini:2019jdz}. Our results for the black saddle on-shell action together with the results in \cite{Freedman:2013oja} suggest that a similar general formula should exist also in matter-coupled supergravity model, like the STU model studied here. Work towards establishing such a formula is in progress \cite{BBCT}. We note that the evaluation of the on-shell action of the black saddle solution and its agreement with the results in the dual field theory relies on the UV/IR map of the supergravity parameters in \eqref{eq:UVIR}. This relation is reminiscent of other UV/IR relations that underly a number of precision results in the context of holography and supersymmetric localization \cite{Freedman:2013oja,Bobev:2013cja,Bobev:2016nua,Bobev:2018hbq,Bobev:2018ugk,Bobev:2018wbt,Bobev:2019bvq}. It is certainly desirable to have a better understanding of the interplay between imposing regularity in the IR region of Euclidean supergravity solutions and the resulting UV/IR map.

\item Another worthy goal is understanding how higher-curvature corrections to supergravity modify the black saddle solutions and their corresponding black hole descendants. This will allow for a holographic calculation of the subleading terms in the large $N$ expansion of the topologically twisted index. This very interesting problem is studied in \cite{Bobev:2020egg}.

\item It is natural to expect that there is a generalization of the supersymmetric black saddle solutions to gauged supergravity theories in other dimensions. Perhaps most relevant in the holographic context is the construction of black saddles in five-dimensional $\mathcal{N}=2$ gauged supergravity. The STU model in five dimensions has a family of static supersymmetric Lorentzian black string solutions studied in detail in \cite{Maldacena:2000mw,Benini:2012cz,Benini:2013cda}. These solutions are holographically dual to the topologically twisted $\mathcal{N}=4$ SYM theory in four dimensions on $T^2\times \Sigma_{\mathfrak{g}}$. The topologically twisted index of this theory was defined in \cite{Benini:2015noa,Benini:2016hjo,Closset:2016arn} and discussed in more detail in \cite{Hosseini:2016cyf}. It will be very interesting to construct these five-dimensional black saddle solutions and study their relation to this topologically twisted index. We should stress that these solutions should be different from the supersymmetric Euclidean solutions discussed in \cite{Cabo-Bizet:2018ehj,Cassani:2019mms} which are relevant for the physics of supersymmetric rotating AdS$_5$ black holes and their relation to the superconformal index of the dual four-dimensional SCFT \cite{Hosseini:2017mds}.\footnote{See also \cite{Bobev:2019zmz,Benini:2019dyp} for a similar discussion in four-dimensional minimal gauged supergravity.}  While there is a conceptual similarity to the black saddles discussed here, the Euclidean solutions in \cite{Cabo-Bizet:2018ehj,Cassani:2019mms,Bobev:2019zmz,Benini:2019dyp} are constructed by analytic continuation of non-extremal black hole solutions, whereas we are interested in a more general class of Euclidean saddle points that generically do not admit any continuation to a black hole.  This is an important technical difference between the supergravity analysis in \cite{Cabo-Bizet:2018ehj,Cassani:2019mms} and the one in this paper.

\item It has recently been proposed in~\cite{Almheiri:2019qdq,Penington:2019kki} that the Page curve for an evaporating black hole can be reproduced from the gravitational path integral by including a new class of saddles, which are referred to as \emph{replica wormholes}.  These replica wormholes are Euclidean geometries that connect multiple copies of the original black hole geometry, and their contribution to the path integral can actually dominate over the original black hole saddle.  The replica wormholes and the black saddles we construct in this work stem from a similar philosophy, namely that Euclidean quantum gravity allows for saddle points that do not admit a Lorentzian interpretation but nonetheless give large contributions to the gravitational path integral.  However, they serve very different purposes; the black saddles are a generalization of black holes for gravitational ensembles that admit no black hole solutions, while the replica wormholes are constructed by gluing together $n$-fold copies of a black hole within a given gravitational ensemble that allows for black holes.  It would be interesting to see if these perspectives can be merged by constructing ``replica black saddles'', i.e. Euclidean saddles that connect multiple copies of a black saddle geometry via the replica trick, as well as study their relation to the dual field theory.

\item The black saddle solutions studied in this work, as well as other similar asymptotically AdS Euclidean solutions constructed recently in the context of holography, provide a precise arena to study Euclidean holography. This is facilitated by the detailed knowledge of the supersymmetric partition function of the dual QFT. It is our hope that these results will find applications beyond their supersymmetric holographic roots and provide novel insights into the path integral of quantum gravity \cite{Gibbons:1976ue,Gibbons:1994cg}.

\end{itemize}

\section*{Acknowledgments}
We are grateful to Pietro Benetti Genolini, Francesco Benini, Pieter Bomans, Davide Cassani, Marcos Crichigno, Fridrik Freyr Gautason, Thomas Hertog, Seyed Morteza Hosseini, Kiril Hristov, Klaas Parmentier, Krzysztof Pilch, Valentin Reys, Chiara Toldo, and Alberto Zaffaroni for valuable discussions. NB is grateful to the participants and organizers of the 2018 ICTP Workshop on Supersymmetric Localization and Holography for a stimulating research environment and in particular to Kiril Hristov and Alberto Zaffaroni for the numerous heated discussions. The work of NB is supported in part by an Odysseus grant G0F9516N from the FWO.  AMC is supported in part by an Odysseus grant G.001.12 from the FWO and by the European Research Council grant no. ERC-2013-CoG 616732 HoloQosmos. The work of VSM is supported by a doctoral fellowship from the FWO. We are also supported by the KU Leuven C1 grant ZKD1118 C16/16/005.

\appendix

\section{Conventions}
\label{app:conventions}

Throughout this paper, we consider four-dimensional Lorentzian and Euclidean spacetimes, with metric signature $(-+++)$ and $(++++)$, respectively.  We denote spacetime indices by Greek letters $\mu,\nu,\ldots$ and flat tangent space indices by Latin letters $a,b,\ldots$.  In Lorentzian signature, the tangent space indices take values $(0,1,2,3)$, while in Euclidean signature they take values $(1,2,3,4)$.  

Our Levi-Civita tensor conventions are
\begin{equation}
	\varepsilon_{\mu\nu\rho\sigma} = \varepsilon_{abcd} e\ind{_\mu^a}e\ind{_\nu^b}e\ind{_\rho^c}e\ind{_\sigma^d}~,
\end{equation}
such that the tensor takes values $\pm e = \pm \sqrt{|g|}$.  We also define the gamma matrix product $\gamma_5$ by
\begin{equation}
	\gamma_5 = \begin{dcases} i \gamma_0 \gamma_1 \gamma_2 \gamma_3 & \text{in Lorentzian signature}~, \\
	\gamma_1 \gamma_2 \gamma_3 \gamma_4 & \text{in Euclidean signature}~,
	\end{dcases}
\end{equation}
such that $\gamma_5$ anti-commutes with all gamma matrices and $\gamma_5^2 = 1$ in both signatures. For a $U(1)$ field strength $F_{\mu\nu}$, we define the dual field strength by
\begin{equation}\label{eq:FtildedefApp}
	\tilde{F}_{\mu\nu} = \begin{dcases}
		- \frac{i}{2} \varepsilon_{\mu\nu\rho\sigma} F^{\rho\sigma} & \text{in Lorentzian signature}~, \\
		\frac{1}{2}\varepsilon_{\mu\nu\rho\sigma} F^{\rho\sigma} & \text{in Euclidean signature}~,
		\end{dcases}
\end{equation}
such that the self-dual and anti-self-dual parts of the field strength are given in both metric signatures by
\begin{equation}
	F^\pm_{\mu\nu} = \frac{1}{2}\left(F_{\mu\nu} \pm \tilde{F}_{\mu\nu}\right)~.
\end{equation}
%

\section{\texorpdfstring{$\mathcal{N}=2$}{N=2} gauged supergravity}
\label{app:n2sugra}

In this appendix we give a brief review of the relevant features of $\mathcal{N}=2$ gauged supergravity that are needed for the derivation of the Euclidean BPS equations in the STU model.  We will primarily follow the notation and conventions in~\cite{Freedman:2012zz,Lauria:2020rhc}, and we refer the reader to these excellent texts for further details.

We consider coupling an $\mathcal{N}=2$ gravity multiplet to $n_V$ vector multiplets. We use $\alpha = 1,\ldots, n_V$ to index the vector multiplets, as well as $I = 0,1,\ldots, n_V$ to index all vector fields (including the graviphoton).  The gravity multiplet is
\begin{equation}
	\{g_{\mu\nu}~,~ \psi_\mu^i~,~ A_\mu^0\}~,
\end{equation}
where $g_{\mu\nu}$ is the metric, $\psi_\mu^i$ is an $SU(2)$ doublet of gravitinos, and $A_\mu^0$ is the graviphoton.  The vector multiplets are given by
\begin{equation}
	\{ A_\mu^\alpha~,~ \lambda^{\alpha i}~,~z^{\alpha}\}~,
\end{equation}
where $A_\mu^\alpha$ is a $U(1)$ vector field, $\lambda^{\alpha i}$ is an $SU(2)$ doublet of gauginos, and $z^\alpha$ is a complex scalar.  Note also that we consider Abelian vector multiplets, which means that the structure constants are $f\ind{_{IJ}^K} = 0$.  The field strengths are then simply given by $F_{\mu\nu}^I = \partial_\mu A_\nu^I - \partial_\nu A_\mu^I$.

The interactions between the vector multiplets are specified by a prepotential $F(L)$, a holomorphic function of some complex scalars $L^I$ with Weyl weight $2$.  These scalars arise in the off-shell superconformal construction of $\mathcal{N}=2$ supergravity.  We denote $M_I \equiv \partial_I F$ to be a derivative of the prepotential with respect to the holomorphic scalars.  $L^I$ and $M_I$ can be presented together as a covariantly holomorphic section 
\begin{equation}
	V = \begin{pmatrix} L^I \\ M_I \end{pmatrix}~, \quad \nabla_{\bar{\alpha}} V \equiv \partial_{\bar{\alpha}} V - \frac{1}{2}(\partial_{\bar{\alpha}} \mathcal{K})V = 0~.
\end{equation}
In gauge-fixing the superconformal symmetries to obtain a Poincar\'e supergravity theory, these scalars become subject to the constraint
\begin{equation}
	\langle V, \bar{V} \rangle = L^I \bar{M}_I - M_I \bar{L}^I = i~.
\end{equation}
This constraint can be solved by defining:
\begin{equation}
	\begin{pmatrix} L^I \\ M_I \end{pmatrix} = e^{\mathcal{K}/2} \begin{pmatrix} X^I \\ F_I \end{pmatrix}~,
\end{equation}
where $X^I$ are complex scalars with Weyl weight $1$ that are functions of the physical scalars $z^\alpha$, and $F_I = \frac{\partial F(X)}{\partial X^I}$.  The K\"ahler potential is then
\begin{equation}
	\mathcal{K}  = -\log\left[ i \left(\bar{X}^I F_I - \bar{F}_I X^I \right) \right]~.
\label{eq:kahler}
\end{equation}
The scalars $X^I$ are not uniquely related to the physical scalars $z^\alpha$. The physical scalars parameterize a special K\"ahler manifold of complex dimension $n_V$ with K\"ahler potential given by (\ref{eq:kahler}), and we have freedom in how we choose the coordinates on this manifold.  The corresponding K\"ahler metric is given by
\begin{equation}
	g_{\alpha \bar{\beta}} = \partial_\alpha \partial_{\bar{\beta}} \mathcal{K}~.
\end{equation}
We also define the scalar kinetic mixing matrix $\mathcal{N}_{IJ}$ by
\begin{equation}
	\mathcal{N}_{IJ} = \bar{F}_{IJ} + i \frac{N_{IK} X^K N_{JL} X^L}{N_{NM} X^N X^M} = \mathcal{R}_{IJ} + i \mathcal{I}_{IJ}~,
\end{equation}
where $N_{IJ} \equiv 2 \,\text{Im}\,F_{IJ}$, and $\mathcal{R},\mathcal{I}$ are the real and imaginary parts of $\mathcal{N}$, respectively.

The Lagrangian for the bosonic fields is
\begin{equation}\begin{aligned}
	e^{-1} \mathcal{L} &= \frac{1}{8\pi G_N}\left( \frac{1}{2} R +\frac{1}{2}\,\text{Im}\,\left[ \mathcal{N}_{IJ} F^{+I}_{\mu\nu} F^{+J\mu\nu}\right] - g_{\alpha \bar{\beta}} \nabla^\mu z^\alpha \nabla_\mu \bar{z}^{\bar{\beta}} - g^2 \mathcal{P}\right)~,
\end{aligned}\end{equation}
where the covariant derivative acting on the scalars is
\begin{equation}
	\nabla_\mu z^\alpha = \partial_\mu z^\alpha + g A_\mu^I k^\alpha_I~,
\end{equation}
where $k_I^\alpha$ are the Killing vectors of the special K\"ahler manifold.  Note that the vector kinetic term can be rewritten in several different ways:
\begin{equation}\begin{aligned}
	e^{-1}\mathcal{L}_\text{vector} &\equiv \frac{1}{2}\,\text{Im}\,\left[ \mathcal{N}_{IJ} F^{+I}_{\mu\nu} F^{+J\mu\nu}\right] \\
	 &= \left(\bar{\mathcal{N}}_{IJ} F^{-I}_{\mu\nu} F^{-J\mu\nu} - \mathcal{N}_{IJ} F^{+I}_{\mu\nu} F^{+J \mu \nu}\right) \\	
	&= \frac{1}{4} \mathcal{I}_{IJ}F^I_{\mu\nu} F^{J\mu\nu} + \frac{1}{4i} \mathcal{R}_{IJ} F^I_{\mu\nu} \tilde{F}^{J\mu\nu}~.
\end{aligned}\end{equation}
The scalar potential is
\begin{equation}
	\mathcal{P} = g_{\alpha \bar{\beta}} k^\alpha_I k^{\bar{\beta}}_J \bar{L}^I L^J + \left(g^{\alpha \bar{\beta}} f^I_\alpha f^J_{\bar{\beta}} - 3 \bar{L}^I L^J\right) \vec{P}_I \cdot \vec{P}_J~,
\end{equation}
where $\vec{P}_I$ are the moment maps and $f_\alpha^I$ are the K\"ahler-covariant derivatives of the symplectic sections:
\begin{equation}
	f^I_\alpha \equiv \nabla_\alpha L^I = \left(\partial_\alpha + \frac{1}{2}\partial_\alpha \mathcal{K}\right) L^I = e^{\mathcal{K}/2}\left(\partial_\alpha + \partial_\alpha \mathcal{K}\right) X^I = e^{\mathcal{K}/2}\nabla_\alpha X^I~.
\end{equation}
A useful rewriting of the potential is
\begin{equation}\begin{aligned}
	\mathcal{P} &= g_{\alpha \bar{\beta}} k^\alpha_I k^{\bar{\beta}}_J \bar{L}^I L^J + \left( U^{IJ} - 3 \bar{L}^I L^J\right) \vec{P}_I\cdot \vec{P}_J~, \\
	U^{IJ} &= g^{\alpha \bar{\beta}} f_{\alpha}^I f_{\bar{\beta}}^J = g^{\alpha \bar{\beta}} \nabla_\alpha L^I \nabla_{\bar{\beta}} \bar{L}^J = - \frac{1}{2}\left(\mathcal{I}^{-1}\right)^{IJ} - \bar{L}^I L^J~.
\end{aligned}\end{equation}

We will now discuss the equations of motion for the bosonic fields in the Lagrangian.  The (trace-reversed) Einstein equation is
\begin{equation}
	R_{\mu\nu} = - \mathcal{I}_{IJ}\left(F^I_{\mu\rho}F\ind{^J_\nu^\rho} - \frac{1}{4} g_{\mu\nu} F^I_{\rho\sigma} F^{J\rho\sigma}\right) + 2 g_{\alpha \bar{\beta}} \nabla_\mu z^\alpha \nabla_\nu \bar{z}^{\bar{\beta}} + g_{\mu\nu} g^2 \mathcal{P}~.
\end{equation}
We are interested in a simple gauging determined by truncating the $\mathcal{N}=8$ gauged supergravity to the STU model. This implies that $n_V=3$ and there is no gauging of the special K\"ahler isometries, in which case we find $k_I^\alpha = 0$.  The Maxwell equation is then
\begin{equation}
	\nabla_\mu\left(\mathcal{I}_{IJ} F^{I\mu\nu} - i \mathcal{R}_{IJ} \tilde{F}^{I\mu\nu}\right) = 0~,
\end{equation}
while the Bianchi identity is simply $\nabla_\mu \tilde{F}^{I\mu\nu} = 0$.  Finally, the scalar equation of motion is
\begin{equation}
	\nabla^\mu\left(g_{\alpha \bar{\beta}} \nabla_\mu \bar{z}^{\bar{\beta}}\right) = -\frac{1}{4}\partial_\alpha \mathcal{I}_{IJ} F^I_{\mu\nu}F^{J\mu\nu} + \frac{i}{4} \partial_\alpha \mathcal{R}_{IJ} F^I_{\mu\nu}\tilde{F}^{J\mu\nu} + \partial_\alpha g_{\delta \bar{\beta}} \nabla^\mu z^\delta \nabla_\mu \bar{z}^{\bar{\beta}} + g^2 \partial_\alpha \mathcal{P}~.
\end{equation}

In the case where $k_I^\alpha = 0$, the supersymmetry variations of the fermionic fields are as follows:
\begin{equation}\begin{aligned}
	\delta \psi_\mu^i &= \mathcal{D}_\mu \epsilon^i + g P_I^{ij} L^I \gamma_\mu \epsilon_j + \frac{1}{4} \mathcal{I}_{IJ} L^I F^{-J}_{ab} \gamma^{ab}\gamma_\mu \varepsilon^{ij} \epsilon_j~, \\
	\delta \lambda^\alpha_i &= -\frac{1}{2} g^{\alpha \bar{\beta}} f_{\bar{\beta}}^I \mathcal{I}_{IJ} F^{-J}_{ab} \gamma^{ab} \varepsilon_{ij}\epsilon^j + \gamma^\mu \nabla_\mu z^\alpha \epsilon_i - 2 g  P_{Iij} g^{\alpha \bar{\beta}} f_{\bar{\beta}}^I \epsilon^j~,
\end{aligned}\end{equation}
where the supercovariant derivative acting on the spinors is
\begin{equation}
	\mathcal{D}_\mu \epsilon^i = \left(\partial_\mu + \frac{1}{4} \omega_\mu^{ab} \gamma_{ab}+ \frac{1}{4}(\partial_\alpha \mathcal{K} \partial_\mu z^\alpha - \partial_{\bar{\alpha}} \mathcal{K} \partial_\mu \bar{z}^{\bar{\alpha}})\right) \epsilon^i + g A_\mu^I P\ind{_I_j^i} \epsilon^j~.
\end{equation}
Note that the moment maps $\vec{P}_I$ here are twice the ones used in \cite{Cacciatori:2008ek}.  Additionally, we define
\begin{equation}
	P\ind{_I_i^j} = \frac{i}{2} \vec{P}_I \cdot \vec{\sigma}\ind{_i^j}~.
\end{equation}
where the Pauli matrices $\left(\sigma_{1,2,3}\right)\ind{_i^j}$ are given by
\begin{equation}
	\sigma_1 = \begin{pmatrix} 0 & 1 \\ 1 & 0 \end{pmatrix}~, \quad \sigma_2 = \begin{pmatrix} 0 & -i \\ i & 0 \end{pmatrix}~, \quad \sigma_3 = \begin{pmatrix} 1 & 0 \\ 0 & -1 \end{pmatrix}~.
\end{equation}
Indices are raised on the left and lowered on the right, such that 
\begin{equation}
	\vec{\sigma}_{ij} = \vec{\sigma}\ind{_i^k} \varepsilon_{kj}~, \quad \vec{\sigma}^{ij} = \varepsilon^{ik}\vec{\sigma}\ind{_k^j}~.
\end{equation}
%

\section{Euclidean BPS conditions}
\label{app:BPS}

In this appendix, we restrict ourselves to the STU model and derive the BPS conditions for the ansatz \eqref{eq:ansatz} in Euclidean signature.

\subsection{STU model}
We work with $n_V = 3$ vector multiplets.  The complex scalars $z^\alpha$ in these vector multiplets parameterize the K\"ahler manifold $\mathcal{M}_V$, which for the STU model is three copies of the Poincar\'e disk:
\begin{equation}
	\mathcal{M}_V = \left[ \frac{SU(1,1)}{U(1)}\right]^3~.
\end{equation}
Accordingly, the scalars must take values on the Poincar\'e disk and thus they satisfy the condition $|z^\alpha| < 1$.  Additionally, we require that the projective scalars are related to the physical scalars $z^\alpha$ by
\begin{equation}\begin{aligned}
	X^0 &= \frac{1}{2\sqrt{2}}(1- z^1)(1- z^2)(1-z^3)~, &\quad& X^1 = \frac{1}{2\sqrt{2}}(1- z^1)(1+z^2)(1+z^3)~, \\
	X^2 &= \frac{1}{2\sqrt{2}}(1 + z^1)(1- z^2)(1+z^3)~, &\quad& X^3 = \frac{1}{2\sqrt{2}}(1+ z^1)(1+z^2)(1-z^3)~.
\end{aligned}\end{equation}
The corresponding prepotential for the scalar fields is given by
\begin{equation}
	F = - 2 i \sqrt{X^0 X^1 X^2 X^3} = -\frac{i}{4}\left(1-(z^1)^2\right)\left(1-(z^2)^2\right)\left(1-(z^3)^2\right)~.
\end{equation}
The K\"ahler potential is
\begin{equation}
	\mathcal{K} = -\log\left[\left(1-|z^1|^2\right)\left(1-|z^2|^2\right)\left(1-|z^3|^2\right)\right]~.
\end{equation}
The K\"ahler metric is hence diagonal and simply given by
\begin{equation}
	g_{\alpha \bar{\beta}} = \text{diag}\left( \frac{1}{\left(1-|z^1|^2\right)^2}~,~\frac{1}{\left(1-|z^2|^2\right)^2}~,~\frac{1}{\left(1-|z^3|^2\right)^2}\right)~.
\end{equation}
We also define the superpotential
\begin{equation}
	\mathcal{V} = 2\left(z^1 z^2 z^3 - 1\right)~.
\end{equation}
The scalar potential can be expressed in terms of this as
\begin{equation}
	\mathcal{P} = \frac{1}{2}e^{\mathcal{K}}\left( g^{\alpha \bar{\beta}} \nabla_\alpha \mathcal{V} \nabla_{\bar{\beta}} \bar{\mathcal{V}} - 3 \mathcal{V} \bar{\mathcal{V}}\right) = 2\left(3 - \sum_{\alpha=1}^3 \frac{2}{1-|z^\alpha|^2}\right)~.
\end{equation}
Finally, the moment maps are
\begin{equation}
	P_I^1 = P_I^2 = 0~, \quad P_I^3 = -1~,
\end{equation}
for any value of $I$.

\subsection{Lorentzian ansatz and supersymmetry variations}

Our metric ansatz is
\begin{equation}\begin{aligned}
	ds^2 &= - e^{2 f_1} dt^2 + e^{2 f_2} dr^2 + e^{2 f_3} ds_{\Sigma_{\mathfrak{g}}}^2~,
\end{aligned}\end{equation}
where $x$ and $y$ are the coordinates on the Riemann surface and $f_1$, $f_2$, and $f_3$ are functions only of $r$.  For a Riemann surface $\Sigma_{\mathfrak{g}}$ of genus $\mathfrak{g}$, we can write the constant curvature metric as
\begin{equation}\label{eq:metappH}
	ds^2_{\Sigma_{\mathfrak{g}}} = H^2\left(dx^2 + dy^2\right)~, \quad H = \begin{dcases} \frac{2}{1+x^2 + y^2} & \mathfrak{g}=0 \\ \sqrt{2\pi} & \mathfrak{g}=1 \\ \frac{1}{y} & \mathfrak{g} > 1 \end{dcases}~.
\end{equation}
We also define the one-form potential
\begin{equation}
	\omega_{\sigg} = \begin{dcases} \frac{2 (x dy - y dx)}{1+x^2+y^2} & \mathfrak{g}=0 \\ \pi(x dy - y dx) & \mathfrak{g} = 1 \\ \frac{dx}{y} & \mathfrak{g} > 1 \end{dcases}~,
\end{equation}
such that $V_{\sigg} \equiv d \omega_{\sigg}= H^2 dx \wedge dy$ is the volume form on the Riemann surface.   We also assume that the gauge fields wrap the time direction and the Riemann surface, such that only $F_{tr}^I$ and $F_{xy}^I$ are non-zero.  Additionally, we take the scalars $z^\alpha$ and the supersymmetry parameters $\epsilon^i$ to have only radial dependence.

The supersymmetry variations in the STU model (wherein we take an Abelian gauging of the isometries) are
\begin{equation}\begin{aligned}
	\delta \psi_\mu^i &= \mathcal{D}_\mu \epsilon^i + \frac{i}{2} g P_I^3 L^I \gamma_\mu \varepsilon^{ik}(\sigma_3)\ind{_k^j} \epsilon_j + \frac{1}{4} \mathcal{I}_{IJ} L^I F^{-J}_{ab}\gamma^{ab} \gamma_\mu \varepsilon^{ij}\epsilon_j~, \\
	\mathcal{D}_\mu \epsilon^i &= \left(\partial_\mu + \frac{1}{4}\omega_\mu^{ab} \gamma_{ab} + \frac{i}{2}\mathcal{A}_\mu \right)\epsilon^i + \frac{i}{2} g P^3_I A_\mu^I (\sigma_3)\ind{_j^i}\epsilon^j~, \\ 
	\delta \lambda_i^\alpha &= - \frac{1}{2}g^{\alpha \bar{\beta}} f_{\bar{\beta}}^I \mathcal{I}_{IJ} F^{-J}_{\mu\nu} \gamma^{\mu\nu} \varepsilon_{ij}\epsilon^j + \gamma^\mu \nabla_\mu z^\alpha \epsilon_i - i g g^{\alpha \bar{\beta}} f_{\bar{\beta}}^I P_I^3 (\sigma_3)\ind{_i^k} \varepsilon_{kj}\epsilon^j~,
\label{eq:var1}
\end{aligned}\end{equation}
where we define the connection $\mathcal{A}_\mu$ by
\begin{equation}
	\mathcal{A}_\mu = - \frac{i}{2}\left(\partial_\alpha \mathcal{K} \partial_\mu z^\alpha - \partial_{\bar{\alpha}} \mathcal{K} \partial_\mu \bar{z}^{\bar{\alpha}}\right)~.
\end{equation}
The corresponding charge conjugate variations are
\begin{equation}\begin{aligned}
	\delta \psi_{i\mu} &= \mathcal{D}_\mu \epsilon_i - \frac{i}{2} g P_I^3 \bar{L}^I \gamma_\mu \varepsilon_{ik}(\sigma_3)\ind{_j^k}\epsilon^j  + \frac{1}{4} \mathcal{I}_{IJ} \bar{L}^I F^{+J}_{ab}\gamma^{ab} \gamma_\mu \varepsilon_{ij}\epsilon^j~, \\
	\mathcal{D}_\mu \epsilon_i &= \left(\partial_\mu + \frac{1}{4}\omega_\mu^{ab} \gamma_{ab} - \frac{i}{2}\mathcal{A}_\mu \right)\epsilon_i - \frac{i}{2} g P^3_I A_\mu^I(\sigma_3)\ind{_i^j}\epsilon_j~, \\ 
	\delta \lambda^{i \bar{\alpha}} &= - \frac{1}{2}g^{\bar{\alpha} {\beta}} f_{{\beta}}^I \mathcal{I}_{IJ} F^{+J}_{\mu\nu} \gamma^{\mu\nu} \varepsilon^{ij}\epsilon_j + \gamma^\mu \nabla_\mu \bar{z}^{\bar{\alpha}} \epsilon^i + i g g^{\bar{\alpha} {\beta}} f_{{\beta}}^I P_I^3 (\sigma_3)\ind{_k^i} \varepsilon^{kj}\epsilon_j~.
\end{aligned}\end{equation}
These conjugate variations will become independent from the variations in (\ref{eq:var1}) when we Euclideanize our theory and so we find it useful to spell them out explicitly.

We now expand the gravitino and gaugino variations explicitly on our ansatz.  We find it useful to expand all (anti-)self-dual field strengths $F^{\pm I}_{\mu\nu}$ in terms of just the ordinary field strength $F^I_{\mu\nu}$.  Useful identities that go a long way towards accomplishing this are
\begin{equation}
	F^\pm_{ab}\gamma^{ab} = 2 F^\pm_{0 a} \gamma^{0a}(1 \mp \gamma_5)~, \qquad F^\pm_{ab} \gamma^{ab} \gamma_\mu = -2 F^\pm_{\mu\nu}\gamma^{\nu} (1 \pm \gamma_5)~.
\end{equation}
At the end of the day, the variations are given by:
\begin{equation}\begin{aligned}
	\delta \psi_t^i &= \frac{1}{2}f_1' e^{f_1 - f_2} \gamma_{01} \epsilon^i + \frac{i}{2}g P_I^3 A_t^I (\sigma_3)\ind{_j^i}\epsilon^j +\frac{i}{2}g e^{f_1} P_I^3 L^I \gamma_0 (\sigma_3)\ind{^i^j} \epsilon_j \\
	&\quad - \frac{i}{2} e^{f_1} \mathcal{I}_{IJ} L^I \left(F_{23}^J - i F_{01}^J\right)\gamma_1 \varepsilon^{ij}\epsilon_j~, \\
	\delta \psi_r^i &= \left(\partial_r + \frac{i}{2} \mathcal{A}_r\right) \epsilon^i + \frac{i}{2} g e^{f_2} P_I^3 L^I \gamma_1 (\sigma_3)\ind{^i^j} \epsilon_j \\
	&\quad - \frac{i}{2} e^{f_2} \mathcal{I}_{IJ} L^I \left(F_{23}^J - i F_{01}^J\right)\gamma_0 \varepsilon^{ij}\epsilon_j~, \\
	\delta \psi_x^i &= \frac{1}{2}\left( \frac{\partial_y H}{H} \gamma_{23} - H f_3' e^{f_3 - f_2} \gamma_{12}\right) \epsilon^i + \frac{i}{2} g P_I^3 A^I_x (\sigma_3)\ind{_j^i}\epsilon^j \\
	& \quad + \frac{i}{2} g H e^{f_3} P_I^3 L^I \gamma_2 (\sigma_3)\ind{^i^j} \epsilon_j - \frac{1}{2} H e^{f_3} \mathcal{I}_{IJ} L^I \left(F_{23}^J - i F_{01}^J\right)\gamma_3 \varepsilon^{ij}\epsilon_j~, \\
	\delta \psi_y^i &= \frac{1}{2}\left(- \frac{\partial_x H}{H} \gamma_{23} - H f_3' e^{f_3 - f_2} \gamma_{13} \right) \epsilon^i + \frac{i}{2} g P_I^3 A^I_y (\sigma_3)\ind{_j^i}\epsilon^j \\
	& \quad + \frac{i}{2} g H e^{f_3} P_I^3 L^I \gamma_3 (\sigma_3)\ind{^i^j} \epsilon_j + \frac{1}{2} H e^{f_3} \mathcal{I}_{IJ} L^I \left(F_{23}^J - i F_{01}^J\right)\gamma_2 \varepsilon^{ij}\epsilon_j~, \\
	\delta \lambda_i^\alpha &= i g^{\alpha \bar{\beta}}f_{\bar{\beta}}^I \bigg{(} \mathcal{I}_{IJ} \left(F_{23}^J - i F_{01}^J\right) \gamma_{01} \varepsilon_{ij} - g P_I^3 (\sigma_3)\ind{_i_j}\bigg{)}\epsilon^j + e^{-f_2} (\partial_r z^\alpha)  \gamma_1 \epsilon_i~.
\end{aligned}\end{equation}
The conjugate variations are similar:
\begin{equation}\begin{aligned}
	\delta \psi_{i t} &= \frac{1}{2}f_1' e^{f_1 - f_2} \gamma_{01} \epsilon_i - \frac{i}{2}g P_I^3 A_t^I (\sigma_3)\ind{_i^j}\epsilon_j +\frac{i}{2}g e^{f_1} P_I^3 \bar{L}^I \gamma_0 (\sigma_3)\ind{_i_j} \epsilon^j \\
	&\quad + \frac{i}{2} e^{f_1} \mathcal{I}_{IJ} \bar{L}^I \left(F_{23}^J + i F_{01}^J\right)\gamma_1 \varepsilon_{ij}\epsilon^j~, \\
	\delta \psi_{ir} &= \left(\partial_r - \frac{i}{2} \mathcal{A}_r\right) \epsilon_i + \frac{i}{2} g e^{f_2} P_I^3 \bar{L}^I \gamma_1 (\sigma_3)\ind{_i_j} \epsilon^j \\
	&\quad + \frac{i}{2} e^{f_2} \mathcal{I}_{IJ} \bar{L}^I \left(F_{23}^J + i F_{01}^J\right)\gamma_0 \varepsilon_{ij}\epsilon^j~, \\
	\delta \psi_{ix} &= \frac{1}{2}\left( \frac{\partial_y H}{H} \gamma_{23} - H f_3' e^{f_3 - f_2} \gamma_{12}\right) \epsilon_i - \frac{i}{2} g P_I^3 A^I_x (\sigma_3)\ind{_i^j}\epsilon_j \\
	& \quad + \frac{i}{2} g H e^{f_3} P_I^3 \bar{L}^I \gamma_2 (\sigma_3)\ind{_i_j} \epsilon^j - \frac{1}{2} H e^{f_3} \mathcal{I}_{IJ} \bar{L}^I \left(F_{23}^J + i F_{01}^J\right)\gamma_3 \varepsilon_{ij}\epsilon^j~, \\
	\delta \psi_{iy} &= \frac{1}{2}\left(- \frac{\partial_x H}{H} \gamma_{23} - H f_3' e^{f_3 - f_2} \gamma_{13} \right) \epsilon_i - \frac{i}{2} g P_I^3 A^I_y (\sigma_3)\ind{_i^j}\epsilon_j \\
	& \quad + \frac{i}{2} g H e^{f_3} P_I^3 \bar{L}^I \gamma_3 (\sigma_3)\ind{_i_j} \epsilon^j + \frac{1}{2} H e^{f_3} \mathcal{I}_{IJ} \bar{L}^I \left(F_{23}^J + i F_{01}^J\right)\gamma_2 \varepsilon_{ij}\epsilon^j~, \\
	\delta \lambda^{i\bar{\alpha}} &= -i g^{\bar{\alpha} {\beta}}f_{{\beta}}^I \bigg{(} \mathcal{I}_{IJ} \left(F_{23}^J + i F_{01}^J\right) \gamma_{01} \varepsilon^{ij} + g P_I^3 (\sigma_3)\ind{^i^j}\bigg{)}\epsilon_j + e^{-f_2} (\partial_r \bar{z}^{\bar{\alpha}})  \gamma_1 \epsilon^i~.
\end{aligned}\end{equation}
%

\subsection{Euclideanization}

We now Euclideanize the supersymmetry variations by Wick rotating the time direction.  That is, we take $t \to -i \tau$ and $x^0 \to -i x^4$.  The corresponding Euclidean metric ansatz is
\begin{equation}\label{eq:metansappEuc}
	ds^2 = e^{2 f_1} d\tau^2 + e^{2 f_2} dr^2 + e^{2 f_3} ds_{\Sigma_{\mathfrak{g}}}^2~.
\end{equation}
The rules for Wick rotating relevant quantities in the variations are as follows:
\begin{equation}
	\gamma_0 \rightarrow i \gamma_4 \,, \qquad A_0 \rightarrow i A_4 \,, \qquad F_{0\mu} \rightarrow i F_{4\mu}\,, \qquad \bar{z}^\alpha \rightarrow \tilde{z}^\alpha \,,
\end{equation}
where we can no longer assume $z^\alpha$ and $\tilde{z}^\alpha$ are complex conjugates.

If we apply this to the Lorentzian variations in the previous subsection, we find the following Euclidean variations:
\begin{equation}\begin{aligned}
	i\delta \psi_{\tau}^i &= \frac{i}{2}f_1' e^{f_1 - f_2} \gamma_{41} \epsilon^i - \frac{1}{2}g P_I^3 A_{\tau}^I (\sigma_3)\ind{_j^i}\epsilon^j -\frac{1}{2}g e^{f_1} P_I^3 L^I \gamma_4 (\sigma_3)\ind{^i^j} \epsilon_j \\
	&\quad - \frac{i}{2} e^{f_1} \mathcal{I}_{IJ} L^I \left(F_{23}^J + F_{41}^J\right)\gamma_1 \varepsilon^{ij}\epsilon_j~, \\
	\delta \psi_r^i &= \left(\partial_r + \frac{i}{2} \mathcal{A}_r\right) \epsilon^i + \frac{i}{2} g e^{f_2} P_I^3 L^I \gamma_1 (\sigma_3)\ind{^i^j} \epsilon_j \\
	&\quad + \frac{1}{2} e^{f_2} \mathcal{I}_{IJ} L^I \left(F_{23}^J + F_{41}^J\right)\gamma_4 \varepsilon^{ij}\epsilon_j~, \\
	\delta \psi_x^i &= \frac{1}{2}\left( \frac{\partial_y H}{H} \gamma_{23} - H f_3' e^{f_3 - f_2} \gamma_{12}\right) \epsilon^i + \frac{i}{2} g P_I^3 A^I_x (\sigma_3)\ind{_j^i}\epsilon^j \\
	& \quad + \frac{i}{2} g H e^{f_3} P_I^3 L^I \gamma_2 (\sigma_3)\ind{^i^j} \epsilon_j - \frac{1}{2} H e^{f_3} \mathcal{I}_{IJ} L^I \left(F_{23}^J + F_{41}^J\right)\gamma_3 \varepsilon^{ij}\epsilon_j~, \\
	\delta \psi_y^i &= \frac{1}{2}\left(- \frac{\partial_x H}{H} \gamma_{23} - H f_3' e^{f_3 - f_2} \gamma_{13} \right) \epsilon^i + \frac{i}{2} g P_I^3 A^I_y (\sigma_3)\ind{_j^i}\epsilon^j \\
	& \quad + \frac{i}{2} g H e^{f_3} P_I^3 L^I \gamma_3 (\sigma_3)\ind{^i^j} \epsilon_j + \frac{1}{2} H e^{f_3} \mathcal{I}_{IJ} L^I \left(F_{23}^J + F_{41}^J\right)\gamma_2 \varepsilon^{ij}\epsilon_j~, \\
	\delta \lambda_i^\alpha &= i g^{\alpha \bar{\beta}}f_{\bar{\beta}}^I \bigg{(} i \mathcal{I}_{IJ} \left(F_{23}^J + F_{41}^J\right) \gamma_{41} \varepsilon_{ij} - g P_I^3 (\sigma_3)\ind{_i_j}\bigg{)}\epsilon^j + e^{-f_2} (\partial_r z^\alpha)  \gamma_1 \epsilon_i~.
\end{aligned}\end{equation}
The Wick-rotated conjugate variations are:
\begin{equation}\begin{aligned}
	i \delta \psi_{i \tau} &= \frac{i}{2}f_1' e^{f_1 - f_2} \gamma_{41} \epsilon_i + \frac{1}{2}g P_I^3 A_{\tau}^I (\sigma_3)\ind{_i^j}\epsilon_j - \frac{1}{2}g e^{f_1} P_I^3 \bar{L}^I \gamma_4 (\sigma_3)\ind{_i_j} \epsilon^j \\
	&\quad + \frac{i}{2} e^{f_1} \mathcal{I}_{IJ} \bar{L}^I \left(F_{23}^J - F_{41}^J\right)\gamma_1 \varepsilon_{ij}\epsilon^j~, \\
	\delta \psi_{ir} &= \left(\partial_r - \frac{i}{2} \mathcal{A}_r\right) \epsilon_i + \frac{i}{2} g e^{f_2} P_I^3 \bar{L}^I \gamma_1 (\sigma_3)\ind{_i_j} \epsilon^j \\
	&\quad - \frac{1}{2} e^{f_2} \mathcal{I}_{IJ} \bar{L}^I \left(F_{23}^J - F_{41}^J\right)\gamma_4 \varepsilon_{ij}\epsilon^j~, \\
	\delta \psi_{ix} &= \frac{1}{2}\left( \frac{\partial_y H}{H} \gamma_{23} - H f_3' e^{f_3 - f_2} \gamma_{12}\right) \epsilon_i - \frac{i}{2} g P_I^3 A^I_x (\sigma_3)\ind{_i^j}\epsilon_j \\
	& \quad + \frac{i}{2} g H e^{f_3} P_I^3 \bar{L}^I \gamma_2 (\sigma_3)\ind{_i_j} \epsilon^j - \frac{1}{2} H e^{f_3} \mathcal{I}_{IJ} \bar{L}^I \left(F_{23}^J - F_{41}^J\right)\gamma_3 \varepsilon_{ij}\epsilon^j~, \\
	\delta \psi_{iy} &= \frac{1}{2}\left(- \frac{\partial_x H}{H} \gamma_{23} - H f_3' e^{f_3 - f_2} \gamma_{13} \right) \epsilon_i - \frac{i}{2} g P_I^3 A^I_y (\sigma_3)\ind{_i^j}\epsilon_j \\
	& \quad + \frac{i}{2} g H e^{f_3} P_I^3 \bar{L}^I \gamma_3 (\sigma_3)\ind{_i_j} \epsilon^j + \frac{1}{2} H e^{f_3} \mathcal{I}_{IJ} \bar{L}^I \left(F_{23}^J - F_{41}^J\right)\gamma_2 \varepsilon_{ij}\epsilon^j~, \\
	\delta \lambda^{i\bar{\alpha}} &= -i g^{\bar{\alpha} {\beta}}f_{{\beta}}^I \bigg{(} i \mathcal{I}_{IJ} \left(F_{23}^J - F_{41}^J\right) \gamma_{41} \varepsilon^{ij} + g P_I^3 (\sigma_3)\ind{^i^j}\bigg{)}\epsilon_j + e^{-f_2} (\partial_r \tilde{z}^{\bar{\alpha}})  \gamma_1 \epsilon^i~.
\label{eq:conj}
\end{aligned}\end{equation}
Note that since we study supergravity solutions with no angular momentum, the spinors $\epsilon^i$ and $\epsilon_i$ in the Euclidean variations above are anti-periodic.  In a fully Euclidean theory of gravity, the metric $g_{\mu\nu}$ and the gauge fields $A_\mu^I$ are in general complex, which means that the conjugate variations (\ref{eq:conj}) should also be modified such that all instances of the metric functions $f_i$ and the gauge fields $A_\mu^I$ are replaced by their complex conjugates.  We will suppress this complex conjugation for notational simplicity, though it is straightforward to introduce this into all conjugate equations where appropriate.

\subsection{Euclidean projectors and BPS equations}

The variations so far are valid for Euclidean backgrounds with the metric ansatz \eqref{eq:metansappEuc}.  We now need to specify the gauge fields. Motivated by the topologically twisted ABJM theory we use the ansatz
\begin{equation}
	A^I =  p^I \omega_{\sigg} + e^I d\tau~,
\end{equation}
where the Wilson lines $e^I(r)$ along the $\tau$-direction are functions of $r$ only, and the $p^I$ are the magnetic charges.  The corresponding field strength is
\begin{equation}
	F^I =  p^I V_{\sigg} - (\partial_r e^I) d\tau \wedge dr~,
\end{equation}
which in turn means that $F^I_{41} = - e^{- f_1 - f_2} \partial_r e^I$ and $F_{23}^I = e^{-2 f_3} p^I$.  

We now plug these expressions into the variations above and set them to zero in order to find BPS solutions.  To make the notation a little simpler in what follows, we  define
\begin{equation}\label{eq:npmdef}
	n^{\pm I} = p^I \pm e^{-f_1 - f_2 + 2 f_3}\partial_r e^I~.
\end{equation}
We also make use of the fact that
\begin{equation}
	\partial_y H = \kappa H \omega_x~, \quad \partial_x H = - \kappa H \omega_y~,
\end{equation}
where $ \kappa = 1$, $\kappa = 0$, and $\kappa = -1$ is the normalized curvature of a Riemann surface with genus $\mathfrak{g} = 0$, $\mathfrak{g}= 1$, and $\mathfrak{g} > 1$, respectively, and $\omega_{x}$ and $\omega_{y}$ are the components of the local one-form potential $\omega_{\sigg}$ on the Riemann surface. The result for the first set of variations is then
\begin{equation}\begin{aligned}
	0 &= f_1' e^{f_1-f_2} \gamma_{41} \epsilon^i + i g P_I^3 e^I (\sigma_3)\ind{_j^i}\epsilon^j + i g e^{f_1} P_I^3 L^I \gamma_4 (\sigma_3)^{ij}\epsilon_j \\
	&\quad - e^{f_1 - 2 f_3} \mathcal{I}_{IJ} L^I n^{-J}\gamma_1 \varepsilon^{ij}\epsilon_j ~, \\
	0 &= \left(2 \partial_r + i \mathcal{A}_r \right)\epsilon^i + i g e^{f_2} P_I^3 L^I \gamma_1 (\sigma_3)^{ij}\epsilon_j + e^{f_2 - 2 f_3} \mathcal{I}_{IJ} L^I n^{-J}\gamma_4 \varepsilon^{ij}\epsilon_j~, \\
	0 &= \left(\kappa \omega_x \gamma_{23} - H f_3' e^{f_3 - f_2} \gamma_{12}\right) \epsilon^i + i g P_I^3 p^I \omega_x (\sigma_3)\ind{_j^i} \epsilon^j \\
	&\quad + i g H e^{f_3} P_I^3 L^I \gamma_2 (\sigma_3)^{ij}\epsilon_j - H e^{-f_3} \mathcal{I}_{IJ} L^I n^{-J} \gamma_3 \varepsilon^{ij}\epsilon_j~, \\
	0 &= \left(\kappa \omega_y \gamma_{23} - H f_3' e^{f_3 - f_2} \gamma_{13}\right) \epsilon^i + i g P_I^3 p^I \omega_y (\sigma_3)\ind{_j^i}\epsilon^j \\
	&\quad + i g H e^{f_3} P_I^3 L^I \gamma_3 (\sigma_3)^{ij} \epsilon_j + H e^{-f_3}\mathcal{I}_{IJ} L^I n^{-J} \gamma_2 \varepsilon^{ij}\epsilon_j~, \\
	0 &= i g^{\alpha \bar{\beta}} f_{\bar{\beta}}^I \left( ie^{-2f_3} \mathcal{I}_{IJ} n^{-J} \gamma_{41} \varepsilon_{ij} - g P_I^3 (\sigma_3)_{ij} \right)\epsilon^j + e^{-f_2}(\partial_r z^\alpha) \gamma_1 \epsilon_i~,
\end{aligned}\end{equation}
while for their conjugates we find
\begin{equation}\begin{aligned}
	0 &= f_1' e^{f_1 - f_2} \gamma_{41}\epsilon_i - i g P_I^3 e^I (\sigma_3)\ind{_i^j}\epsilon_j + i g e^{f_1} P_I^3 \bar{L}^I \gamma_4 (\sigma_3)_{ij}\epsilon^j \\
	&\quad + e^{f_1 - 2 f_3}\mathcal{I}_{IJ} \bar{L}^I n^{+J} \gamma_1 \varepsilon_{ij}\epsilon^j~, \\
	0 &= \left(2 \partial_r - i \mathcal{A}_r\right) \epsilon_i + i g e^{f_2} P_I^3 \bar{L}^I \gamma_1 (\sigma_3)_{ij}\epsilon^j - e^{f_2 - 2 f_3} \mathcal{I}_{IJ} \bar{L}^I n^{+J} \gamma_4 \varepsilon_{ij}\epsilon^j~, \\
	0 &= \left(\kappa \omega_x \gamma_{23} - H f_3' e^{f_3 - f_2} \gamma_{12}\right)\epsilon_i - i g P_I^3 p^I \omega_x (\sigma_3)\ind{_i^j} \epsilon_j \\
	&\quad + i g H e^{f_3} P_I^3 \bar{L}^I \gamma_2 (\sigma_3)_{ij}\epsilon^j - H e^{-f_3} \mathcal{I}_{IJ}\bar{L}^I n^{+J} \gamma_3 \varepsilon_{ij}\epsilon^j~, \\
	0 &= \left(\kappa \omega_y \gamma_{23} - H f_3' e^{f_3 - f_2} \gamma_{13}\right)\epsilon_i - i g P_I^3 p^I \omega_y (\sigma_3)\ind{_i^j} \epsilon_j \\
	&\quad + i g H e^{f_3} P_I^3 \bar{L}^I \gamma_3 (\sigma_3)_{ij}\epsilon^j + H e^{-f_3} \mathcal{I}_{IJ}\bar{L}^I n^{+J} \gamma_2 \varepsilon_{ij}\epsilon^j~, \\
	0 &= - i g^{\bar{\alpha}\beta}f_{\beta}^I\left( i e^{-2f_3}\mathcal{I}_{IJ}n^{+J}\gamma_{41}\varepsilon^{ij} + g P_I^3 (\sigma_3)^{ij}\right)\epsilon_j + e^{-f_2}(\partial_r \tilde{z}^{\bar{\alpha}})\gamma_1 \epsilon^i~.
\end{aligned}\end{equation}
The terms in the equations proportional to $\omega_x$ (or $\omega_y$) must cancel among themselves.  This immediately implies that
\begin{equation}\begin{aligned}
	0 &= \kappa \gamma_{23} \epsilon^i + i g P_I^3 p^I (\sigma_3)\ind{_j^i}\epsilon^j~, \\
	0 &= \kappa \gamma_{23} \epsilon_i - i g P_I^3 p^I (\sigma_3)\ind{_i^j}\epsilon_j~,
\end{aligned}\end{equation}
which can only be solved by imposing projectors of the form
\begin{equation}\begin{aligned}
	\gamma_{23} \epsilon^i &= - i \xi (\sigma_3)\ind{_j^i} \epsilon^j~, \\
	\gamma_{23} \epsilon_i &= i \xi (\sigma_3)\ind{_i^j} \epsilon_j~,
\end{aligned}\end{equation}
with $\xi = \pm 1$.  It is useful to rewrite these relations as
\begin{equation}\begin{aligned}
	\gamma_{41} \epsilon^i &= - i \xi (\sigma_3)\ind{_j^i} \epsilon^j~, \\
	\gamma_{41} \epsilon_i &= - i \xi (\sigma_3)\ind{_i^j} \epsilon_j~.
\end{aligned}\end{equation}
Note that the two projectors are a priori independent, but we must choose them such that the resulting constraints are consistent:
\begin{equation}
	0 = \kappa - \xi g P_I^3 p^I~, \quad\text{or equivalently} \quad \sum_I p^I = -\frac{\kappa}{\xi g}~.
\end{equation}
This is the supergravity analogue of the topological twist condition in (\ref{eq:toptwistcond}).  The remaining BPS equations become the following:
\begin{equation}\begin{aligned}
	0 &= \left(f_1' e^{-f_2}  - \xi g e^{-f_1} P_I^3 e^I \right) \epsilon^i + i g P_I^3 L^I \gamma_1 (\sigma_3)^{ij}\epsilon_j + e^{- 2 f_3} \mathcal{I}_{IJ} L^I n^{-J}\gamma_4 \varepsilon^{ij}\epsilon_j ~, \\
	0 &= e^{-f_2}\left(2 \partial_r + i \mathcal{A}_r \right)\epsilon^i + i g  P_I^3 L^I \gamma_1 (\sigma_3)^{ij}\epsilon_j + e^{- 2 f_3} \mathcal{I}_{IJ} L^I n^{-J}\gamma_4 \varepsilon^{ij}\epsilon_j~, \\
	0 &= f_3' e^{-f_2} \gamma_1 \epsilon^i + i \left( g P_I^3 L^I - \xi e^{-2 f_3} \mathcal{I}_{IJ}L^I n^{-J}\right)(\sigma_3)^{ij} \epsilon_j~, \\
	0 &= -i g^{\alpha \bar{\beta}} f_{\bar{\beta}}^I \left( \xi e^{-2f_3} \mathcal{I}_{IJ} n^{-J} + g P_I^3 \right)(\sigma_3)_{ij} \epsilon^j + e^{-f_2}(\partial_r z^\alpha) \gamma_1 \epsilon_i~,
\end{aligned}\end{equation}
plus the conjugates
\begin{equation}\begin{aligned}
	0 &= \left(f_1' e^{- f_2} + \xi g e^{-f_1} P_I^3 e^I\right) \epsilon_i + i g  P_I^3 \bar{L}^I \gamma_1 (\sigma_3)_{ij}\epsilon^j - e^{- 2 f_3}\mathcal{I}_{IJ} \bar{L}^I n^{+J} \gamma_4 \varepsilon_{ij}\epsilon^j~, \\
	0 &= e^{-f_2}\left(2 \partial_r - i \mathcal{A}_r\right) \epsilon_i + i g  P_I^3 \bar{L}^I \gamma_1 (\sigma_3)_{ij}\epsilon^j - e^{- 2 f_3} \mathcal{I}_{IJ} \bar{L}^I n^{+J} \gamma_4 \varepsilon_{ij}\epsilon^j~, \\
	0 &= f_3' e^{-f_2} \gamma_1 \epsilon_i + i \left( g P_I^3 \bar{L}^I - \xi e^{-2f_3} \mathcal{I}_{IJ}\bar{L}^I n^{+J}\right) (\sigma_3)_{ij}\epsilon^j~, \\
	0 &= - i g^{\bar{\alpha}\beta}f_{\beta}^I\left( \xi e^{-2f_3}\mathcal{I}_{IJ}n^{+J} + g P_I^3 \right)(\sigma_3)^{ij} \epsilon_j + e^{-f_2}(\partial_r \tilde{z}^{\bar{\alpha}})\gamma_1 \epsilon^i~.
\end{aligned}\end{equation}
Comparison of the first two equations in both sets tells us that the spinors obey the differential equations
\begin{equation}\begin{aligned}
	\left(\partial_r + \frac{i}{2}\mathcal{A}_r - \frac{f_1'}{2} + \frac{1}{2}\xi g e^{f_2-f_1} P_I^3 e^I\right)\epsilon^i &= 0~, \\
	\left(\partial_r - \frac{i}{2}\mathcal{A}_r - \frac{f_1'}{2} - \frac{1}{2}\xi g e^{f_2-f_1} P_I^3 e^I\right) \epsilon_i &= 0~.
\end{aligned}\end{equation}
To solve the remaining equations, we need to impose the projectors
\begin{equation}\begin{aligned}
	\gamma_4 \epsilon^i &= i M \varepsilon^{ij}\epsilon_j~, \\
	\gamma_4 \epsilon_i &= i \widetilde{M} \varepsilon_{ij}\epsilon^j~,
\end{aligned}\end{equation}
or equivalently we could write them as
\begin{equation}\begin{aligned}
	\gamma_1 \epsilon^i &= - \xi M (\sigma_3)^{ij}\epsilon_j~, \\
	\gamma_1 \epsilon_i &= \xi \widetilde{M} (\sigma_3)_{ij}\epsilon^j~,
\end{aligned}\end{equation}
where $M$ and $\widetilde{M}$ are functions of $r$ that satisfy $M \widetilde{M} = 1$.  The BPS conditions, with these projectors, become:
\begin{equation}\begin{aligned}
	0 &= \left( M(f_1' e^{-f_2} - \xi g e^{-f_1} P_I^3 e^I) - i \xi g P_I^3 L^I - i e^{-2 f_3} \mathcal{I}_{IJ} L^I n^{-J}\right) \epsilon^i~, \\
	0 &= \left(M f_3' e^{-f_2} - i \xi g P_I^3 L^I + i e^{-2 f_3} \mathcal{I}_{IJ}L^I n^{-J}\right)(\sigma_3)^{ij} \epsilon_j~, \\
	0 &= \left( \widetilde{M} e^{-f_2} \partial_r z^\alpha - i g^{\alpha \bar{\beta}} f_{\bar{\beta}}^I(e^{-2 f_3} \mathcal{I}_{IJ} n^{-J} + \xi g P_I^3)\right) (\sigma_3)_{ij}\epsilon^j~,
\end{aligned}\end{equation}
as well as the conjugate conditions
\begin{equation}\begin{aligned}
	0 &= \left( \widetilde{M}(f_1' e^{-f_2} + \xi g e^{-f_1} P_I^3 e^I) + i \xi g P_I^3 \bar{L}^I + i e^{-2 f_3} \mathcal{I}_{IJ} \bar{L}^I n^{+J}\right) \epsilon_i~, \\
	0 &= \left(\widetilde{M} f_3' e^{-f_2} + i \xi g P_I^3 \bar{L}^I - i e^{-2f_3} \mathcal{I}_{IJ}\bar{L}^I n^{+J}\right) (\sigma_3)_{ij}\epsilon^j~, \\
	0 &= \left( M e^{-f_2} \partial_r \tilde{z}^{\bar{\alpha}} + i g^{\bar{\alpha} {\beta}} f_{{\beta}}^I(e^{-2 f_3} \mathcal{I}_{IJ} n^{+J} + \xi g P_I^3)\right) (\sigma_3)^{ij}\epsilon_j~.
\end{aligned}\end{equation}
The only way to satisfy these with non-vanishing spinors is to set the terms in parentheses equal to zero.  Writing these equations explicitly, we conclude that the Euclideanized BPS conditions are as follows:
\begin{equation}\begin{aligned}
	0 &= M(f_1' e^{-f_2} - \xi g e^{-f_1} P_I^3 e^I) - i \xi g P_I^3 L^I - i e^{-2 f_3} \mathcal{I}_{IJ} L^I n^{-J}~, \\
	0 &= \widetilde{M}(f_1' e^{-f_2} + \xi g e^{-f_1} P_I^3 e^I) + i \xi g P_I^3 \bar{L}^I + i e^{-2 f_3} \mathcal{I}_{IJ} \bar{L}^I n^{+J}~, \\
	0 &= M f_3' e^{-f_2} - i \xi g P_I^3 L^I + i e^{-2 f_3} \mathcal{I}_{IJ}L^I n^{-J}~, \\
	0 &= \widetilde{M} f_3' e^{-f_2} + i \xi g P_I^3 \bar{L}^I - i e^{-2f_3} \mathcal{I}_{IJ}\bar{L}^I n^{+J}~, \\
	0 &= \widetilde{M} e^{-f_2} \partial_r z^\alpha - i g^{\alpha \bar{\beta}} f_{\bar{\beta}}^I(e^{-2 f_3} \mathcal{I}_{IJ} n^{-J} + \xi g P_I^3)~, \\
	0 &= M e^{-f_2} \partial_r \tilde{z}^{\bar{\alpha}} + i g^{\bar{\alpha} {\beta}} f_{{\beta}}^I(e^{-2 f_3} \mathcal{I}_{IJ} n^{+J} + \xi g P_I^3)~.
\end{aligned}\end{equation}
Note that the terms involving the electric parameters $e^I$ show up with opposite signs in the BPS and conjugate BPS conditions, as a consequence of how Wick rotation affects the gauge field along the time direction.

In Lorentzian signature, the Maxwell equations relate the electric charges $q_I$ to the magnetic charges $p^I$ and electric parameters $e^I$ as:
\begin{equation}
	q_I = e^{-f_1-f_2+2f_3} \mathcal{I}_{IJ} \partial_r e^J + \mathcal{R}_{IJ} p^J~.
\end{equation}
In Euclidean signature, the Wick-rotation $x^0 \to - i x^4$ changes these charge relations.  In particular, it sends $e^I \to i e^I$ and $q_I \to i q_I$.  The electric charges in Euclidean signature are thus defined as follows:
\begin{equation}
	q_I = e^{-f_1-f_2+2f_3} \mathcal{I}_{IJ} \partial_r e^J - i \mathcal{R}_{IJ} p^J~.
\end{equation}
This is also precisely what can be obtained directly from the Euclidean Maxwell equations.  We then conclude that the Euclidean electric charges are related to the quantities $n^{\pm I}$ in \eqref{eq:npmdef} as follows: 
\begin{equation}\begin{aligned}
	\mathcal{I}_{IJ} n^{+J} &= q_I + i \bar{\mathcal{N}}_{IJ} p^J~, \\
	\mathcal{I}_{IJ} n^{-J} &= - q_I - i \mathcal{N}_{IJ} p^J~.
\end{aligned}\end{equation}
Using the identity $\mathcal{N}_{IJ} L^J = M_I$, we also find that
\begin{equation}\begin{aligned}
	\mathcal{I}_{IJ} \bar{L}^I n^{+J} &= \bar{L}^I q_I  + i\bar{M}_I p^I ~, \\
	\mathcal{I}_{IJ} L^I n^{-J} &= - L^I q_I  - i M_I  p^I~.
\end{aligned}\end{equation}
In terms of the proper Euclidean electric charges, the BPS conditions can now be written as:
\begin{equation}\begin{aligned}
	0 &= Mf_1' e^{-f_2} - M \xi g e^{-f_1} P_I^3 e^I - i \xi g P_I^3 L^I + e^{-2 f_3} (i L^I q_I - M_I p^I)~, \\
	0 &= \widetilde{M} f_1' e^{-f_2} + \widetilde{M} \xi g e^{-f_1} P_I^3 e^I + i \xi g P_I^3 \bar{L}^I + e^{-2 f_3} (i \bar{L}^I q_I - \bar{M}_I p^I)~, \\
	0 &= M f_3' e^{-f_2} - i \xi g P_I^3 L^I - e^{-2 f_3} (i L^I q_I - M_I p^I)~, \\
	0 &= \widetilde{M} f_3' e^{-f_2} + i \xi g P_I^3 \bar{L}^I - e^{-2f_3} (i \bar{L}^I q_I - \bar{M}_I p^I)~, \\
	0 &= \widetilde{M} e^{-f_2} \partial_r z^\alpha + g^{\alpha \bar{\beta}} f_{\bar{\beta}}^I \left(e^{-2 f_3} (i q_I - \mathcal{N}_{IJ} p^J) - i \xi g P_I^3 \right)~, \\
	0 &= M e^{-f_2} \partial_r \tilde{z}^{\bar{\alpha}} + g^{\bar{\alpha} {\beta}} f_{{\beta}}^I\left(e^{-2 f_3} (i q_I - \bar{\mathcal{N}}_{IJ}p^J) + i \xi g P_I^3\right)~.
\end{aligned}\end{equation}
If we further use the identity
\begin{equation}
	\nabla_\alpha M_J = \nabla_\alpha\left(L^I \mathcal{N}_{IJ}\right) = \left(\nabla_\alpha L^I\right)\bar{\mathcal{N}}_{IJ} = f_{\alpha}^I \bar{\mathcal{N}}_{IJ}~,
\end{equation}
we can finally write the BPS equations as follows:
\begin{equation}\begin{aligned}
	0 &= Mf_1' e^{-f_2} - M \xi g e^{-f_1} P_I^3 e^I - i \xi g P_I^3 L^I + e^{-2 f_3} (i L^I q_I - M_I p^I)~, \\
	0 &= \widetilde{M} f_1' e^{-f_2} + \widetilde{M} \xi g e^{-f_1} P_I^3 e^I + i \xi g P_I^3 \bar{L}^I + e^{-2 f_3} (i \bar{L}^I q_I - \bar{M}_I p^I)~, \\
	0 &= M f_3' e^{-f_2} - i \xi g P_I^3 L^I - e^{-2 f_3} (i L^I q_I - M_I p^I)~, \\
	0 &= \widetilde{M} f_3' e^{-f_2} + i \xi g P_I^3 \bar{L}^I - e^{-2f_3} (i \bar{L}^I q_I - \bar{M}_I p^I)~, \\
	0 &= \widetilde{M} e^{-f_2} \partial_r z^\alpha + g^{\alpha \bar{\beta}} \nabla_{\bar{\beta}} \left(e^{-2 f_3} (i \bar{L}^I q_I - \bar{M}_I p^I) - i \xi g P_I^3 \bar{L}^I \right)~, \\
	0 &= M e^{-f_2} \partial_r \tilde{z}^{\bar{\alpha}} + g^{\bar{\alpha} {\beta}} \nabla_{\beta} \left(e^{-2 f_3} (i L^I q_I - M_I p^I) + i \xi g P_I^3 L^I\right)~.
\end{aligned}\end{equation}
By defining $g_I \equiv \xi g P_I^3$ and rewriting the holomorphic sections as $L^I = e^{\mathcal{K}/2} X^I$ and $M_I = e^{\mathcal{K}/2} F_I$, these equations become precisely the BPS conditions we used throughout the main body of the paper, as given in (\ref{eq:BPS}).

\section{Deriving the on-shell action}
\label{app:moreholorenorm}

An important ingredient in the calculation of the on-shell action of the Euclidean black saddle solutions of the form \eqref{eq:ansatz} in Section~\ref{sec:holorenorm} is the fact that the on-shell action (\ref{eq:sbulk}) is a total derivative for any solution to the bulk equations of motion.  This feature allows to derive an expression for the on-shell action that only relies on solving the BPS equations perturbatively around the UV and IR regions of the spacetime.  In this appendix, we prove explicitly that the on-shell action takes the total derivative form presented in~\eqref{eq:sbulk}.

First, let us repeat some of the expressions from the main body of the text for clarity.  The Euclidean action~\eqref{eq:SEucl} that we are interested in evaluating on-shell is
\begin{equation}
	S = \frac{1}{8\pi G_N}\int d^4x\,\sqrt{g}\,\left[ - \frac{1}{2}R - \frac{1}{4} \mathcal{I}_{IJ} F^I_{\mu\nu}F^{J\mu\nu} + \frac{i}{4} \mathcal{R}_{IJ} F^I_{\mu\nu} \tilde{F}^{J\mu\nu} + g_{\alpha \bar{\beta}} \nabla^\mu z^\alpha \nabla_\mu \tilde{z}^{\bar{\beta}} + g^2 \mathcal{P}\right]~,
\label{eq:SEucl2}
\end{equation}
while the ansatz in~\eqref{eq:ansatz} is
\begin{equation}\begin{aligned}\label{eq:ansatz2}
	ds^2 &= e^{2 f_1(r)} d\tau^2 + e^{2 f_2(r)} dr^2 + e^{2 f_3(r)} ds_{\sigg}^2~, \\
	A^I &= e^I(r) d\tau + p^I \omega_{\sigg}~, \\
	z^\alpha &= z^\alpha(r)~, \\
	\tilde{z}^{{\alpha}} &= \tilde{z}^{{\alpha}}(r)~.
\end{aligned}\end{equation}
This is the most general ansatz we can write down that is compatible with the symmetries of the dual topologically twisted ABJM theory, and so it should be sufficient to work within this ansatz when computing the topologically twisted index holographically.

Since the metric functions $f_{1,2,3}$, the scalars $z^\alpha$ and $\tilde{z}^\alpha$, and the Wilson lines $e^I$ are all functions only of the radial coordinate $r$, the integration over the $\tau$ coordinate and the Riemann surface coordinates $x$ and $y$ is trivial. The evaluation of the on-shell action therefore reduces to a one-dimensional integral over $r$.  To simplify the integrand, we first note that the Ricci scalar on our ansatz evaluates to
\begin{equation}
	R = 2 e^{-2 f_3} \kappa - 2 e^{-2 f_2}\left( (f_1')^2-f_1' f_2' + 2f_1' f_3' - 2 f_2' f_3' + 3 (f_3')^2 + f_1'' + 2 f_3''\right)~, \\
\end{equation}
while the vector field kinetic term evaluates to
\begin{equation}
	- \frac{1}{4}\mathcal{I}_{IJ} F_{\mu\nu}^I F^{J\mu\nu} - \frac{1}{4i} \mathcal{R}_{IJ} F^I_{\mu\nu} \tilde{F}^{J\mu\nu}  = -e^{-4 f_3} V_\text{BH} - e^{-f_1 - f_2 - 2 f_3} q_I \partial_r e^I~,
\end{equation}
where the function $V_\text{BH}$ is defined by
\begin{equation}
	V_\text{BH} \equiv \frac{1}{2}\begin{pmatrix}
		p^I & q_I
	\end{pmatrix}
	\begin{pmatrix}
		\mathcal{I}_{IJ} + \mathcal{R}_{IK} \left(\mathcal{I}^{-1}\right)^{KL}\mathcal{R}_{LJ}\,\, & -i \mathcal{R}_{IK} \left(\mathcal{I}^{-1}\right)^{KJ} \\
		-i \left(\mathcal{I}^{-1}\right)^{IK} \mathcal{R}_{KJ} & -\left(\mathcal{I}^{-1}\right)^{IJ}
	\end{pmatrix}
	\begin{pmatrix}
		p^J \\
		q_J
	\end{pmatrix}~,
\end{equation}
and should be thought of as the Euclidean analogue of the usual black hole potential that is extremized in the black hole attractor mechanism~\cite{Ferrara:1997tw}.  The Euclidean bulk action evaluated on our ansatz therefore reduces to
\begin{equation}\begin{aligned}
	S_\text{bulk} &= \frac{\text{Vol}(\Sigma_{\mathfrak{g}})\beta_\tau}{8\pi G_N}\int dr\,\bigg{[} - e^{f_1 + f_2}\kappa - q_I \partial_r e^I\\
	&\quad + e^{f_1 - f_2 + 2 f_3}\left( (f_1')^2-f_1' f_2' + 2f_1' f_3' - 2 f_2' f_3' + 3 (f_3')^2 + f_1'' + 2 f_3''\right) \\ 
	&\quad + e^{f_1 + f_2 + 2 f_3}\left( - e^{-4 f_3} V_\text{BH} + e^{-2 f_2} g_{\alpha \bar{\beta}} \partial_r z^\alpha \partial_r \tilde{z}^{\bar{\beta}} + g^2 \mathcal{P}\right) \bigg{]}~,
\label{eq:sbulkeapp}
\end{aligned}\end{equation}
where $\beta_\tau$ denotes the periodicity of the coordinate $\tau$ and $\text{Vol}(\Sigma_{\mathfrak{g}})$ is the volume of the Riemann surface.

Now, we bring the expression in \eqref{eq:sbulkeapp} on-shell by using the equations of motion.  In particular, if we evaluate the full Einstein equation on our ansatz, we can solve for all quantities that involve the scalars purely in terms of metric functions.  To do this, we first write the Einstein equation as
\begin{equation}
	R_{\mu\nu} = \hat{T}_{\mu\nu}~,
\end{equation}
where the trace-reversed energy-momentum tensor $\hat{T}_{\mu\nu} = T_{\mu\nu} - \frac{1}{2} g_{\mu\nu} T\ind{_\rho^\rho}$ of the STU model is given by
\begin{equation}
	\hat{T}_{\mu\nu} = - \mathcal{I}_{IJ}\left(F^I_{\mu\rho}F\ind{^J_\nu^\rho} - \frac{1}{4} g_{\mu\nu} F^I_{\rho\sigma}F^{J\rho\sigma}\right) + 2 g_{\alpha \bar{\beta}} \nabla_\mu z^\alpha \nabla_\nu \tilde{z}^{\bar{\beta}} + g_{\mu\nu} g^2 \mathcal{P}~.
\end{equation}
Evaluating this on our ansatz, we find the following non-zero components:
\begin{equation}\begin{aligned}
	\hat{T}_{\tau\tau} &= e^{2 f_1 - 4 f_3} V_\text{BH}  + e^{2 f_1} g^2 \mathcal{P}~, \\
	\hat{T}_{rr} &= e^{2 f_2 - 4 f_3} V_\text{BH} + 2 g_{\alpha \bar{\beta}} \partial_r z^\alpha \partial_r \tilde{z}^{\bar{\beta}} + e^{2 f_2} g^2 \mathcal{P}~, \\
	\hat{T}_{xx} &= \hat{T}_{yy} = H(x,y)^2\left( e^{-2 f_3} V_\text{BH} + e^{2 f_3} g^2 \mathcal{P}\right)~,
\end{aligned}\end{equation}
where $H(x,y)$ is defined in \eqref{eq:metappH}. There are therefore three independent components of the Einstein equation.  This gives us exactly enough equations to solve for $V_\text{BH}$, $g_{\alpha \bar{\beta}} \partial_r z^\alpha \partial_r \tilde{z}^{\bar{\beta}}$, and $\mathcal{P}$ in terms of metric functions.  The result is:
\begin{equation}\begin{aligned}
	V_\text{BH} &= - \frac{1}{2}e^{2 f_3} \kappa - \frac{1}{2}e^{-2 f_2 + 4 f_3}\left( (f_1' - f_3')(f_1' - f_2' + 2 f_3') + f_1'' - f_3''\right)~, \\
	g_{\alpha \bar{\beta}} \partial_r z^\alpha \partial_r \tilde{z}^{\bar{\beta}} &= (f_1'+f_2'- f_3')f_3' - f_3''~,\\
	g^2 \mathcal{P} &= \frac{1}{2} e^{-2 f_3}\kappa - \frac{1}{2} e^{-2 f_2}\left( (f_1' + f_3')(f_1' - f_2' + 2 f_3') + f_1'' + f_3''\right)~.
\end{aligned}\end{equation}
Plugging these relations back into the Euclidean supergravity action (\ref{eq:sbulkeapp}), we find that it reduces drastically to the following:
\begin{equation}\begin{aligned}
	S_\text{bulk} &= \frac{\text{Vol}(\Sigma_{\mathfrak{g}})\beta_\tau}{8\pi G_N} \int dr\,\left[e^{f_1 - f_2 + 2 f_3}\left( f_1' (f_1 - f_2' + 2 f_3') + f_1''\right) - q_I \partial_r e^I\right] \\
	&= \frac{\text{Vol}(\Sigma_{\mathfrak{g}})\beta_\tau}{8\pi G_N}  \int dr\, \left[ e^{f_1 - f_2 + 2 f_3} f_1' - q_I e^I \right]'~,
\end{aligned}\end{equation}
where in the last line we used that the charges $q_I$ are conserved, due to the Maxwell equations.  Therefore we arrive precisely at the all-important relation in (\ref{eq:sbulk}).

\bibliography{ads4_bib}
\bibliographystyle{JHEP}

\end{document}